\documentclass[pre,aps,10pt]{revtex4}

\usepackage{times}
\usepackage{amsmath}
\usepackage{graphicx}
\usepackage{amssymb}
\usepackage{color}
\usepackage{amsthm}
\usepackage{epstopdf}

\newcommand{\beq}{\begin{equation}}
\newcommand{\eeq}{\end{equation}}
\newcommand{\e}{\mathrm{e}}

\newtheorem{remark}{Remark}
\newtheorem{theorem}{\bf Theorem}[section]
\newtheorem{lemma}{\bf Lemma}[section]
\newtheorem{proposition}{\bf Proposition}[section]

\begin{document}

\title{An energy-based stability criterion for solitary traveling waves in Hamiltonian lattices}

\author{%%%% Author details
Haitao Xu$^{1,2}$, Jes\'us Cuevas--Maraver$^{3,4}$, Panayotis G. Kevrekidis$^{5}$ and Anna Vainchtein$^{6}$}

%%%%%%%%% Insert author address here
\address{$^{1}$Institute for Mathematics and its Applications, University of
  Minnesota, Minneapolis, MN 55455 USA\\
$^{2}$Center for Mathematical Science, Huazhong University of Science and Technology, Wuhan, Hubei 430074, China\\
$^{3}$Grupo de F\'{\i}sica No Lineal, Departamento de F\'{\i}sica Aplicada I, Universidad de Sevilla. Escuela Polit{\'e}cnica Superior, C/ Virgen de \'Africa, 7, 41011-Sevilla, Spain\\
$^{4}$Instituto de Matem\'aticas de la Universidad de Sevilla (IMUS), Edificio Celestino Mutis. Avda. Reina Mercedes s/n, 41012-Sevilla, Spain\\
$^{5}$Department of Mathematics and Statistics, University of
Massachusetts, Amherst, MA 01003-9305, USA\\
$^{6}$Department of Mathematics, University of Pittsburgh, Pittsburgh, PA 15260, USA}

%%%% Subject entries to be placed here %%%%
%\subject{applied mathematics, analysis}

%%%% Keyword entries to be placed here %%%%
%\keywords{solitary traveling wave, Hamiltonian lattice, stability}

%%%% Insert corresponding author and its email address}
%\corres{Haitao Xu\\
%\email{pocketxumk3t@gmail.com}}
\begin{abstract}
  In {this} work, we revisit a criterion, originally proposed
  in [Nonlinearity {\bf 17}, 207 (2004)], for the stability of
  {solitary} traveling waves in Hamiltonian, infinite-dimensional
  lattice dynamical systems. We discuss the implications of
  this criterion from the point of view of stability theory, both
  at the level of the spectral analysis of the advance-delay
  differential equations {in} the co-traveling frame, as well as
  at that of the Floquet problem arising when considering the
  traveling wave as a periodic orbit modulo a shift. We establish
  the correspondence of these perspectives {for} the pertinent
  eigenvalue {and} Floquet multiplier  and provide explicit expressions
  for their dependence on the {velocity of the traveling wave} in the vicinity of the critical point.
  Numerical results are used to corroborate the relevant predictions in
  {two different} models, where the stability
  may change {twice. Some} extensions, generalizations and
  future directions of {this investigation} are {also discussed}.
\end{abstract}

\maketitle

\section{Introduction}
The interplay of dispersion and nonlinearity in many spatially discrete physical and biological systems often gives rise to solitary traveling waves \cite{toda, DEGM84, Remoissenet96, Scott03, DauxoisPeyrard10, Ablowitz11}. In particular, such waves were experimentally observed in granular materials \cite{CFF97,nesterenko,sen,review}, and
electrical transmission lines \cite{HirotaSuzuki73,KMR88,Remoissenet96}, among others.
%and optical fibers \cite{MSG80,WHK88}.
Following their discovery \cite{ZK65} in the Fermi-Pasta-Ulam (FPU) system \cite{FPU55,Campbell05, Gallavotti07}, which revolutionized  nonlinear science, there has been an enormous amount of literature on solitary traveling waves in Hamiltonian lattices. {Numerous} significant developments include
the discovery of the integrable Toda lattice \cite{toda}, existence results \cite{FW94,SW97,PR11}, fundamental studies of the low-energy near-sonic regime \cite{pegof99,pegof02,pegof04a,pegof04b,McMillan02,Iooss00}, where the waves are well described by the Korteweg-de Vries (KdV) equation, as well as the development of numerical \cite{eilbeck,mertens,english} and quasicontinuum \cite{Collins81,Nesterenko83,Rosenau86,Wattis93,Rosenau03} approximations of such solutions.

An important issue that then naturally arises concerns establishing the necessary and sufficient conditions for the stability of solitary traveling waves. Stability results have been obtained in some special cases such as Toda lattice \cite{mizu,wayne} or the near-sonic FPU problem \cite{pegof99,pegof02,pegof04a,pegof04b}, where the system's integrability or near-integrability make certain specialized techniques available. However, obtaining rigorous stability criteria in the general case of a non-integrable Hamiltonian lattice away from the KdV limit remains a challenge. While numerical simulations in many systems suggest stability of solitary traveling waves, some recent studies \cite{mertens1,mertens2,at} demonstrate the possibility of unstable waves when the energy (Hamiltonian) $H$ of the wave decreases with its velocity $c$.

In a recent paper~\cite{ourTW},  inspired by and extending the fundamental
work of~\cite{pegof04a}, we examined the sufficient (but not necessary) condition for a change in the wave's spectral stability occurring when the function $H(c)$ changes its monotonicity. More specifically, we showed that when $H'(c_0)=0$ for some critical velocity $c_0$, a pair of eigenvalues associated with the traveling wave (in the co-traveling frame of reference in which the
wave is steady) cross zero and emerge on the real axis when $c$ is either above or below $c_0$, thus creating instability. While this energy-based criterion first appeared in \cite{pegof04a}, where it was motivated by the study of the FPU problem in the near-sonic limit, in \cite{ourTW} we provided a concise proof as well as an explicit leading-order calculation for these two near-zero eigenvalues. In addition, we tested the criterion numerically for a range of lattice problems. This was accomplished by using two complementary approaches. The first approach involves the analysis of the spectrum of the linearized operator associated with traveling waves as stationary solutions of the corresponding advance-delay partial differential equation in the co-traveling frame. The second approach is based on representing lattice traveling waves as periodic orbits modulo shift and computing the corresponding Floquet multipliers. When the wave becomes unstable, a Floquet multiplier exits into the real line from the unit circle. Importantly, through this computation, we identified the potentially unstable mode and characterized the growth rate of the instability when the mode becomes unstable. The examples in \cite{ourTW} illustrate that the results of the two numerical approaches match and agree with the analytical predictions. It should also be noted that the Floquet multiplier approach connects the established energy-based criterion to the one recently obtained for discrete breathers \cite{dmp}.

In this paper we present a significantly more detailed, systematic description of the analysis leading to the results in \cite{ourTW} and consider several non-trivial extensions, {including an equivalent algorithm for examining the energy criterion through the Floquet multiplier approach,
  an examination of the higher-dimensional generalized kernel of the linearized operator and the consideration of {non-generic but
  mathematically interesting} case when the first and
  second derivatives of the energy function vanish simultaneously. Moreover,
  we discuss in detail {how the} near-zero eigenvalues in the spectral problem {depend to the leading order on the velocity of the solitary traveling wave near the critical point} and investigate the corresponding conditions, through which we show the connection between the eigenvalues and the symmetries of the system.}

We then illustrate these findings numerically by considering two examples, both of which involve a non-monotone $H(c)$, with $H'(c)<0$ associated with unstable solitary traveling waves. One of the examples involves the $\alpha$-FPU lattice with exponentially decaying long-range Kac-Baker interactions \cite{mertens2,Neuper94,Gaididei95,Gaididei97} and was also studied in~\cite{ourTW} for different parameter values. The other example concerns a smooth regularization of the FPU problem with piecewise quadratic potential \cite{at}. Unlike the original problem, for which our analysis framework is not available due to the lack of regularity, the regularized model allows us to compute the eigenvalues and Floquet multipliers. Both examples demonstrate the power of the energy-based criterion, an excellent matching of the results obtained by the two complementary approaches and a very good overall agreement with the theoretical predictions. In addition, the regularized model helps explain the numerical stability results
found earlier~\cite{at,ourTW} for the problem with biquadratic potential.

The paper is organized as follows. In Sec~\ref{sec:prelim}, we formulate the problem, introduce the function spaces in which we will work,
and discuss some relevant properties of the Hamiltonian lattice
dynamical system. The energy-based criterion is derived in Sec.~\ref{sec:stab} by considering the eigenvalue problem for the linearized operator, while the alternative perspective based on the Floquet analysis is described in Sec.~\ref{sec:Floquet}. Detailed asymptotic analysis of the eigenvalues splitting from zero near the stability threshold is provided in Sec.~\ref{sec:split}. Sec.~\ref{sec:examples} illustrates and verifies the results numerically via two examples. We discuss the results and identify future challenges in Sec.~\ref{sec:conclude}. Results of more technical nature, numerical procedures used in the computations and discussion of the connection to other energy-based criteria are included in the Supplementary Material.

\section{Problem formulation and preliminaries}
\label{sec:prelim}
We consider a one-dimensional Hamiltonian lattice with infinite-dimensional vectors $q=(...,q_{-1},q_0,q_1,...)^T$ and $p=(...,p_{-1},p_0,p_1,...)^T$ denoting its generalized coordinates and momenta, respectively. The dynamics of the lattice is governed by
\begin{equation}
\label{eq0}
\frac{d}{dt}{x}(t)=\mathcal{A}({x}(t))=J \frac{\partial\mathcal{H}}{\partial x},
\qquad x(t)=\left(\begin{array}{c}
  {q(t)} \\
  {p(t)}
\end{array} \right),
\qquad J=\left(\begin{array}{cc}
0 & I \\
-I & 0
\end{array} \right)
\end{equation}
where $\mathcal{H}$ is the Hamiltonian energy density of the system, $I$ is the identity matrix, and $J$ is skew-symmetric. We assume that the lattice is homogeneous, i.e., \eqref{eq0} is invariant under integer shifts. To make it mathematically precise, we introduce the shift operators
\[
S_{\pm}y_j=y_{j\pm1}, \quad
\bar{S}_{\pm}y=(..., S_{\pm}y_{-1}, S_{\pm}y_0, S_{\pm}y_1,...)^T, \qquad
\tilde{S}_{\pm}\left(\begin{array}{c}
  {y} \\
  {z}
\end{array} \right)=\left(\begin{array}{c}
  {\bar{S}_{\pm}y} \\
  {\bar{S}_{\pm}z}
\end{array} \right),
\]
where $y=(...,y_{-1},y_0,y_1,...)^T$, $z=(...,z_{-1},z_0,z_1,...)^T$. We are interested in the systems that satisfy
\[
\mathcal{A}(\tilde{S}_{\pm}x)=\tilde{S}_{\pm}\mathcal{A}(x) .
\]
If $\mathcal{A}$ is at least $C^1$, differentiation yields that
\beq
D\mathcal{A}(\tilde{S}_{\pm}x)=\tilde{S}_{\pm}D\mathcal{A}(x) \tilde{S}_{\mp},
\label{eq:DA_prop}
\eeq
where $D$ is the differential operator. If the Hamiltonian density depends on displacements only through strain variables $r=(\bar{S}_+-1)q$ ($r_j=q_{j+1}-q_j$), the problem can be alternatively formulated
as
\beq
\label{eq1}
\frac{d}{dt}{y}(t)=J_1 \frac{\partial\mathcal{H}}{\partial y}, \quad y(t)=\left(\begin{array}{c}
  {r(t)} \\
  {p(t)}
\end{array} \right), \quad J_1=\left(\begin{array}{cc}
0 & \bar{S}_+-I \\
I-\bar{S}_- & 0
\end{array} \right),
\eeq
where $J_1$ is skew-symmetric. Below we focus on the formulation \eqref{eq0}, but our arguments can also be applied to \eqref{eq1}.

 We now assume that the system~(\ref{eq0}) has a family of solitary traveling wave solutions $x_{tw}(t;c)$ parameterized by nonzero velocity $c$. These are localized solutions of the form
 \beq
 x_{tw}(t;c)=\left(\begin{array}{c}
  {q_{tw}(t;c)} \\
  {p_{tw}(t;c)}
\end{array} \right), \quad q_{tw,j}(t;c)=\hat{q}(j-ct), \quad p_{tw,j}(t;c)=\hat{p}(j-ct),
\label{eq:TWansatz}
\eeq
with
\beq
\lim_{j\to \pm \infty} q_{tw,j}(t;c)=q_{tw,\pm\infty}(c), \quad \lim_{j\to \pm \infty} p_{tw,j}(t;c)=0
\label{eq:BCs_gen}
\eeq
for any fixed $t$ and $c$. In what follows, we assume for simplicity that $q_{tw,\pm\infty}(c)=0$; note, however, that if either one of these values is nonzero (a kink-type solution), the strain formulation \eqref{eq1} can be considered instead, with the (similarly defined) asymptotics $r_{tw,\pm\infty}(c)=0$.

Note that the traveling wave solutions that we
consider~\eqref{eq:TWansatz} satisfy
\[
x_{tw}(t+\frac{1}{c};c)=\tilde{S}_{-}x_{tw}(t;c).
\]
As a result, in order to fully define a traveling wave $x_{tw,j}(t;c)$ for all sites $j\in\mathbb{Z}$ and all time $t\in\mathbb{R}$, it is sufficient to specify it for one site $j=j_0$ and all time $t\in\mathbb{R}$, or alternatively for all sites $j\in\mathbb{Z}$ and one period $t\in [t_0, t_0+\frac{1}{c})$.

%and thus can be fully identified for all times $t$ if either $x_{tw}([0,\frac{1}{c});c)$ is known or $x_{tw,j}(\mathbb{R};c)$ is %known for any single $j\in\mathbb{Z}$.
In other words, a traveling wave solution is \emph{periodic modulo shift} with period $T=1/c$. With this in mind, it is convenient to consider the map $\mathcal{P}_T: \left(\begin{array}{c}
  {q} \\
  {p}
\end{array} \right) \mapsto  \left(\begin{array}{c}
  {\phi} \\
  {\psi}
\end{array} \right)$, where $q$ and $p$ are infinite-dimensional functions defined on $[0,T)$ while $\phi$ and $\psi$ are functions from $\mathbb{R}$ to $\mathbb{R}$, with $\phi({j}T+t)=q_j(t)$ and $\psi({j}T+t)=p_j(t)$ for any $j\in\mathbb{Z}$ and $t\in [0,T)$. In particular, we have
$\mathcal{P}_{\frac{1}{c}} x_{tw}([0,\frac{1}{c}))=\left(\begin{array}{c}
  {q_{tw,0}}(\mathbb{R}) \\
  {p_{tw,0}}(\mathbb{R})
\end{array} \right)=x_{tw,0}(\mathbb{R})$. Similarly, we define another map, $\tilde{\mathcal{P}}_T:  \left(\begin{array}{c}
  {\phi(t)} \\
  {\psi(t)}
\end{array} \right) \mapsto  \left(\begin{array}{c}
  {q(t)} \\
  {p(t)}
\end{array} \right)$, where $\phi$ and $\psi$ are functions from $\mathbb{R}$ to $\mathbb{R}$, while $q_j(t)=e^{jT\partial t}\phi(t)=\phi(jT+t)$ and $p_j(t)=e^{jT\partial t}\psi(t)=\psi({j}T+t)$ for $j\in\mathbb{Z}$ and $t\in [0,T)$. Here $e^{s\partial_{\tau}}$ is the translation operator such that $(e^{s\partial_{\tau}}\Gamma)(\tau_0)=\Gamma(\tau_0+s)$. It is easy to see that $\tilde{\mathcal{P}}_T=({\mathcal{P}}_T)^{-1}$ and $\tilde{\mathcal{P}}_{1/c}(x_{tw,0}(\mathbb{R}))=x_{tw}([0,\frac{1}{c}))$.

Since our solutions are spatially localized, we can consider a finite-energy space such as
\[
D^1([0,T]):=\{\left(\begin{array}{c}
  {q(t)} \\
  {p(t)}
\end{array} \right),\; t\in [0, T]\; |\; \int_0^T \sum_{j\in\mathbb{Z}} ( |q_j(t)|^2+|p_j(t)|^2+|q'_j(t)|^2+|p'_j(t)|^2 ) dt < \infty \}.
\]
One can check that $\mathcal{P}_T$ maps $D^1([0,T])$ to
\[
H^1(\mathbb{R},\mathbb{R}^2):=\{\left(\begin{array}{c}
  {\phi(t)} \\
  {\psi(t)}
\end{array} \right),\; t\in\mathbb{R}\; |\; \int_{\mathbb{R}} ( |\phi(t)|^2+|\psi(t)|^2+|\phi'(t)|^2+|\psi'(t)|^2 ) dt < \infty \}.
\]
For the solutions that are exponentially localized in space, one can consider weighted spaces such as
\[
D^1_a([0,T]):=\{\left(\begin{array}{c}
  {q(t)} \\
  {p(t)}
\end{array} \right),\; t\in [0, T]\; |\; \int_0^T \sum_{j\in\mathbb{Z}} ( |q_j(t)|^2+|p_j(t)|^2+|q'_j(t)|^2+|p'_j(t)|^2 )e^{2at} dt < \infty \},
\]
which will be mapped to
\[
H^1_a(\mathbb{R},\mathbb{R}^2):=\{\left(\begin{array}{c}
  {\phi(t)} \\
  {\psi(t)}
\end{array} \right),\; t\in\mathbb{R}\; | \;\int_{\mathbb{R}} ( |\phi(t)|^2+|\psi(t)|^2+|\phi'(t)|^2+|\psi'(t)|^2 )e^{2at} dt < \infty \}
\]
by $\mathcal{P}_T$.

\section{Energy-based criterion for spectral stability of solitary traveling waves}
\label{sec:stab}
In what follows we assume that $x_{tw}(t;c)$ is smooth in $c$ and define
\[
\tau=c t, \quad X(\tau)=x(t).
\]
In view of \eqref{eq0}, $X(\tau)$ satisfies
\begin{equation}
\label{eq2}
\frac{d}{d\tau}{X}(\tau)=\frac{1}{c}\mathcal{A}({X}(\tau))=\frac{1}{c} J \frac{\partial\mathcal{H}}{\partial X}.
\end{equation}
Note that for given $c$,
\[
X_{tw}(\tau;c)=\left(\begin{array}{c}
  {Q_{tw}(\tau;c)} \\
  {P_{tw}(\tau;c)}
\end{array} \right)=x_{tw}(t;c)
\]
and its derivatives $\partial_{\tau} (X_{tw}(\tau;c))$ and $\partial_{c} (X_{tw}(\tau;c))$ are traveling waves with velocity $1$, or, equivalently, these functions are periodic in $\tau$ modulo shift with period $1$. Here we assume $\lim_{j\to\pm\infty}\partial_{c}^k (X_{tw,j}(\tau;c))=0$ for $k=1,2,...,p$ (usually $p=2$ or $p=3$ is sufficient) such that $\partial_{c}^k (X_{tw,j}(\tau;c))$ for small $k$ are still localized.
Observe also that $\Theta=\mathcal{P}_1 X_{tw}$ {(where $\mathcal{P}_1=\mathcal{P}_T|_{T=1}$)}
satisfies
\begin{equation}
\label{eq3}
\frac{d}{d\tau}{\Theta}(\tau)=\frac{1}{c}(\mathcal{P}_1)^{-1}\mathcal{A}(\mathcal{P}_1^{-1}{\Theta})=\frac{1}{c}\tilde{\mathcal{A}}({\Theta}).
\end{equation}

To explore spectral stability of a solitary traveling wave solution $X_{tw}$, we linearize \eqref{eq2} around it with $X=X_{tw}+\epsilon Y$, obtaining
\begin{equation}
\label{eq4}
\frac{d}{d\tau}{Y}(\tau)=\frac{1}{c}D\mathcal{A}({X_{tw}}(\tau))Y(\tau)=\frac{1}{c} J \frac{\partial^2\mathcal{H}}{\partial X^2}\bigg|_{X=X_{tw}}Y(\tau).
\end{equation}
We consider perturbations in the co-traveling frame of the form $Y(\tau)=e^{\lambda \tau}Z_{tw}(\tau)$, where $Z_{tw}$ is a traveling wave with velocity $1$, i.e., $Z_{tw}(\tau)=\tilde{S}_+ Z_{tw}(\tau+1)$. This yields the eigenvalue problem for the corresponding linear operator $\mathcal{L}$:
\begin{equation}
\label{eq5}
\lambda Z_{tw}=\mathcal{L} Z_{tw}, \qquad \mathcal{L}=\frac{1}{c}D\mathcal{A}({X_{tw}})-\frac{d}{d\tau}= \frac{1}{c} J \frac{\partial^2\mathcal{H}}{\partial X^2}\bigg|_{X=X_{tw}}-\frac{d}{d\tau}.
\end{equation}
Note that existence of an eigenvalue with ${\rm Re}(\lambda) > 0$ corresponds to instability.
Equivalently, one can linearize \eqref{eq3} around $\Theta$ using perturbation term $e^{\lambda \tau}\Gamma(\tau)$ to obtain
\begin{equation}
\label{eq6}
\lambda \Gamma= \tilde{\mathcal{L}}\Gamma, \qquad \tilde{\mathcal{L}}=\frac{1}{c}D\tilde{\mathcal{A}}(\Theta)-\frac{d}{d\tau}
\end{equation}
{{as} an alternative representation {of the eigenvalue problem}.
%in $H^1(\mathbb{R}, \mathbb{R}^2)$ or $L^2(\mathbb{R}, \mathbb{R}^2)$}.

We now discuss the choice of the function space for the eigenvalue problem \eqref{eq5} (or, equivalently, \eqref{eq6}). If we consider $\mathcal{L}$ as a closed operator in
\[
D^0([0,1]):=\{ x(t)=\left(\begin{array}{c}
  {q(t)} \\
  {p(t)}
\end{array} \right),\; t\in [0, 1]\; |\; \int_0^1 \sum_{j\in\mathbb{Z}} ( |q_j(t)|^2+|p_j(t)|^2) dt < \infty \},
\]
with its domain being $D^1([0,1])$ (which corresponds to $\tilde{\mathcal{L}}$ with domain in $H^1(\mathbb{R}, \mathbb{R}^2)$), the essential spectrum of $\mathcal{L}$ usually contains the imaginary axis (see e.g. the FPU case~\cite{pegof04a}) and the zero eigenvalue (whose existence is discussed below) is of course embedded in it. Since in this work we are interested in the motion of the zero eigenvalue and its connection to the stability, we can either obtain numerical insight on the
(potential) effect of the essential spectrum from spectral computations in the case of a finite domain
or attempt to separate the zero eigenvalue from the essential spectrum by
working with weighted spaces, in order to carry out the relevant calculations.

In what follows, we pursue the first approach. According to the numerical simulations for examples presented in Sec.~\ref{sec:examples}, the contribution from the essential spectrum to the motion of the zero eigenvalue appears to be {\it negligible} at the leading order.
Motivated by this a posteriori justification in what follows, we consider $\mathcal{L}$ in $D^1([0,1])$ and ignore the effects from the essential spectrum. While it is certainly worthwhile to analyze in detail these effects, which may result in higher-order contributions, such analysis is outside of the purview of the present study.

As an alternative strategy, one can move the essential spectrum away from the origin by considering weighted spaces such as $D_a^1([0,1])$. In this case we can carry out the same calculations and obtain the results presented below, provided two important modifications are made.
The first one concerns the inner product $\langle \cdot, \cdot \rangle$ in $D^0([0,1])$.
One can define the function $\mathcal{B}(Z, \tilde{Z})=\langle Z,\tilde{Z} \rangle_{D^0([0,1])}$ that is continuous on $D_a^0([0,1]) \times D_{-a}^0([0,1])$ (since it is usually assumed that $Z, \tilde{Z}\in D_{|a|}^0([0,1])=D_a^0([0,1]) \cap D_{-a}^0([0,1])$) and use this function $\mathcal{B}(\cdot, \cdot)$ to replace the inner products in $D^0([0,1])$. In particular, $\mathcal{B}(Z, J^{-1}Z)=0$ corresponds to the conserved symplectic form~\cite{pegof04a}. As an example showing how $\mathcal{B}$ will be used, the inner product $\langle e_1, J^{-1}e_0 \rangle_{D^0([0,1])}$ appearing in \eqref{eq:solv} and \eqref{eq:crit} below will become $\mathcal{B}( e_1, J^{-1}e_0 )$.
Another issue is that for the same statements to hold in the framework of weighted spaces, our assumptions need to be suitably modified. For instance, our analysis below requires that the traveling wave solution $X_{tw}$, its derivatives $\partial_{\tau}X_{tw}$ and $\partial_c^{k}X_{tw}$ and generalized eigenfunctions of $\mathcal{L}$ for $\lambda=0$ are in $D^0([0,1])$, but the use of weighted spaces requires these functions to be in $D_a^0([0,1])$ (or $D_{|a|}^0([0,1])$). This imposes a strong exponential decay condition, which may not be satisfied by many of the existing solitary wave solutions in lattices. For example, solutions in the example with long-range interactions presented in Sec.~\ref{sec:examples} usually have algebraic decay. For this reason as well as those mentioned above, $\mathcal{L}$ with domain $D^1([0,1])$ is considered in the present work. Similar issues in a slightly different but related setting (a continuum KdV system near a point of soliton stability change) have been brought up in~\cite{raynor}.

We now consider the properties of the linear operator $\mathcal{L}$.
%defined in \eqref{eq5}.
Observe that $\mathcal{L}$ is a densely defined unbounded operator on the Hilbert space $D^0([0,1])$ with its domain being $D^1([0,1])$. Thus the adjoint of $\mathcal{L}$ can be defined, and indeed we find that
\beq
\mathcal{L}^*=-J^{-1} \mathcal{L} J,
\label{eq:adjoint}
\eeq
so that $\mathcal{L}J$ and $J\mathcal{L}$ are self-adjoint (note that $J^{-1}=-J$). Observe also that the spectrum of $\mathcal{L}$ is $2\pi i$-periodic, since $e^{\lambda \tau}Z_{tw}(\tau)=e^{(\lambda+i2n\pi )\tau}(Z_{tw}(\tau)e^{-i2n\pi\tau})$ and $(Z_{tw}(\tau)e^{-i2n\pi\tau})$ is also a traveling wave with velocity $1$. This symmetry is due to the fact that the shift operators $\tilde{S}_{\pm}$ commute with the multiplier $e^{2n\pi\tau}$ for $n\in\mathbb{Z}$. Due to the time-translation invariance of the solution $X_{tw}$, an important property of $\mathcal{L}$ is that its kernel contains
\beq
e_0=\partial_{\tau}X_{tw},
\label{eq:e0}
\eeq
as can be directly verified by differentiating \eqref{eq2} in $\tau$ and evaluating the result at $X=X_{tw}$. Here we assume the generic situation when $\ker(\mathcal{L})={\rm span}\{e_0\}$. Similarly, we have $\ker(\mathcal{L}^*)={\rm span}\{J^{-1}e_0\}$. Next, we observe that $\langle e_0, J^{-1} e_0 \rangle=0$, where $\langle \cdot , \cdot \rangle$ denotes the inner product in $D^0([0,1])$. This implies that $e_0\in(\ker(\mathcal{L}^*))^{\perp}={\rm im}(\mathcal{L})$, and thus there exists $e_1\in D^1([0,1])$ such that $e_0=\mathcal{L}e_1$, so that the algebraic multiplicity of the eigenvalue $\lambda=0$ for $\mathcal{L}$ is at least two. In fact, differentiating \eqref{eq3} in $c$, we obtain $c\mathcal{L}(\partial_{c}X_{tw})=\partial_{\tau}X_{tw}$, so
\beq
e_1=c\partial_c X_{tw}
\label{eq:e1}
\eeq
satisfies $\mathcal{L}e_1=e_0$ and thus represents a generalized eigenvector of $\mathcal{L}$ associated with the zero eigenvalue.

As we have seen, the spectrum of $\mathcal{L}$ always contains at least a double eigenvalue at zero. Following a similar argument, one finds that the algebraic multiplicity of the eigenvalue $\lambda=0$ is higher than two if and only if there exists $e_2\in D^1([0,1])$ such that $\mathcal{L}e_2=e_1$, which is equivalent to
\beq
\langle e_1, J^{-1} e_0 \rangle=0.
\label{eq:solv}
\eeq
We will now prove that this can only happen when the derivative of the $c$-dependent conserved Hamiltonian of the system,
\[
H(c)=\int_0^1  {\cal H}\big|_{X_{tw}(\tau;c)} d \tau,
\]
is zero. Indeed, direct calculation shows that
\beq
\begin{split}
0 &= \langle e_1, J^{-1} e_0 \rangle
= \langle c \partial_{c} X_{tw}, J^{-1} \partial_{\tau}X_{tw} \rangle
= \bigg\langle c\partial_{c} X_{tw},  \frac{1}{c}\frac{\partial \mathcal{H}}{\partial X}\bigg|_{X=X_{tw}} \bigg\rangle\\
&= \int_0^1 (\partial_{c}X_{tw}) \cdot \biggl(\frac{\partial \mathcal{H}}{\partial X}\bigg|_{X=X_{tw}}\biggr) d \tau
=  \int_0^1  \mathcal{H}'(c)|_{X=X_{tw}(\tau;c)} d\tau = H'(c).
\end{split}
\label{eq:crit}
\eeq
Assuming that $H'(c)=0$, we note that since $\langle e_2, J^{-1} e_0 \rangle=\langle e_2, J^{-1}\mathcal{L}e_1 \rangle=\langle J^{-1}\mathcal{L}e_2, e_1 \rangle=\langle J^{-1}e_1, e_1 \rangle=0$, $e_2$ belongs $(\ker(\mathcal{L^*}))^{\perp}={\rm im}(\mathcal{L})$, and hence there exists $e_3$ such that $\mathcal{L}e_3=e_2$. In other words, if the algebraic multiplicity of $\lambda=0$ is higher than two, then it must be at least four. By similar arguments, the algebraic multiplicity of the zero eigenvalue is always even, due to the symmetries associated with the Hamiltonian nature of the system.

Equation \eqref{eq:crit} implies that if $H'(c)=0$ for some $c=c_0$, the zero eigenvalue is at least quadruple. Assume the generic case when $H''(c_0)\neq 0$. As soon as $c$ deviates from the critical velocity $c_0$, the above solvability condition \eqref{eq:solv} fails, and hence the second pair of eigenvalues will start to move away from zero and generically emerge on the real axis when $c$ is either below or above the threshold value $c_0$, indicating spectral \emph{instability} of the corresponding traveling waves. This is confirmed by the relationship we derive in Sec.~\ref{sec:split} below between the sign of $H'(c)$ and the leading-order behavior of $\lambda$ near zero (see Theorem~\ref{lemma1}). Thus the condition $H'(c)=0$ is a sufficient condition for a change of the stability of a solitary traveling wave in Hamiltonian systems under consideration. {It should be noted that this condition is not necessary for the onset of instability. Indeed, a} variety of different instability mechanisms may be available
to the Hamiltonian system, including the Hamiltonian Hopf scenario, {in which}
two eigenvalues may collide and give rise to a quartet associated with
an oscillatory instability, or the period doubling {case}, where
a pair of multipliers (see the Floquet analysis below) may exit
the unit circle at $-1$. However, we will not systematically {explore} these
{instability mechanisms} in what follows.

We conclude this section with some technical results and notations that will be used below in Sec.~\ref{sec:split}. If $H'(c)\neq 0$, the $2$-by-$2$ matrix $M$ with entries given by $M_{j,k}=\langle J^{-1} e_{j}, e_{k}\rangle$, $j,k=0,1$, is invertible. More precisely,
\[
M=\left(\begin{array}{cc}
  0 & -\alpha_1 \\
  \alpha_1 & 0
\end{array} \right), \quad \alpha_1=\langle J^{-1}e_1, e_0\rangle \neq 0.
\]
Let $G_{2,0}={\rm span}\{ e^{i2n\pi\tau}e_k\; | \; n\in\mathbb{Z}, \; k=0,1 \}$, \; $G_2={\rm cl}(G_{2,0})$ and
\[
G_2^{\#}=\{ X\in D^0([0,1])\; | \;\langle X, J^{-1} e^{i2n\pi\tau}e_k \rangle=0, \; n\in\mathbb{Z}, k=0,1 \}.
\]
Then $D^0([0,1])$ has the direct-sum decomposition $D^0([0,1])=G_2\bigoplus G_2^{\#}$ (see \cite{pegof04a}).

When $H'(c)= 0$ and the zero eigenvalue is exactly quadruple, then $\langle e_0, J ^{-1}e_1\rangle=\langle e_0, J ^{-1}e_2\rangle=\langle e_1, J^{-1} e_2\rangle=0$ but $\langle e_0, J^{-1} e_3\rangle\neq 0$.
Then
\beq
\begin{split}
&M=(\langle J^{-1}e_j, e_k \rangle)_{j,k=0,\dots,3}=\left(\begin{array}{cccc}
  0 & 0 & 0  & -\alpha_1\\
  0 & 0 & \alpha_1 & 0\\
  0 & -\alpha_1 & 0 & -\alpha_2 \\
  \alpha_1 & 0 & \alpha_2 & 0
\end{array} \right), \\
&\alpha_1=\langle J^{-1 }e_3, e_0 \rangle=\langle J^{-1 }e_1, e_2 \rangle\neq 0, \quad \alpha_2=\langle J^{-1} e_3, e_2 \rangle.
\end{split}
\label{eq:M}
\eeq
 Observe that $M$ is again invertible because $\alpha_1\neq 0$ and we can explicitly find
 \[
 M^{-1}=\frac{1}{\alpha_1}\left(\begin{array}{cccc}
  0 & -\frac{\alpha_2}{\alpha_1} & 0  & 1\\
  \frac{\alpha_2}{\alpha_1} & 0 & -1 & 0\\
  0 & 1 & 0 & 0 \\
  -1 & 0 & 0 & 0
\end{array} \right).
\]
 In this case, it is convenient to define $G_{4,0}={\rm span}\{ e^{i2n\pi\tau}e_k | n\in\mathbb{Z}, k=0,1,2,3 \}$, $G_4={\rm cl}(G_{4,0})$ and $G_4^{\#}=\{ X\in D^0([0,1]) | \langle X, J^{-1} e^{i2n\pi\tau}e_k \rangle=0, n\in\mathbb{Z}, k=0,1,2,3 \}$. Then $D^0([0,1])$ has the direct-sum decomposition $D^0([0,1])=G_4\bigoplus G_4^{\#}$.

{
In fact, one can easily extend the results above to the case where $H'(c)=0$ and the algebraic multiplicity of the zero eigenvalue is $n$, {which, as we recall, must be an even number. In this case} $\mathcal{L}^n e_{n-1}=\mathcal{L}^{n-1} e_{n-2}=\mathcal{L}^{n-2} e_{n-3}=...=\mathcal{L} e_0=0$, $\alpha_1=\langle J^{-1 }e_{n-1}, e_0 \rangle \neq 0$, {and} $M=(\langle J^{-1}e_j, e_k \rangle)_{j,k=0,\dots,n-1}$ is invertible. Similarly, we define $G_{n,0}={\rm span}\{ e^{i2n\pi\tau}e_k | n\in\mathbb{Z}, k=0,1,2,..,n-1 \}$, $G_n={\rm cl}(G_{n,0})$ and $G_n^{\#}=\{ X\in D^0([0,1]) | \langle X, J^{-1} e^{i2n\pi\tau}e_k \rangle=0, n\in\mathbb{Z}, k=0,1,2,...,n \}$, {and observe that} $D^0([0,1])=G_n\bigoplus G_n^{\#}$.
}

\section{Floquet analysis of the spectral stability of solitary traveling waves}
\label{sec:Floquet}
In this section we consider an alternative approach for studying the spectral stability of solitary traveling waves that is analytically equivalent to but numerically different from the eigenvalue problem \eqref{eq5}. It is based on the idea that the considered traveling waves on a lattice are periodic orbits modulo shifts, and solitary traveling waves in particular can thus be identified with localized time-periodic solutions known as discrete breathers. As a result, one expects an equivalence between the analysis of Floquet multipliers for the monodromy matrix associated with the relevant periodic orbit \cite{aubry,FlachPR2008,arnold} and the eigenvalue problem \eqref{eq5}, as discussed below.

We begin by defining the evolution operator $\mathcal{E}(\tau_0)$ such that $\mathcal{E}(\tau_0) X_0=X(\tau_0)$, where
\beq
\label{eq_floquet_1}
\begin{split}
c\frac{d}{d\tau}X(\tau)&=\mathcal{A}(X(\tau)), \quad \tau\in[0,\tau_0] \\
X(0)&=X_0.
\end{split}
\eeq
Then traveling wave solutions of \eqref{eq2} satisfy $\mathcal{E}(1)X_{tw}(\tau)=\tilde{S}_{-}X_{tw}(\tau)$ for any $\tau$ in $[0,1]$, and in particular,
\[
\tilde{S}_{+}\mathcal{E}(1)X_{tw}(0)=X_{tw}(0).
\]
In other words, such solutions can be viewed as the fixed points of the map
\beq
\mathcal{F}=\tilde{S}_{+}\mathcal{E}(1)-I.
\label{eq:map}
\eeq
To study their stability, we consider another evolution operator $\mathcal{E}_2(\tau_1,\tau_0)$ satisfying $\mathcal{E}_2(\tau_1,\tau_0) Y_0=Y(\tau_1)$, where
\beq
\label{eq_floquet_2}
\begin{split}
c\frac{d}{d\tau}Y(\tau)&=D\mathcal{A}(X_{tw}(\tau))Y(\tau), \quad \tau\in[\tau_0,\tau_1] \\
Y(\tau_0)&=Y_0.
\end{split}
\eeq
Let $X_0=X_{tw}(0)+\epsilon Y_0$, where $\epsilon Y_0$ is a small perturbation of the fixed point $X_{tw}(0)$ of the map $\mathcal{F}$ in \eqref{eq:map}. Substituting $X_0$ into \eqref{eq_floquet_1}, we find that to the leading order
\[
\mathcal{F}(X_0)=(\tilde{S}_{+}\mathcal{E}(1)-I)X_0=(\tilde{S}_{+}\mathcal{E}_2(1,0)-I)Y_0.
\]
Suppose now that $Y_0$ satisfies $\mathcal{E}_2(\tau,0)Y_0=e^{\lambda \tau}Z_{tw}(\tau)$, where $Z_{tw}$ is a traveling wave. Then one can study the spectrum of $\tilde{S}_{+}\mathcal{E}_2(1,0)$ to analyze the spectral stability of the solitary traveling wave solution $X_{tw}$. The proposition stated below shows that this spectrum is closely related to the spectrum of the linear operator $\mathcal{L}$ defined in \eqref{eq5}.

\begin{proposition}
$Y_0$ is an eigenfunction of $\tilde{S}_+ \mathcal{E}_2(\tau_1,\tau_0)$ associated with the eigenvalue $\mu=e^{\lambda}$ if and only if $Z(\tau)=e^{-\lambda \tau}\mathcal{E}_2(\tau,0)Y_0$ is an eigenfunction of $\mathcal{L}$ associated with the eigenvalue $\lambda$.
\end{proposition}

The proof can be found in Sec.~I.A of the Supplementary Material. Note that $|\mu|>1$ implies instability and corresponds to the case ${\rm Re}(\lambda)>0$. The relationship between $\lambda$ and $\mu$ also explains why the spectrum of $\mathcal{L}$ is $2\pi i$-periodic. We now state the result concerning the generalized eigenfunctions:

\begin{proposition}
If $Z_0(\tau)$ is an eigenfunction of $\mathcal{L}$ associated with the eigenvalue $\lambda$ and $(\mathcal{L}-\lambda I)^k Z_{k}=(\mathcal{L}-\lambda I)^{k-1} Z_{k-1}=...=(\mathcal{L}-\lambda I) Z_1=Z_0$, then $Y_0=Z_0(0)$ is an eigenfunction of $\tilde{S}_+ \mathcal{E}_2(\tau_1,\tau_0)$ associated with the eigenvalue $\mu=e^{\lambda}$, and there exist $\{Y_j\}_{1\leq j\leq k}$ such that $(\tilde{S}_+ \mathcal{E}_2(\tau_1,\tau_0) -\mu I )^k Y_k=(\tilde{S}_+ \mathcal{E}_2(\tau_1,\tau_0) -\mu I )^{k-1} Y_{k-1}=...=(\tilde{S}_+ \mathcal{E}_2(\tau_1,\tau_0) -\mu I ) Y_1 =Y_0$. The converse is also true.
\end{proposition}

See Sec.~I.B of the Supplementary Material for the proof. These results can be used to compute $\{e_0, e_1,e_2,e_3\}$ for $\mathcal{L}$ at $c=c_0$ such that $H'(c_0)=0$ without solving the eigenvalue problem \eqref{eq5}. To do so, we first find $\{Y_0, Y_1, Y_2, Y_3 \}$ such that $(\tilde{S}_+ \mathcal{E}_2(\tau_1,\tau_0)- I)Y_0=0$ and $(\tilde{S}_+ \mathcal{E}_2(\tau_1,\tau_0) - I )^3 Y_3=(\tilde{S}_+ \mathcal{E}_2(\tau_1,\tau_0) -I )^2 Y_2=(\tilde{S}_+ \mathcal{E}_2(\tau_1,\tau_0) - I ) Y_1 =Y_0$ and let $\tilde{Y}_j=\sum_{i=0}^3 \Omega^{-1}_{j i}{Y_{i}}$ for $0\leq j\leq k$, with $\Omega$ satisfying
\[
\Omega\left(\begin{array}{cccc}
  0 & 0 & 0  & 0\\
  1 & 0 & 0 & 0\\
  \frac{1}{2} & 1 & 0 & 0 \\
  \frac{1}{6} & \frac{1}{2} & 1 & 0
\end{array} \right)=\left(\begin{array}{cccc}
  0 & 0 & 0  & 0\\
  1 & 0 & 0 & 0\\
  0 & 1 & 0 & 0 \\
  0 & 0 & 1 & 0
\end{array} \right)\Omega.
\]
For example, one can use
\[
\Omega=\left(\begin{array}{rrrr}
  1 & 0 & 0  & 0\\
  0 & 1 & 0 & 0\\
  0 & \frac{1}{2} & 1 & 0 \\
  0 & \frac{1}{3} & -1 & 1
\end{array} \right).
\]
Let $\tilde{Y}_j(\tau)=\mathcal{E}_2(\tau,0)\tilde{Y}_j$ for $0\leq j\leq 3$ and set
\[
\begin{split}
&\tilde{Z}_0(\tau)=Y_0{\tau}, \quad \tilde{Z}_1(\tau)=Y_1(\tau)-\tau \tilde{Z}_0(\tau), \quad \tilde{Z}_2(\tau)=Y_2(\tau)-\tau \tilde{Z}_1(\tau)-\frac{1}{2}\tau^2 \tilde{Z}_0, \\
&\tilde{Z}_3(\tau)=Y_3(\tau)-\tau \tilde{Z}_2(\tau)-\frac{1}{2}\tau^2 \tilde{Z}_1-\frac{1}{6}\tau^3\tilde{Z}_0.
\end{split}
\]
We then seek $e_i$ in the form $e_i=\sum_{j=0}^{3}F_{i,j}\tilde{Z}_j$, $0\leq i\leq 3$ and find $F=(F_{i,j})_{0\leq i,j\leq 3}$ such that
\[
e_0=\partial_{\tau}X_{tw}(\tau;c_0), \quad \mathcal{L}^3 e_3=\mathcal{L}^2 e_2=\mathcal{L} e_1=e_0.
\]
In particular, let $F_{0,0}=\frac{\partial_{\tau}X_{tw}(\tau;c_0)}{\tilde{Z}_0(\tau)}$, then $F_{i,j}=\delta_{i,j}F_{0,0}$ for $0\leq i, j\leq 3$.

Suppose $Z_{tw}(\tau)$ is a traveling wave defined on $\mathbb{R}$ (although it is sufficient to define it on $[0,1]$) and we define $\tilde{\mathcal{E}}_2(\tau_0)$ such that $(\tilde{\mathcal{E}}_2(\tau_0)Z_{tw})(\tau)=\mathcal{E}_2(\tau+\tau_0,\tau)Z_{tw}(\tau)$. It can be checked that $\tilde{\mathcal{E}}_2(\tau_0)Z_{tw}$ is still a traveling wave. Since $e^{-s\partial_{\tau}}\mathcal{E}_2(\tau+\tau_0+s,\tau+s)e^{s\partial_{\tau}}=\mathcal{E}_2(\tau+\tau_0,\tau)$, it follows that
$e^{-s\partial_{\tau}}\tilde{\mathcal{E}}_2(\tau_0+s)e^{s\partial_{\tau}}=\tilde{\mathcal{E}}_2(\tau_0)$. In fact, similar to the arguments in~\cite{pegof04a}, one can show $e^{-s\partial_{\tau}}\tilde{\mathcal{E}}_2(s)$ is a group with the infinitesimal generator $\frac{1}{c}D\mathcal{A}(X_{tw})-\partial_{\tau}=\mathcal{L}$. In other words,
\[
e^{-s\partial_{\tau}}\tilde{\mathcal{E}}_2(s)={\rm exp}(s(\frac{1}{c}D\mathcal{A}(X_{tw})-\partial_{\tau}))={\rm exp}({s}\mathcal{L}).
\]

\section{Splitting of the zero eigenvalue}
\label{sec:split}
Recall that the spectrum of $\mathcal{L}$ has a double eigenvalue for any velocity $c$ in the family of solitary traveling wave solutions such that $H'(c) \neq 0$. We now consider the situation when there exists a critical nonzero velocity $c_0$ such that $H'(c_0)=0$ but $H''(c_0) \neq 0$. As shown in Sec.~\ref{sec:stab}, in this case the multiplicity of the zero eigenvalue of $\mathcal{L}$ at $c=c_0$ is at least four, so that we generically expect the onset of instability due to the additional two or more eigenvalues splitting away from zero and appearing on the real axis. We start by considering the case when the eigenvalue is exactly quadruple, i.e., there exist $\{ e_0, e_1, e_2, e_3 \}$ such that $\mathcal{L}^4 e_3=\mathcal{L}^3 e_2=\mathcal{L}^2 e_1 =\mathcal{L} e_0=0$ and $\langle J^{-1}e_0, e_3 \rangle \neq 0$, and analyze the fate of the two eigenvalues splitting away from zero when $c \neq 0$.

Considering the neighborhood of the critical velocity $c=c_0$, we write $c=c_0+\epsilon$ and expand
\beq
X_{tw}(\tau;c)=U_0+\epsilon U_1+\epsilon^2 U_2+\epsilon^3 U_3+\dots,
\label{eq:X_expand}
\eeq
where we define $U_0=X_{tw}(\tau;c_0)$, \; $U_1=(\partial_{c}X_{tw}(\tau;c))|_{c=c_0}$, \; $U_2=\frac{1}{2}({\partial^2_{c}}X_{tw}(\tau;c))|_{c=c_0}$
and $U_3=\frac{1}{6}({\partial^3_{c}}X_{tw}(\tau;c))|_{c=c_0}$;
%{\bf isn't it more appropriate to use the symbols $\partial_{cc}^2,
%  \partial_{ccc}^3$ etc. ??}
%(\hx{Yes, it would be natural to just use these symbols. The goal of using $U_j$'s is just to shorten equations and I feel they do help.})
recall that we assumed that $X_{tw}$ is smooth in $c$.
Accordingly, the operator $\mathcal{L}$ has the expansion
\beq
\mathcal{L}=\mathcal{L}_0+\epsilon\mathcal{L}_1+\epsilon^2\mathcal{L}_2+\epsilon^3\mathcal{L}_3\dots, \quad \mathcal{L}_k=\frac{1}{k!}(\frac{\partial}{\partial c})^k \mathcal{L}|_{c=c_0}, \; k=0,1,2,\dots
\label{eq:L_expand}
\eeq

Recall that $\{e_0,e_1,e_2,e_3\}$ are the eigenfunction and generalized eigenfunctions of $\mathcal{L}_0$ associated with $\lambda=0$ and satisfying \eqref{eq:M}. In particular, $e_0=\partial_{\tau}U_0$ and $e_1=c_0 U_1$. In what follows, we will use $G_4=\text{cl}({\rm span}\{e_0,e_1,e_2,e_3\})$ and define the $4 \times 4$ matrices $K$, $L$, {$B$ and $F$} with the entries
\beq
K_{jk}=\langle J^{-1}e_j, \mathcal{L}_1 e_k \rangle, \quad L_{jk}=\langle J^{-1}e_j, \mathcal{L}_2 e_k \rangle,
\quad B_{jk}=\langle J^{-1}e_j, \mathcal{L}_3 e_k \rangle, \quad
F_{jk}=\langle J^{-1}e_j, \mathcal{L}_4 e_k \rangle,
\label{eq:KLB}
\eeq
where $j,k=0,\dots,3$. Note that these matrices are symmetric because $J^{-1}=-J=J^T$ and $J\mathcal{L}$ is self-adjoint.

We now consider the behavior of the two additional eigenvalues in the neighborhood of $c_0$. Below, we will show the following result:
\begin{proposition}
\label{rmk_ev}
If the generalized kernel of $\mathcal{L}_0$ is exactly four-dimensional, the essential spectrum does not contain zero, and $H''(c_0)\neq0$, 
then the splitting eigenvalues are $\lambda=\mathcal{O}(\epsilon^{1/2})$.
\end{proposition}
We remark that this proposition holds in the framework with weighted spaces (in the setting with unweighted spaces where zero is usually contained in the essential spectrum, the conditions are usually not satisfied and one needs to investigate whether the effects of the essential spectrum can be neglected). The result follows from the fact that two of the four eigenvalues of $\mathcal{L}$ are always zero. It suffices to calculate the leading-order terms of the eigenvalues for the perturbed operator $\mathcal{L}$ at $c=c_0+\epsilon$. By restricting the operator in the invariant subspace $G_4$, the question reduces to the perturbation of the matrix
\beq
A_0=\left(\begin{array}{cccc}
  0 & 1 & 0  & 0\\
  0 & 0 & 1 & 0\\
  0 & 0 & 0 & 1 \\
  0 & 0 & 0 & 0
\end{array} \right)
\label{eq:A0}
\eeq
with two constraints that hold for any $c$,
\begin{equation}
\mathcal{L}(\partial_{\tau} X_{tw}(\tau;c))=0
\label{eq:constraint1}
\end{equation}
and
\begin{equation}
\mathcal{L}(c\partial_{c} X_{tw}(\tau;c))=\partial_{\tau} X_{tw}(\tau;c).
\label{eq:constraint2}
\end{equation}
 Note that the characteristic polynomial of the unperturbed matrix $A_0$ is $\lambda^4=0$. For the matrix $A_0$ with $O(\epsilon)$ perturbation, the characteristic polynomial is
 \beq
 \lambda^4+a_3 \lambda^3+a_2 \lambda^2+a_1\lambda +a_0=0,
 \label{eq:poly}
 \eeq
 where the coefficients $a_j$ are at most $O(\epsilon)$. Moreover, due to two existing constraints \eqref{eq:constraint1} and \eqref{eq:constraint2}, two of the eigenvalues are always zero, so we have $\lambda^2(\lambda^2+a_3 \lambda+a_2)=0$. Thus, either $\lambda\sim \epsilon^{1/2}$ (if $a_2\neq 0$) or $\lambda\sim\epsilon$ (if $a_2=0$). However, as we show below, $a_2$ is proportional to $H''(c_0)$ and thus nonzero by assumption.\\

\subsection{Reduced eigenvalue problem without constraints}
To be more specific, let $Z|_{G_4}=(e_0, e_1, e_2, e_3)
\left(\begin{array}{c}
d_0 \\
d_1 \\
d_2 \\
d_3
\end{array} \right)
$ and $P=\mathcal{L}|_{G_4}$ where $G_4$ is defined near the end of Sec.~\ref{sec:stab}. Since $D^0([0,1])=G_4\bigoplus G_4^{\#}$, here we consider the decomposition of $Z$ such that $Z|_{G_4}\in G_4$ and $Z-Z|_{G_4}\in G_4^{\#}$ while $\mathcal{L}|_{G_4}$ is the operator $\mathcal{L}$ restricted on the generalized kernel $G_4$. Then the reduced eigenvalue problem is
\begin{equation}
\label{eq_reduced_ev}
P(e_0, e_1, e_2, e_3)
\left(\begin{array}{c}
d_0 \\
d_1 \\
d_2 \\
d_3
\end{array} \right)
=\lambda (e_0, e_1, e_2, e_3)
\left(\begin{array}{c}
d_0 \\
d_1 \\
d_2 \\
d_3
\end{array} \right).
\end{equation}
Let $A=A_0+\epsilon A_1+\epsilon^2 A_2+\epsilon^3 A_3\dots$ be the matrix representation of $P$ in the basis $G_4$, with the unperturbed matrix $A_0$ given by \eqref{eq:A0}. Then \eqref{eq_reduced_ev} can be written as
\begin{equation}
\label{eq_reduced_ev_2}
A
\left(\begin{array}{c}
d_0 \\
d_1 \\
d_2 \\
d_3
\end{array} \right)
=\lambda
\left(\begin{array}{c}
d_0 \\
d_1 \\
d_2 \\
d_3
\end{array} \right).
\end{equation}
Projecting \eqref{eq_reduced_ev} onto $J^{-1}e_j$ for $j=0,\dots,3$, we obtain
\begin{equation}
\label{eq_reduced_ev_mat}
Q
\left(\begin{array}{c}
d_0 \\
d_1 \\
d_2 \\
d_3
\end{array} \right)
=\lambda M
\left(\begin{array}{c}
d_0 \\
d_1 \\
d_2 \\
d_3
\end{array} \right),
\end{equation}
where $M$ is given by \eqref{eq:M} and $Q=(\langle J^{-1}e_j, P e_k \rangle)=(\langle J^{-1}e_j, \mathcal{L} e_k \rangle)$, $j,k=0,\dots,3$, can be written as $Q=Q_0+\epsilon Q_1+\epsilon^2 Q_2+\epsilon^3 Q_3+\dots$. Note that $Q_i=M A_i$ for $i=0,1,\dots$ are symmetric, and in particular, $Q_1=K$, $Q_2=L$ and $Q_3=B$, where we recall \eqref{eq:KLB}. Considering the eigenvalue problem \eqref{eq_reduced_ev_2}, we obtain the following result:

\begin{proposition}
\label{rmk_ev_free}
For the eigenvalue problem \eqref{eq_reduced_ev_2}) without any constraints imposed,
\begin{itemize}
\item $\lambda=\mathcal{O}(\epsilon^{1/2})$ if and only if
\beq
A_{1,30}=0,
\label{eq:cond1}
\eeq
\beq
A_{1,31}+A_{1,20}=0
\label{eq:cond2}
\eeq
and
\beq
(A_{1,32}+A_{1,21}+A_{1,10})^2+(A_{1,20}A_{1,33}+A_{1,10}A_{1,32}+A_{1,00}A_{1,31}-A_{2,30})^2\neq 0;
\label{eq:cond3}
\eeq
\item $\lambda=\mathcal{O}(\epsilon^{1/3})$ if and only if \eqref{eq:cond1} holds but \eqref{eq:cond2} fails;
\item $\lambda=\mathcal{O}(\epsilon^{1/4})$ if and only if \eqref{eq:cond1} fails.
\end{itemize}
\end{proposition}

Observe, however, that the condition \eqref{eq:cond2} always holds due to the symmetry of $K$, which implies that $K_{01}=-\alpha_1 A_{1,31}$ is equal to $K_{10}=\alpha_1 A_{1,20}$ (recall that $\alpha_1 \neq 0$). This eliminates the second possibility ($\lambda=\mathcal{O}(\epsilon^{1/3})$). As we will show below, the first constraint \eqref{eq:constraint1} ensures that \eqref{eq:cond1} holds, thus eliminating the third possibility ($\lambda=\mathcal{O}(\epsilon^{1/4})$). Meanwhile, the second constraint \eqref{eq:constraint2} and our assumption that $H''(c_0) \neq 0$ together ensure that \eqref{eq:cond3} holds, so we have $\lambda=\mathcal{O}(\epsilon^{1/2})$.

\subsection{Reduced eigenvalue problem with constraints and the leading approximation for the splitting eigenvalues}
\label{sec_splitting_sqrt_1}
In this subsection we use the constraints \eqref{eq:constraint1} and \eqref{eq:constraint2} to show that the splitting eigenvalues are of $\mathcal{O}(\epsilon^{1/2})$ and compute their leading approximation.

Since \eqref{eq:constraint1} holds for any $c$, expansion around $c=c_0$ using \eqref{eq:X_expand} and \eqref{eq:L_expand} shows that
\begin{eqnarray*}
0&=&\mathcal{L}_0 (\partial_{\tau} U_0)\\
0&=&\mathcal{L}_0 (\partial_{\tau} U_1)+\mathcal{L}_1 (\partial_{\tau} U_0)\\
0&=&\mathcal{L}_0 (\partial_{\tau} U_2)+\mathcal{L}_1 (\partial_{\tau} U_1)+\mathcal{L}_2 (\partial_{\tau} U_0)\\
0&=&\mathcal{L}_0 (\partial_{\tau} U_3)+\mathcal{L}_1 (\partial_{\tau} U_2)+\mathcal{L}_2 (\partial_{\tau} U_1)+\mathcal{L}_3 (\partial_{\tau} U_0).
%\\
%0&=&\mathcal{L}_0 (\partial_{\tau} U_4)+\mathcal{L}_1 (\partial_{\tau} U_3)+\mathcal{L}_2 (\partial_{\tau} U_2)+\mathcal{L}_3 (\partial_{\tau} U_1)+\mathcal{L}_4 (\partial_{\tau} U_0)
\end{eqnarray*}
Using
\beq
\partial_{\tau} U_j=\sum_{k=0}^3 (g_{jk}e_k)+(\partial_{\tau} U_j)^{\#}, \quad (\partial_{\tau} U_j)^{\#} \in G_4^{\perp},
\label{eq:dU}
\eeq
we can rewrite the above equations as
\begin{eqnarray}
\label{eqn_U_expand1_1_2}
0&=&Q_0 g_0\\
\label{eqn_U_expand1_2_2}
0&=&Q_0 g_1+Q_1 g_0\\
\label{eqn_U_expand1_3_2}
0&=&Q_0 g_2 +Q_1 g_1+Q_2 g_0\\
\label{eqn_U_expand1_4_2}
0&=&Q_0 g_3+Q_1 g_2+Q_2 g_1+Q_3 g_0.
%\\
%\label{eqn_U_expand1_5_2}
%0&=&Q_0 g_4+Q_1 g_3+Q_2 g_2+Q_3 g_1+Q_4 g_0
\end{eqnarray}
Since $e_0=\partial_\tau U_0$, we have $g_0=(g_{00}, g_{01}, g_{02}, g_{03})^T=(1,0,0,0)^T$. {Recalling that $Q_0=M A_0$,  $Q_1=K$, $Q_2=L$ and $Q_3=B$},
we then use \eqref{eqn_U_expand1_2_2} to find that
\beq
K_{00}=0,
\label{cond_1}
\eeq
which corresponds to \eqref{eq:cond1}, and
\beq
g_1=-A_0^T M^{-1} Q_1 g_0 + g_{10} g_0,
\label{eq:g1}
\eeq
which yields
\beq
\label{eq:g1comp}
g_{11}=-\frac{K_{30}}{\alpha_1}+\frac{\alpha_2}{\alpha_1^2}K_{10} \qquad
g_{12}=\frac{K_{20}}{\alpha_1} \qquad
g_{13}=-\frac{K_{10}}{\alpha_1}.
\eeq
Using \eqref{eqn_U_expand1_3_2}, we then obtain
\beq
\label{cond_2}
\sum_{j=1}^3 g_{1j}K_{0j}+ L_{00}=0,
\eeq
which corresponds to
\beq
A_{1,20}A_{1,33}+A_{1,10}A_{1,32}+A_{1,00}A_{1,31}-A_{2,30}=0,
\label{eq:cond2_A}
\eeq
and the coefficients $g_2=-A_0^T M^{-1} Q_1 g_1 -A_0^T M^{-1} Q_2 g_0 +g_{20} g_0$, given in detail by (21) in Sec.~II of the Supplementary Material.

Finally, \eqref{eqn_U_expand1_4_2} yields
\beq
\label{cond_3}
\sum_{j=1}^3 g_{2j}K_{0j}+ \sum_{j=0}^3 g_{1j}L_{0j}+B_{00}=0
\eeq
and the formulas (22) for the coefficients $g_3=-A_0^T M^{-1} Q_1 g_2 -A_0^T M^{-1} Q_2 g_1-A_0^T M^{-1} Q_3 g_0 +g_{30} g_0$.

We now turn to the constraint \eqref{eq:constraint2}, which again holds for any $c$. Expanding both sides in $\epsilon$ and using \eqref{eq:X_expand}, \eqref{eq:L_expand}, we obtain
\begin{eqnarray*}
\label{eqn_U_expand2_1}
\mathcal{L}_0 (c_0 U_1)&=&\partial_{\tau}U_0 \\
\label{eqn_U_expand2_2}
\mathcal{L}_0 (2c_0 U_2+U_1) + \mathcal{L}_1 (c_0 U_1) &=& \partial_{\tau}U_1 \\
\label{eqn_U_expand2_3}
\mathcal{L}_0 (3 c_0U_3+2U_2) + \mathcal{L}_1 (2c_0 U_2+U_1) +\mathcal{L}_2 (c_0 U_1) &=& \partial_{\tau}U_2 \\
\label{eqn_U_expand2_4}
\mathcal{L}_0 (4 c_0U_4+3U_3)+\mathcal{L}_1 (3 c_0U_3+2U_2) + \mathcal{L}_2 (2c_0 U_2+U_1) +\mathcal{L}_3 (c_0 U_1) &=& \partial_{\tau}U_3.
\end{eqnarray*}
Using
\beq
U_j=\sum_{k=0}^3 (f_{jk}e_k)+(U_j)^{\#}, \quad (U_j)^{\#} \in G_4^{\perp},
\label{eq:U}
\eeq
we have
\begin{eqnarray}
\label{eqn_U_expand2_1_2}
c_0 Q_0 f_1&=&M g_0 \\
\label{eqn_U_expand2_2_2}
Q_0 (2c_0 f_2+f_1)+Q_1 c_0 f_1&=& M g_1 \\
\label{eqn_U_expand2_3_2}
Q_0 (3c_0 f_3+2 f_2)+Q_1 (2 c_0 f_2+f_1) + Q_2 c_0 f_1&=& M g_2 \\
\label{eqn_U_expand2_4_2}
Q_0 (4c_0 f_4+3 f_3)+Q_1 (3 c_0 f_3+2 f_1) + Q_2 (2c_0 f_2 +f_1) + Q_3 c_0 f_1&=& M g_3.
\end{eqnarray}
As shown in Sec.~II of the Supplementary Material, this yields the coefficients $f_{ij}$, $i,j=1,\dots,3$ in \eqref{eq:U}. In particular, we have
\beq
f_{23}=\frac{K_{20}-K_{11}-c_0 f_{10}K_{10}}{2\alpha_1 c_0}.
\label{eq:f23}
\eeq

Recall that Proposition~\ref{rmk_ev_free} states that $\lambda=\mathcal{O}(\epsilon^{1/2})$ if and only if \eqref{eq:cond1}, \eqref{eq:cond2} and \eqref{eq:cond3} hold. The above analysis shows that \eqref{eq:cond1} is imposed by the constraint \eqref{eq:constraint1}. As we discussed above, \eqref{eq:cond2} also holds, due to the symmetry of $K$. Observe also that \eqref{eq:cond2_A} derived above means that \eqref{eq:cond3} holds if and only if
\beq
A_{1,32}+A_{1,21}+A_{1,10} \neq 0.
\label{eq:cond3_new}
\eeq
We now show that this inequality holds due to our assumption that $H''(c_0) \neq 0$, so that by Proposition~\ref{rmk_ev_free} we have $\lambda=\mathcal{O}(\epsilon^{1/2})$. Indeed, writing $H(c)=H(c_0)+\epsilon H'(c_0)+\frac{\epsilon^2}{2}H''(c_0)+\frac{\epsilon^3}{6}H'''(c_0)+o(\epsilon^3)$
and recalling \eqref{eq2}, \eqref{eq:M}, \eqref{eq:X_expand}, \eqref{eq:dU}, \eqref{eq:g1comp}, \eqref{eq:U} and \eqref{eq:f23}, we obtain
\beq
\label{eq_H_2nd_d}
\begin{split}
H''(c_0)&=2\langle\nabla\mathcal{H}, U_2 \rangle + \langle U_1, \nabla^2\mathcal{H} U_1 \rangle\\
&= 2 c_0 \langle J^{-1}e_0, U_2 \rangle + c_0\langle U_1, J^{-1}\mathcal{L}_0 U_1 \rangle + c_0\langle U_1, J^{-1} \partial_{\tau}U_1 \rangle \\
&= 2 c_0 f_{23} \langle J^{-1}e_0, e_3 \rangle + \langle U_1, J^{-1} e_0 \rangle + c_0 f_{10}g_{13} \langle e_0, J^{-1}e_3 \rangle +g_{12}\langle e_1, J^{-1}e_2 \rangle \\
&= 2 c_0 \frac{K_{20}-K_{11}-c_0 f_{10} K_{10}}{2\alpha_1 c_0} (-\alpha_1) +0 -c_0 f_{10}K_{10}-K_{20}\\
&= K_{11}-2K_{20}=\alpha_1(A_{1,21}+A_{1,32}+A_{1,10}),
\end{split}
\eeq
where
\beq
\nabla\mathcal{H}=\frac{\partial \mathcal{H}}{\partial X}\bigg|_{X=X_{tw}(\tau;c_0)}, \quad
\nabla^2\mathcal{H}=\frac{\partial^2 \mathcal{H}}{\partial X^2}\bigg|_{X=X_{tw}(\tau,c_0)}.
\label{eq:nabla}
\eeq
Here we used the fact that $K_{11}=\alpha_1 A_{1,21}$ and $K_{02}=-\alpha_1 A_{1,32}=-\alpha_1 A_{1,10}=K_{20}$, where the last equality takes into account \eqref{eq:cond1}. Recall also that $\alpha_1$ defined in \eqref{eq:M} is nonzero.

Substituting $\lambda=\epsilon^{1/2}\lambda_1+\epsilon \lambda_2+\epsilon^{3/2}\lambda_3+\dots$ and $Z=Z_0+\epsilon^{1/2} Z_1+\epsilon Z_2+\epsilon^{3/2} Z_3+\dots$ into $\lambda Z=\mathcal{L}Z$, we obtain
\begin{eqnarray*}
\label{eqn_lambda1_expand0}
0&=&\mathcal{L}_0 Z_0\\
\label{eqn_lambda1_expand1}
\lambda_1 Z_0&=&\mathcal{L}_0 Z_1\\
\label{eqn_lambda1_expand2}
\lambda_1 Z_1+\lambda_2 Z_0&=&\mathcal{L}_0 Z_2+\mathcal{L}_1 Z_0\\
\label{eqn_lambda1_expand3}
\lambda_1 Z_2+\lambda_2 Z_1+\lambda_3 Z_0&=&\mathcal{L}_0 Z_3+\mathcal{L}_1 Z_1\\
\label{eqn_lambda1_expand4}
\lambda_1 Z_3+\lambda_2 Z_2+\lambda_3 Z_1+\lambda_4 Z_0&=&\mathcal{L}_0 Z_4+\mathcal{L}_1 Z_2+\mathcal{L}_2 Z_0.
\end{eqnarray*}
With $Z_j=\sum_{k=0}^3 (C_{jk}e_k)+(Z_j)^{\#}$, where $(Z_j)^{\#} \in G_4^{\perp}$, we then have
\begin{eqnarray}
\label{eqn_lambda1_expand0_2}
0&=&A_0 C_0\\
\label{eqn_lambda1_expand1_2}
\lambda_1 C_0&=&A_0 C_1\\
\label{eqn_lambda1_expand2_2}
\lambda_1 C_1+\lambda_2 C_0&=&A_0 C_2+M^{-1} Q_1 C_0\\
\label{eqn_lambda1_expand3_2}
\lambda_1 C_2+\lambda_2 C_1+\lambda_3 C_0&=&A_0 C_3+M^{-1} Q_1 C_1\\
\label{eqn_lambda1_expand4_2}
\lambda_1 C_3+\lambda_2 C_2+\lambda_3 C_1+\lambda_4 C_0&=&A_0 C_4+M^{-1}Q_1 C_2+M^{-1}Q_2 C_0,
\end{eqnarray}
where we recall that $M$ defined in \eqref{eq:M} is skew-symmetric, while $Q_0=MA_0$, $Q_1=K$, $Q_2=L$ and $Q_3=B$ are symmetric.
Setting $C_{00}=1$, we find that
\begin{eqnarray*}
\label{eqn_lambda1_expand0_3}
C_0&=&C_{00}g_0=g_0, \\
\label{eqn_lambda1_expand1_3}
C_1&=&\lambda_1 A_0^T C_0+C_{10}g_0=\lambda_1 A_0^T g_0+C_{10}g_0, \\
\label{eqn_lambda1_expand2_3}
C_2&=&A_0^T (\lambda_1 C_1+\lambda_2 C_0-M^{-1} K C_0)+C_{20}g_0,\\
%&=&\lambda_1^2 (A_0^T)^2  g_0+\lambda_1 C_{10} A_0^T g_0+\lambda_2 A_0^T g_0-A_0^T M^{-1}Kg_0 +C_{20}g_0, \\
\label{eqn_lambda1_expand3_3}
C_3&=&A_0^T(\lambda_1 C_2+\lambda_2 C_1+\lambda_3 C_0-M^{-1} K C_1)+C_{30}g_0.
%&=&\lambda_1^3 (A_0^T)^3 g_0 +\lambda_1^2 (A_0)^T C_{10}g_0+2\lambda_1\lambda_2(A_0^T)^2 g_0\\
%& &+\lambda_1 [C_{20} A_0^T g_0-(A_0^T)^2 M^{-1}K g_0-A_0^T M^{-1}K A_0^{T} g_0] \\
%& &+\lambda_2 A_0^T C_{10} g_0 + \lambda_3 A_0^T g_0 -A_0^T M^{-1} K C_{10}g_0+C_{30}g_0
\end{eqnarray*}

{
  Due to the special form of $A_0$, we first check that the last row of each system in \eqref{eqn_lambda1_expand0_2}--\eqref{eqn_lambda1_expand3_2}
  %, \eqref{eqn_lambda1_expand1_2},
  %\eqref{eqn_lambda1_expand2_2}\eqref{eqn_lambda1_expand3_2}
  is satisfied. The first two are straightforward. Observe that the last row of \eqref{eqn_lambda1_expand2_2} is equivalent to the last row of $\lambda_1 C_1 +\lambda_2 C_0=M^{-1}K C_0$, or $0=M^{-1}K C_0$, since the last components in $C_0$ and $C_1$ are zero, and yields \eqref{cond_1}. In addition, the last row of~\eqref{eqn_lambda1_expand3_2} is equivalent to the last row of $\lambda_1 C_2=M^{-1}K C_1$, which can be checked using \eqref{eqn_U_expand2_2_2}.
}
The last row of \eqref{eqn_lambda1_expand4_2} then yields
\[
\alpha_1 \lambda_1^4 + (2 K_{20}-K_{11})\lambda_1^2+\biggl(L_{00}+\frac{K_{20}^2}{\alpha_1}+\frac{\alpha_2 K_{10}^2}{\alpha_1^2}
-\frac{2 K_{10}K_{30}}{\alpha_1}\biggr)=0,
\]
where the last term is zero by \eqref{cond_2}, as can be seen after substituting \eqref{eq:g1comp} and taking into account the symmetry of $K$, and the coefficient in front of $\lambda_1^2$ in the second term is $-H''(c_0)$ by \eqref{eq_H_2nd_d} (for further discussion about the connection between the last row of the equations and the constraints \eqref{eq:constraint1}--\eqref{eq:constraint2}, see Sec.~II of the Supplementary Material).

We thus obtain the particularly simple final form:
\[
\lambda_1^2(\alpha_1 \lambda_1^2-H''(c_0))=0,
\]
which has a double root at zero, corresponding to two eigenvalues that are always zero, and a nonzero pair of roots satisfying $\lambda_1^2=H''(c_0)/\alpha_1$. Clearly, $H''(c_0)\alpha_1<0$ thus corresponds to a pair of imaginary $\lambda_1$ and $H''(c_0)\alpha_1>0$ yields a pair of real ones, implying the transition to instability as $c$ passes through $c_0$. To be more specific, if $H'(c_0)=0$ and $H''(c_0)\alpha_1>0$, as $c$ increases from $c_1<c_0$ to $c_0$, there will be a pair of eigenvalues $\lambda\approx \pm i \sqrt{\frac{H''(c_0)}{\alpha_1}(c_0-c)}$ moving towards each other along the imaginary axis and colliding at the origin at $c=c_0$. As we continue increasing $c$ to $c_2>c_0$, a pair of zero eigenvalues will split and move on the real axis as $\lambda\approx \pm \sqrt{\frac{H''(c_0)}{\alpha_1}(c-c_0)}$. If $H''(c_0)\alpha_1<0$, this process will be reversed as $c$ increases.

We summarize these results in the following Theorem:
\begin{theorem}
\label{lemma1}
Consider a lattice Hamiltonian system $\frac{dx}{dt}=J \frac{\partial\mathcal{H}}{\partial x}$ and assume that
\begin{itemize}
\item it has a family of traveling wave solutions $x_{tw}(t;c)$ parameterized by velocity $c$ that are smooth in $c$;
\item for $\tau=ct$ and $X_{tw}(\tau; c)=x_{tw}(t;c)$, we have $\partial_{\tau} (X_{tw,j}(\tau;c))$ and $\partial_{c}^k (X_{tw,j}(\tau;c))$ in $D^0([0,1])$ for $k=1,2$ (if weighted spaces are used, these functions should be in $D^0_a([0,1])$);
\item there exists velocity $c_0$ such that the system's Hamiltonian $H(c)=\int_0^1 \mathcal{H}|_{X=X_{tw}(\tau;c)} d\tau$ satisfies $H'(c_0)=0$ and $H''(c_0)\neq0$;
\item the generalized kernel of $\mathcal{L}_{0}=\frac{1}{c}J \frac{\partial^2\mathcal{H}}{\partial X^2}|_{X=X_{tw}(\tau;c_0)}$ is of dimension $4$, with $\mathcal{L}_0^4 e_3=\mathcal{L}_0^3 e_2=\mathcal{L}_0^2 e_1=\mathcal{L}_0 e_0=0$.
\item the contribution from the essential spectrum to the zero eigenvalue is at a higher order (this assumption is not needed if weighted spaces are used);
\end{itemize}
Then there exist $c_1<c_0$ and $c_2>c_0$ such that for $c\in(c_1,c_2)$, a pair of eigenvalues of $\mathcal{L}$ is given by
\[
\lambda=\pm \sqrt{\frac{H''(c_0)}{\alpha_1}(c-c_0)}+O(|c-c_0|),
\]
where $\alpha_1=\langle J^{-1}e_3, e_0 \rangle$. In particular, if $H''(c_0)\alpha_1<0$, the solution $x_{tw}(t;c)$ is unstable for $c\in(c_1,c_0)$. If $H''(c_0)\alpha_1>0$, the solution $x_{tw}(t;c)$ is unstable for $c\in(c_0,c_2)$.
\end{theorem}

\subsection{Higher-dimensional generalized kernel}
\label{sec:HD_kernel}
We now generalize Theorem~\ref{lemma1} to the case when the generalized kernel of $\mathcal{L}_0$ has dimension larger than or equal to $4$: ${\rm gker}(\mathcal{L}_{0})=n$, $n \geq 4$. Recall that $n$ must be even.
\begin{lemma}
\label{lemma2}
Suppose all the assumptions in Theorem~\ref{lemma1} hold except that ${\rm dim}({\rm gker}(\mathcal{L}_0))=n \geq 4$, where $n$ is even, and
\[
\mathcal{L}_0^n e_{n-1}=\mathcal{L}_0^{n-1} e_{n-2}=...=\mathcal{L}_0^2 e_1=\mathcal{L}_0 e_0=0.
\]
Then there exist $c_1<c_0$ and $c_2>c_0$ such that for $c\in(c_1,c_2)$, the leading-order terms of nonzero eigenvalues of $\mathcal{L}$ are
\[
\lambda=e^{j\frac{2\pi i}{(n-2)}} \biggl(\frac{H''(c_0)}{\alpha_1}(c-c_0)\biggr)^{1/(n-2)}+o(|c-c_0|^{1/(n-2)}), \quad j=0,1,\dots,n-3.
\]
where $\alpha_1=\langle J^{-1}e_{n-1}, e_0 \rangle$.
\end{lemma}
The proof can be found in Sec.~I.C of the Supplementary Material.\\

The degenerate case when $H'(c_0)=H''(c_0)=0$ but $H'''(c_0) \neq 0$ is briefly discussed in Sec.~III of the Supplementary Material.

\section{Examples}
\label{sec:examples}
In this section we illustrate the results of our stability analysis by considering two examples of lattice dynamical systems that have solitary wave solutions
which change stability. Details of the numerical procedures used to obtain these solutions and analyze their stability can be found in Sec.~IV of the Supplementary Material.

As our general setup, we consider a lattice of coupled nonlinear oscillators with a generic nearest-neighbor potential and all-to-all harmonic long-range interactions. In principle, the methodology can capture nonlinear long-range interactions, but here we consider linear ones for simplicity. Such a system is described by the following Hamiltonian:
\begin{equation}
\label{eq:Ham}
H=\sum_{n=-\infty}^\infty\left[\frac{1}{2}\dot q_n^2+V(q_{n+1}-q_n)+\frac{1}{4}\sum_{m=-\infty}^\infty \Lambda(m) (q_n-q_{n+m})^2\right],
\end{equation}
where $q_n(t)$ represents the displacement of $n$th particle from the equilibrium position at time $t$, $\dot{q}_n(t) \equiv q_n'(t)$, $V(r)$ is a generic potential governing the nonlinear interactions between nearest neighbors, and $\Lambda(m)$ are long-range interaction coefficients, which decay as $|m| \rightarrow \infty$; in the absence of such interactions, $\Lambda(m)=0$. For instance, $\Lambda(m)=\rho(\e^\gamma-1)\e^{-\gamma|m|}(1-\delta_{m,0})$, where $\rho>0$ and $\gamma>0$ are constants, corresponds to the Kac-Baker interaction \cite{mertens2}, and $\Lambda(m)=\rho |m|^{-s}(1-\delta_{m,0})$ corresponds to the dipole-dipole (Coulomb) interaction for $s=5$ ($s=3$) between charged particles on a lattice \cite{mertens1}. The dynamics of the system is governed by
\begin{equation}
\label{eq:dyn1}
    \ddot q_n-V'(q_{n+1}-q_n)+V'(q_n-q_{n-1})+\sum_{m=1}^\infty\Lambda(m)(2q_n-q_{n+m}-q_{n-m})=0,
\end{equation}
Since the solitary solutions we consider are kink-like in terms of displacement, it is more convenient to rewrite \eqref{eq:dyn1} in terms of the strain variables $r_n=q_{n+1}-q_{n}$, obtaining
\begin{equation}
\label{eq:dyn}
   \ddot r_n+2V'(r_n)-V'(r_{n+1})-V'(r_{n-1})+\sum_{m=1}^\infty\Lambda(m)(2r_n-r_{n+m}-r_{n-m})=0.
\end{equation}
Solitary traveling wave solutions of \eqref{eq:dyn} have the form
\[
r_n(t)=\phi(\xi), \quad \xi=n-ct,
\]
where $c$ is the velocity of the wave. Such solutions thus satisfy the advance-delay differential equation
\begin{equation}\label{eq:ade}
\begin{split}
    &c^2\phi''(\xi)+2V'(\phi(\xi))-V'(\phi(\xi+1))-V'(\phi(\xi-1))\\
    &+\sum_{m=1}^\infty\Lambda(m)(2\phi(\xi)-\phi(\xi+m)-\phi(\xi-m))=0.
\end{split}
\end{equation}
In addition, the solutions must vanish at infinity:
\begin{equation}
\label{eq:BCs}
\lim_{\xi\rightarrow \pm \infty}\phi(\xi)=0.
\end{equation}

As our first example of a system where stability of a solitary traveling wave changes with velocity, we consider a smooth regularization of the model
studied in \cite{at}. The original model is an FPU lattice \eqref{eq:Ham} with only short-range interactions ($\Lambda(m)=0$) and a biquadratic potential,
\begin{equation}
\label{eq:potentialat}
    V(r)=\begin{cases}
             \frac{r^2}{2}, & |r|\leq r_c \\
             \frac{\chi}{2} (|r|-r_c)^2 + r_c |r| - \frac{r_c^2}{2},  & |r|>r_c,
           \end{cases}
\end{equation}
which allows construction of explicit solitary traveling waves \cite{at}. Here $r_c>0$ is a fixed transition strain separating the quadratic regimes, and $\chi>1$ is the elastic modulus for $|r|>r_c$, while the other modulus is rescaled to one. As shown in \cite{at}, the system has a family of solitary traveling wave solutions $r_n(t)=\phi(n-ct)$ with velocities $c$ in the range $1<c<\sqrt{\chi}$. As $c$ approaches the lower sonic limit, $c \rightarrow 1$, solutions delocalize, with $r_n(t) \rightarrow r_c>0$ for all $n$. As a result,
essentially by construction, the energy $H(c)$ of the system tends to infinity in this limit. This is in contrast to the smoother FPU systems, where solutions delocalize to zero as they approach the sonic limit, with $H(c) \rightarrow 0$ and are well described by the KdV equation \cite{pegof99}. In this case, however, the system is linear for $|r|<r_c$ and thus cannot have nonlinear waves delocalizing at zero. As the velocity $c$ increases away from $c=1$, the wave becomes more localized, and the energy $H(c)$ decreases for $c$ just above the sonic limit. However, its amplitude also increases, so eventually the energy starts growing with $c$ and tends to infinity as $c \rightarrow \sqrt{\chi}$. As as result, $H(c)$ has a minimum at velocity $c=c_0$, with $H'(c_0)=0$, $H'(c)<0$ for $1<c<c_0$ and $H'(c)>0$ for $c_0<c<\sqrt{\chi}$. According to our energy-based criterion, we thus expect to see stability change at $c=c_0$. This is confirmed by
direct numerical (time evolution)
simulations in \cite{at,ourTW}, which indicate that the solitary traveling waves are unstable when $1<c<c_0$ and stable for $c_0<c<\sqrt{\chi}$.

The potential \eqref{eq:potentialat} has a discontinuous second derivative:
\begin{equation}
\label{eq:potentialat2nd}
    V''(r)=\begin{cases}
             1, & |r|\leq r_c \\
             \chi,  & |r|>r_c,
           \end{cases}
\end{equation}
so our analysis framework is not directly applicable in this case. Instead, we consider a smooth regularization of this potential, with $V''(r)$ in the form
\begin{equation}
\label{eq:potentialreg2nd}
    V''(r)=1+\frac{\chi-1}{\pi}\left(\arctan\frac{r^2-r_c^2}{\varepsilon^2}+\arctan\frac{r_c^2}{\varepsilon^2}\right),
\end{equation}
where $\varepsilon>0$ is the regularization parameter. The expression for $V(r)$ is quite cumbersome and thus not included here. Fig.~\ref{fig:regular1} shows the shape of $V'(r)$ and $V''(r)$ for different values of $\varepsilon$,
including $\varepsilon=0$ when $V(r)$ reduces to \eqref{eq:potentialat}.
\begin{figure}
\begin{center}
\begin{tabular}{cc}
\includegraphics[width=6.0cm]{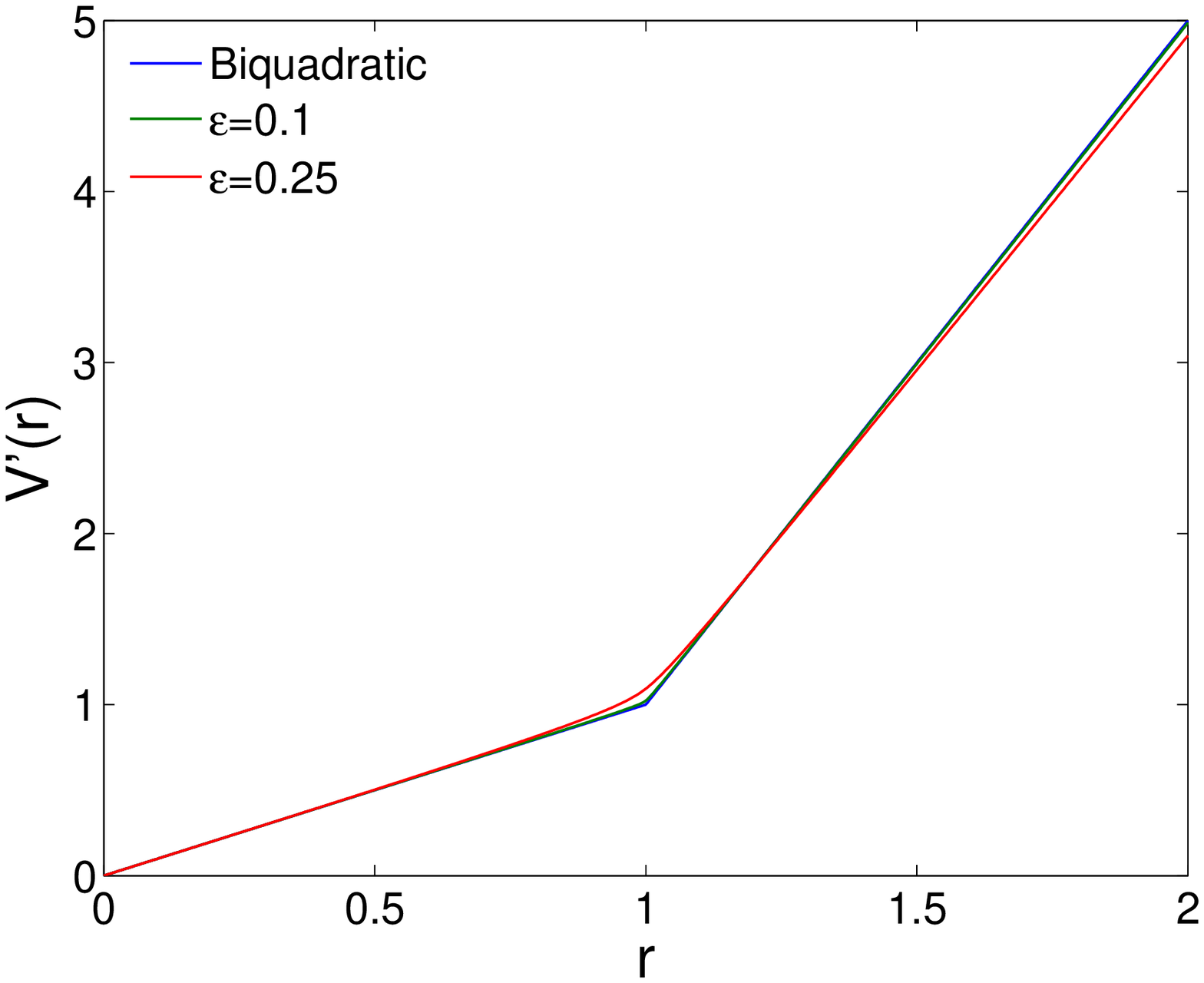} &
\includegraphics[width=6.0cm]{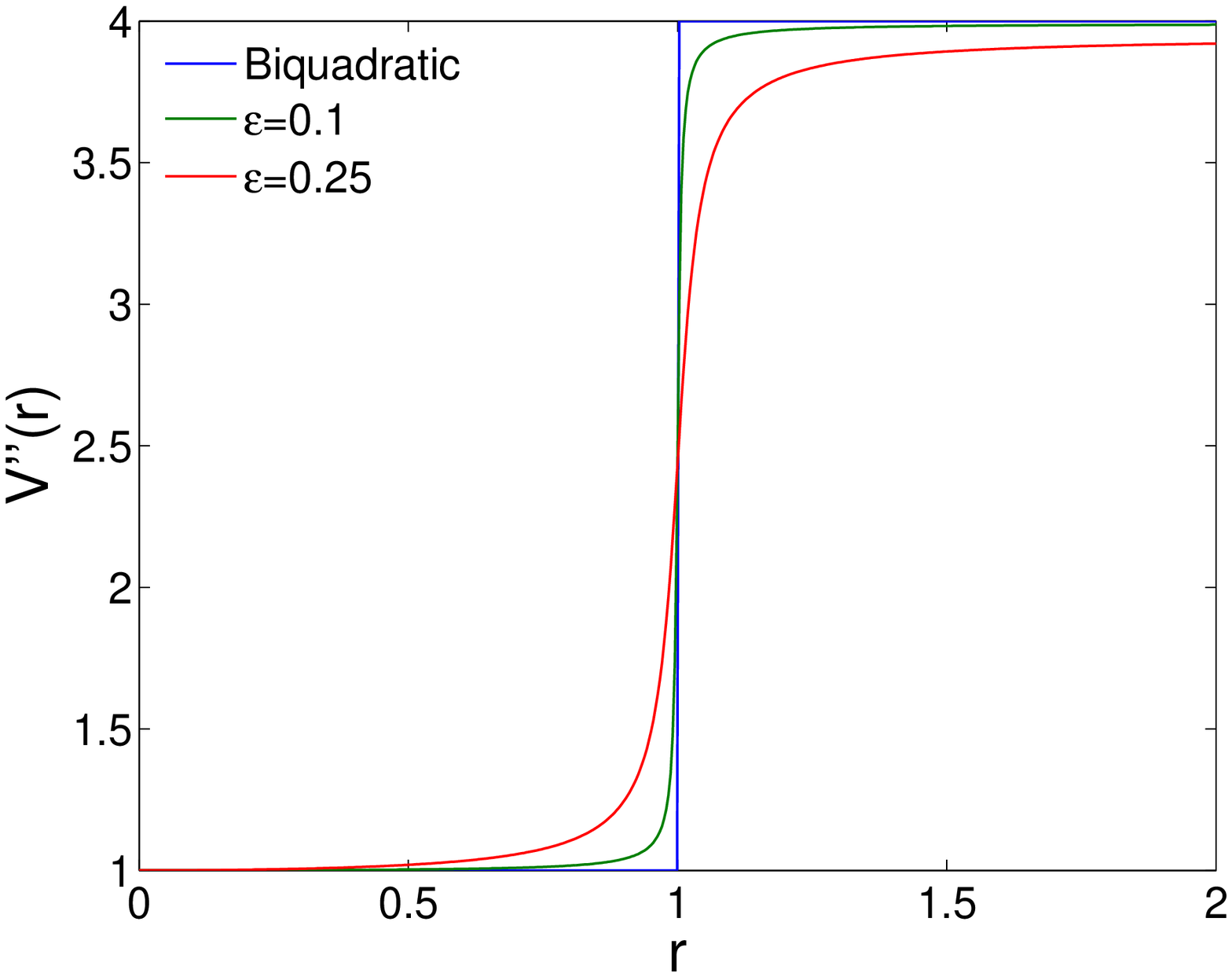}
\end{tabular}
\end{center}
\caption{First (left panel) and second (right panel) derivatives of the regularized potential $V(r)$ for different values of the parameter $\varepsilon$. Here $r_c=1$ and $\chi=4$, and only $r \geq 0$ is shown due to symmetry.}
\label{fig:regular1}
\end{figure}
Note that for all $\varepsilon \geq 0$ we have the same lower sonic limit: $c_s=V''(0)=1$. Increasing $\varepsilon$ from zero smoothens the corners of $V'(r)$ ar $r=\pm r_c$ and decreases its slope at $|r|>r_c$.

We used spectral methods (see Sec.~IV of the Supplementary Material for details) to construct a family of solitary traveling waves for the regularized system, with typical profiles shown for the strain variable in the left panel of Fig.~\ref{fig:profiles}.
\begin{figure}
\begin{center}
\begin{tabular}{cc}
\includegraphics[width=6.0cm]{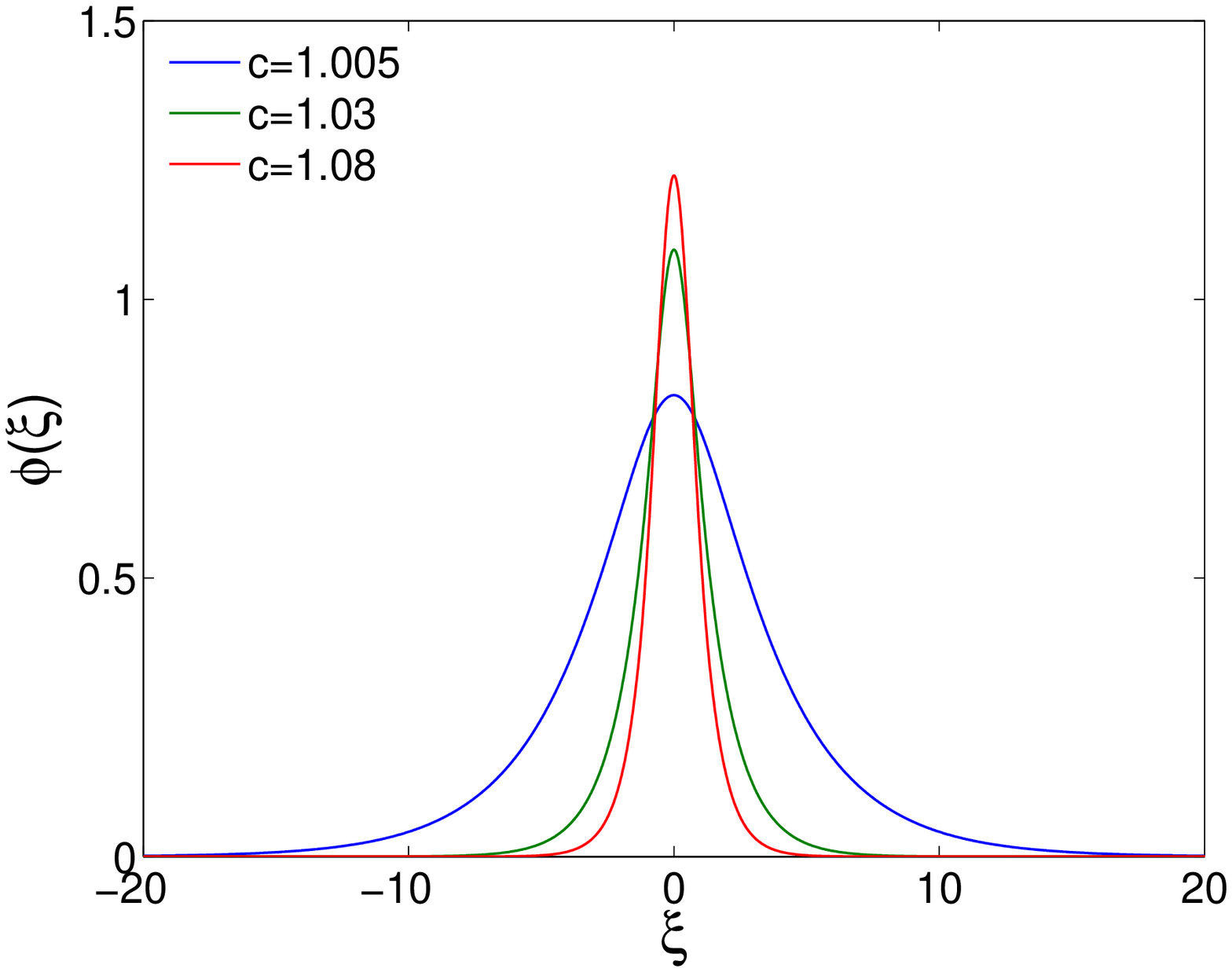} &
\includegraphics[width=6.0cm]{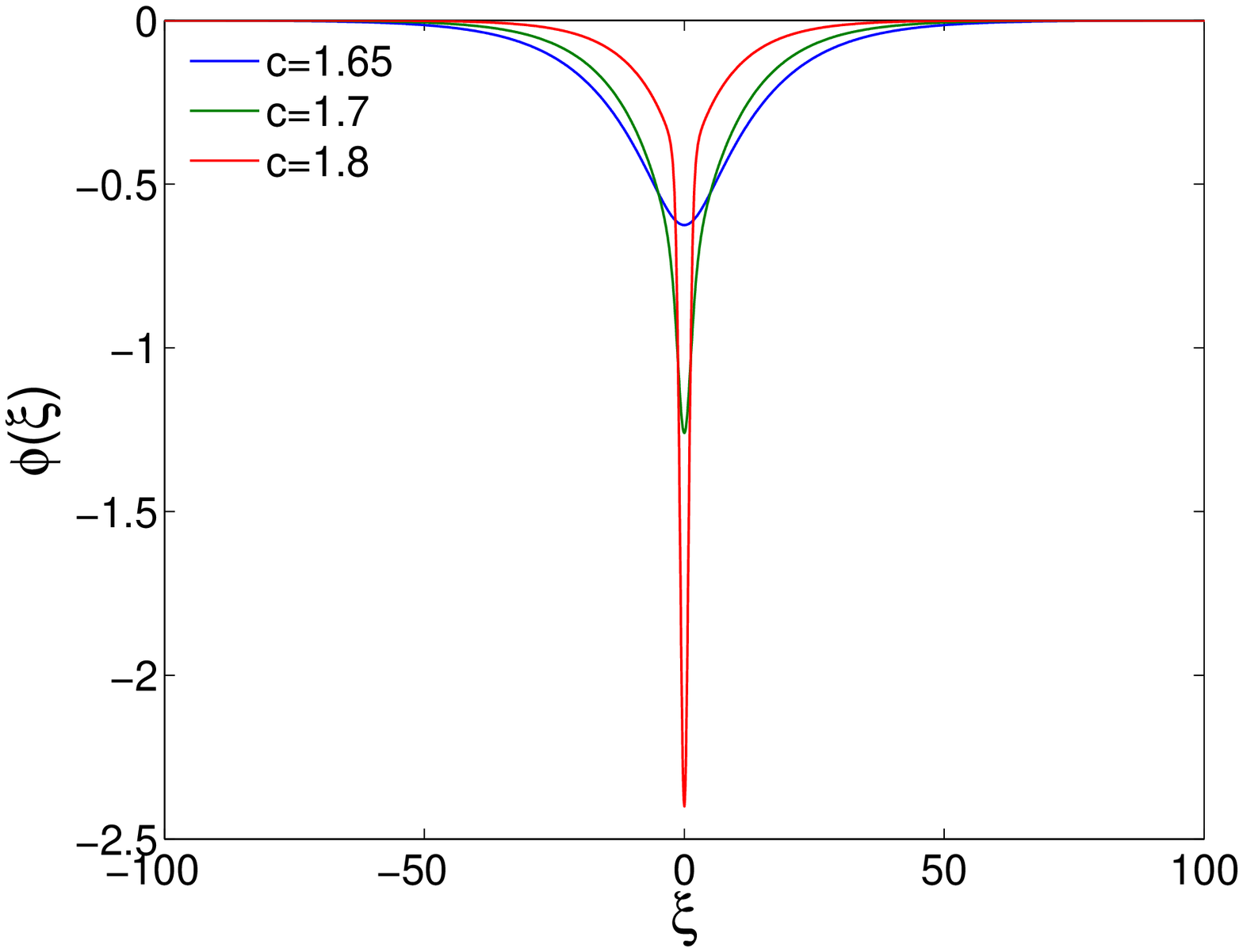}
\end{tabular}
\end{center}
\caption{Typical profiles $\phi(\xi)$ of solitary traveling waves for the regularized potential with $\varepsilon=0.25$, $\chi=4$ and $r_c=1$ (left panel) and the $\alpha$-FPU model with Kac-Baker long-range interactions, for $\gamma=0.18$ and $\rho=0.02$ (right panel).}
\label{fig:profiles}
\end{figure}
Importantly, the smooth regularized potential makes possible the regular near-sonic KdV limit, with solutions delocalizing to zero and $H(c) \rightarrow 0$ as $c \rightarrow 1$. For small enough $\varepsilon$, we have a narrow interval $1<c<c_{0,max}$, where $H(c)$ rapidly increases from zero. The energy then reaches a local maximum value at $c=c_{0,max}$ ($H'(c_{0,max})=0$, $H''(c_{0,max})<0$) and decreases for $c_{0,max}<c<c_{0,min}$ to a local minimum value at $c=c_{0,min}$ ($H'(c_{0,min})=0$, $H''(c_{0,min})>0$) and then starts increasing again. As $\varepsilon$ tends to zero, we have $c_{0,max} \rightarrow 1$, $c_{0,min} \rightarrow c_0$, and the local maximum value $H(c_{0,max})$ tends to infinity. This is illustrated in the left panel of Fig.~\ref{fig:regular2}.
\begin{figure}
\begin{tabular}{cc}
\includegraphics[width=6.0cm]{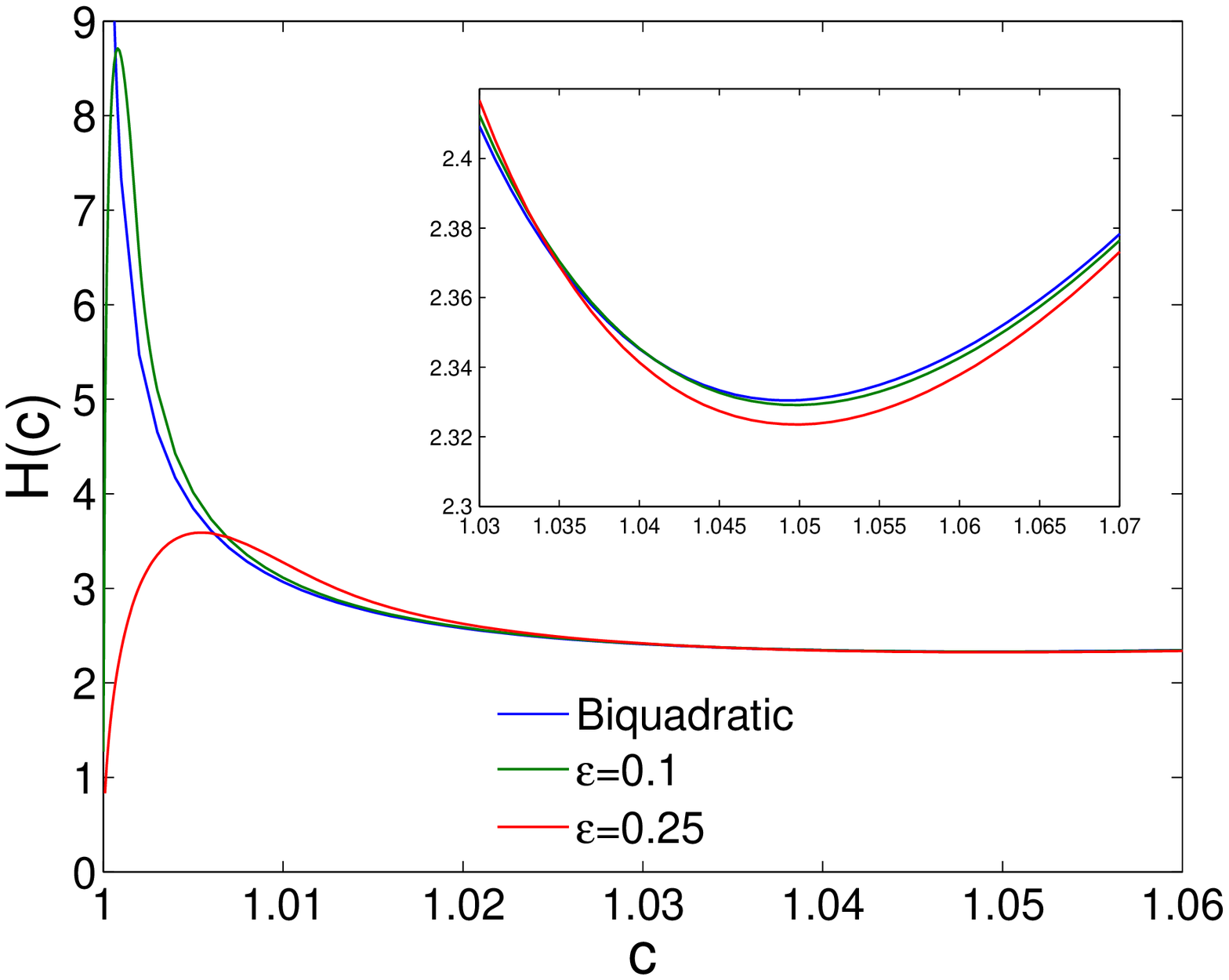} &
\includegraphics[width=6.0cm]{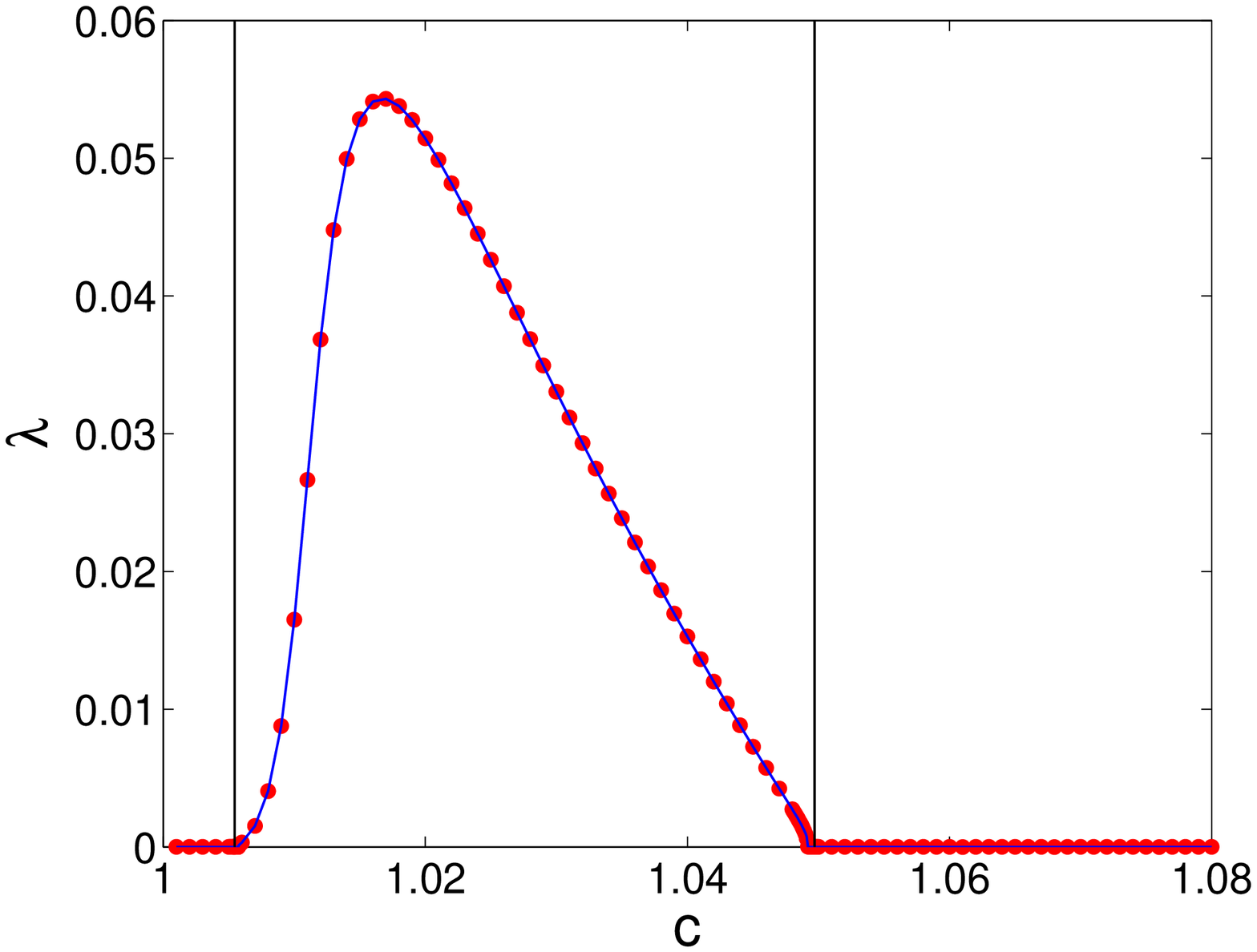}
\end{tabular}
\caption{Traveling solitary waves for the regularized potential with $\varepsilon=0.25$, $\chi=4$, $r_c=1$. The left panel shows the energy-velocity dependence $H(c)$, with $H'(c)>0$ implying (spectral) stability, and $H'(c)<0$ implying instability. The right panel confirms this by showing the {maximum real} eigenvalue obtained by diagonalizing the linearization operator $\mathcal{L}$ (dots) and transforming the relevant Floquet multiplier $\mu$ into a corresponding eigenvalue (for comparison) via the relation $\lambda=\log(\mu)$ (solid curve). {Vertical lines in the right panel indicate the {values of $c$ where} monotonicity of $H(c)$ {changes}.}}
\label{fig:regular2}
\end{figure}

The regularization also allows to analyze the spectral stability of the solitary traveling wave solutions using the two approaches described in Sections~\ref{sec:stab}-\ref{sec:split}, co-traveling steady-state eigenvalue analysis and calculation of Floquet multipliers for periodic orbits (modulo shift). Fig.~\ref{fig:regular2} showcases the power of the stability criterion and illustrates the complementary nature of the two approaches. For the parameter values $\varepsilon=0.25$, $\chi=4$ and $r_c=1$, we have $c_{0,max}=1.00546$ and $c_{0,min}=1.0497$ such that $H'(c)<0$ for $c_{0,max}<c<c_{0,min}$ and $H'(c)>0$ otherwise. In the velocity interval $(c_{0,max},c_{0,min})$, an eigenvalue of the operator $\mathcal{L}$ crosses through $\lambda=0$ {into the positive real axis} (dots in right panel of Fig.~\ref{fig:regular2}), indicating instability. In fact, it can be shown~\cite{pegof04a} that the stability problem in the co-traveling frame also possesses eigenvalues $\lambda + i (2 \pi j )$, where $j \in \mathbb{Z}$, as demonstrated in the left panel of Fig.~\ref{fig:regular3}.
Notice that the finite domain nature of the computation leads to a
number of {\it spurious} unstable eigenvalues in the figure.
However, we have systematically checked (see also Sec.~IV of the Supplementary Material for
relevant details) that
their real part decreases with the system length and hence expect such
spurious instabilities to be absent in the infinite domain limit.
Finally, the solid curve in the right panel of the Fig.~\ref{fig:regular2} shows the Floquet multiplier calculation associated with the time $T=1/c$ map of the corresponding periodic orbit, transformed (in order to compare with the steady state eigenvalue approach) according to the relation $\lambda=\log(\mu)$. Confirming the complementary picture put forth, we find that in this case a pair of Floquet multipliers crosses through $(1,0)$ and into the real axis for the exact same parametric interval. The right panel of Fig.~\ref{fig:regular3} depicts the typical Floquet spectrum for an unstable traveling wave. Notice that there is a slight mismatch between the bifurcation loci predicted by the local extrema of $H(c)$ (denoted by $c_{0,max}$ and $c_{0,min}$) and the values predicted by the eigenvalues of the linearization operator $\mathcal{L}$ (denoted by $c_{0,max}^l$ and $c_{0,min}^l$, respectively). When the system length $l$ is increased, $c_{0,max}$ and $c_{0,min}$ remain invariant but $c_0^l$ approaches the corresponding $c_0$ values. This is a consequence of the fact that for small $l$ the solution at the boundaries is larger than the machine precision.
\begin{figure}
\begin{center}
\begin{tabular}{cc}
\includegraphics[width=6.0cm]{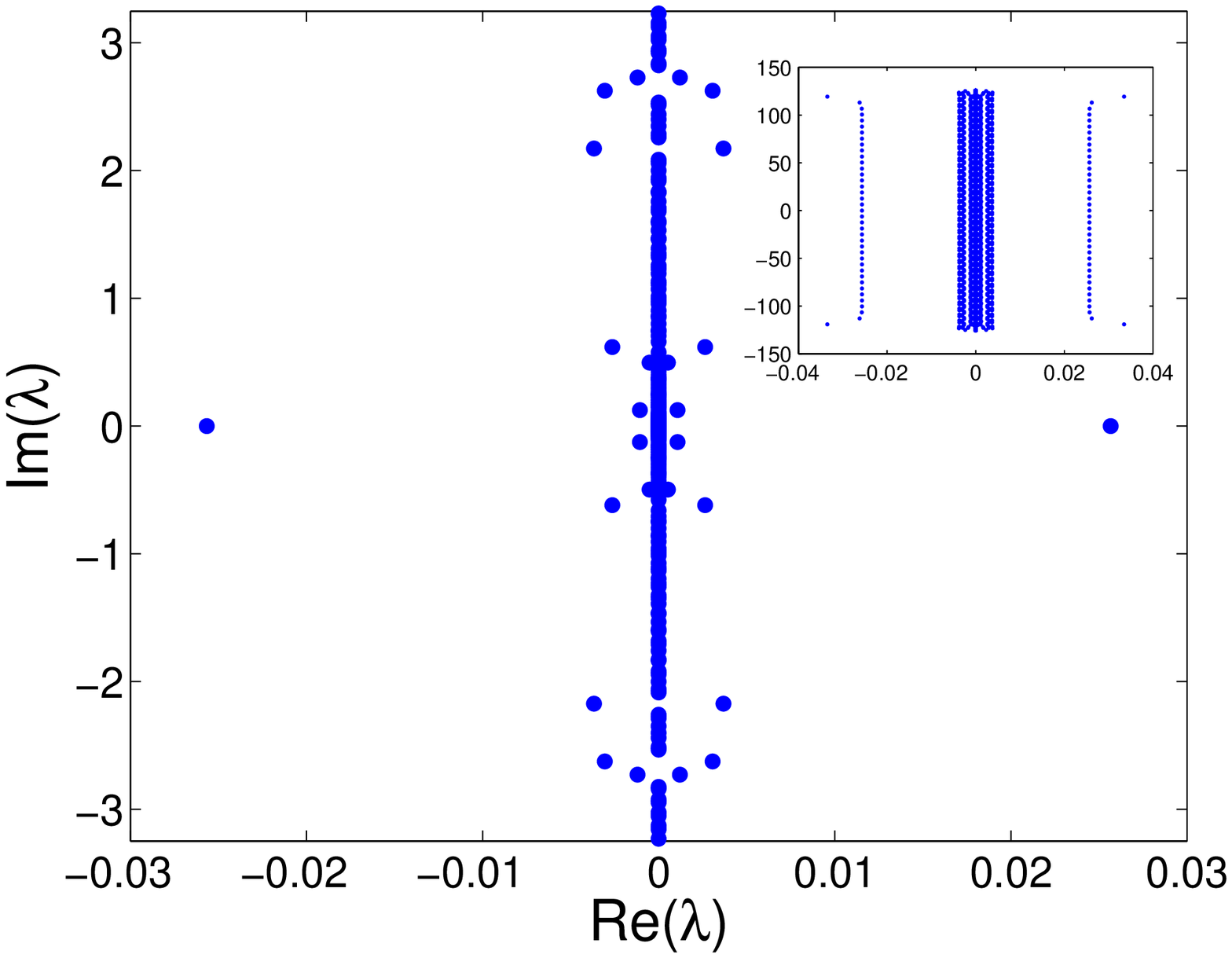} &
\includegraphics[width=6.0cm]{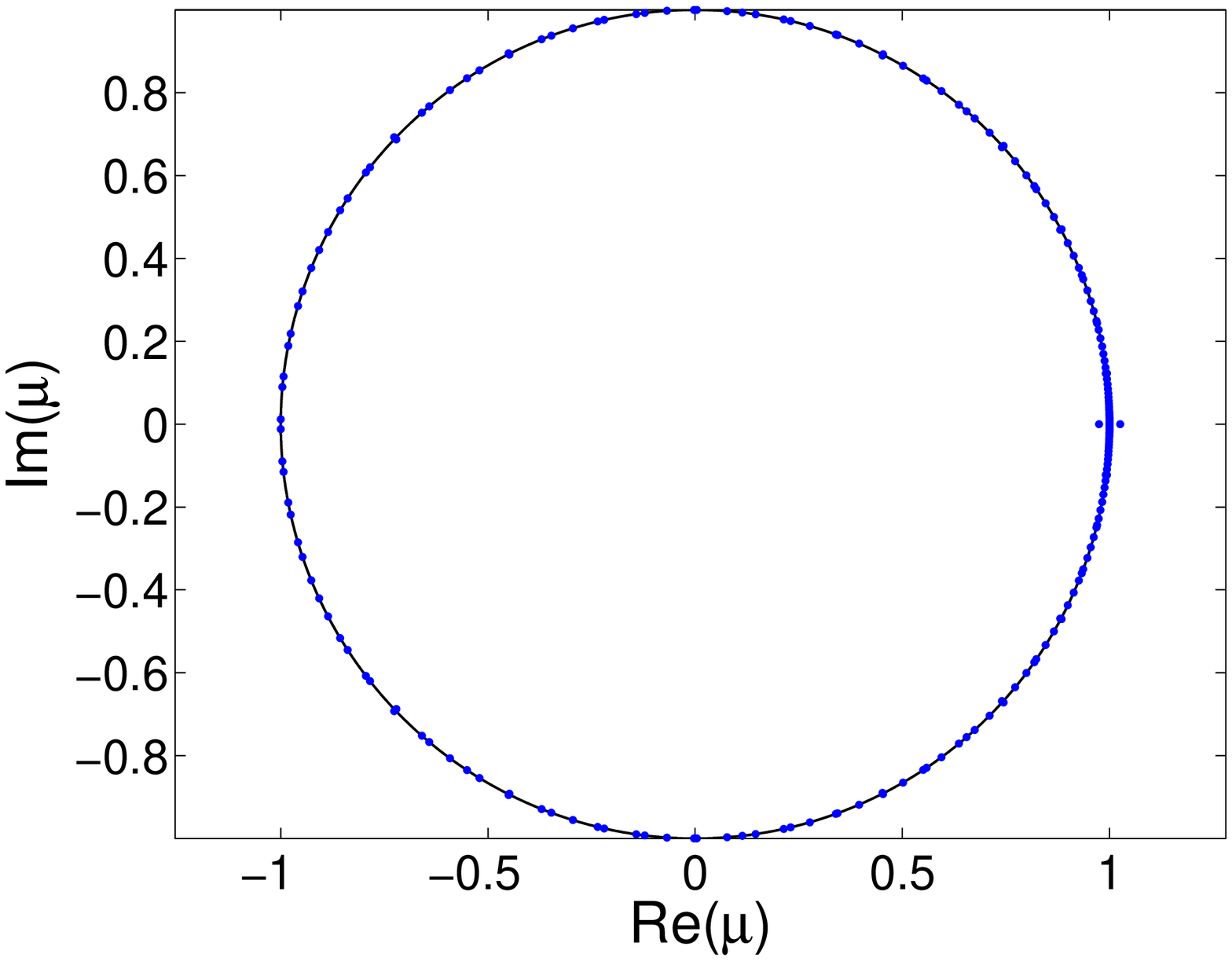}
\end{tabular}
\end{center}
\caption{Stability analysis of the unstable solitary traveling wave with velocity $c=1.034$ for the regularized potential with $\varepsilon=0.25$, $\chi=4$, $r_c=1$. The left panel shows the spectrum of the linearization operator $\mathcal{L}$ for $\mathrm{Im}(\lambda)\in(-\pi,\pi )$; the inset corresponds to the full spectrum. In the right panel, the Floquet multipliers spectrum is shown.}
\label{fig:regular3}
\end{figure}

To connect these results with the theoretical analysis of \eqref{eq5}, Fig.~\ref{fig:regular4} shows the dependence of $\lambda^2$ with respect to $c$ near $c_0^l$ for both bifurcations, which according to Theorem~\ref{lemma1}, must be linear in the vicinity of $c_0^l$ with the slope
\beq
\beta_t=H''(c_0)/\alpha_1,
\label{eq:beta}
\eeq
    where $\alpha_1$ is defined in \eqref{eq:M}. {{To compute $\beta_t$} in \eqref{eq:beta}, we use $c_0$, i.e. the value of $c$ at which $H'(c)=0$; however, for determining $\alpha_1$ we calculate the eigenmodes at $c_0^l$, as this value corresponds to the point where we are closer to a quadruplet of zero eigenvalues}. Table~\ref{tab:regular} shows the numerical value of the slope $\beta$ (obtained by the best linear fit) and its theoretical prediction \eqref{eq:beta} for both bifurcations, at $c_{0,max}$ and $c_{0,min}$, where $\alpha_1$ is computed numerically. There is a mismatch of $\sim1\%$ at the second bifurcation, at $c_{0,min}$. This mismatch is likely due to the fact that $\alpha_1$ in Lemma~\ref{lemma1} cannot be computed at the precise value of $c_0^l$ in the numerical setup. The theoretical prediction $\beta_t$ at the first bifurcation
    %, at $c_{0,max}$, is far from the
    at $c_{0,max}$ is less accurate in comparison to the numerical value of
    $\beta$ (although it has the same order of magnitude). This is
    likely caused by the fact that the solitary traveling wave is rather wide at the bifurcation (recall that the solitary waves delocalize when $c$ approaches the sonic limit $c_s=1$, as illustrated in Fig.~\ref{fig:profiles}), and $\mathcal{L}e_1$ and $\mathcal{L}^2e_2$ are not precisely in the kernel of $\mathcal{L}$, causing an error in computing $\alpha_1$ and therefore $\beta_t$. We note that, although the value $l=100$ has been fixed for the length of the system in the calculations leading to Fig.~\ref{fig:regular3} and Table~\ref{tab:regular}, we used different values of the mesh size $\Delta\xi$ (see Sec.~IV of the Supplementary Material for details) at the two bifurcations,
$\Delta\xi=0.025$ at the first one and $\Delta\xi=0.0125$ at the second. These choices were motivated by the fact that at high $c$ the solitary wave must be resolved with a finer grid in order to \emph{feel} the regularization of the potential. Better results for the theoretical predictions at the first bifurcation could be obtained {by substantially increasing the system's length}; however, this would render the linearization problem rather unwieldy.
\begin{figure}
\begin{center}
\begin{tabular}{cc}
\includegraphics[width=6.0cm]{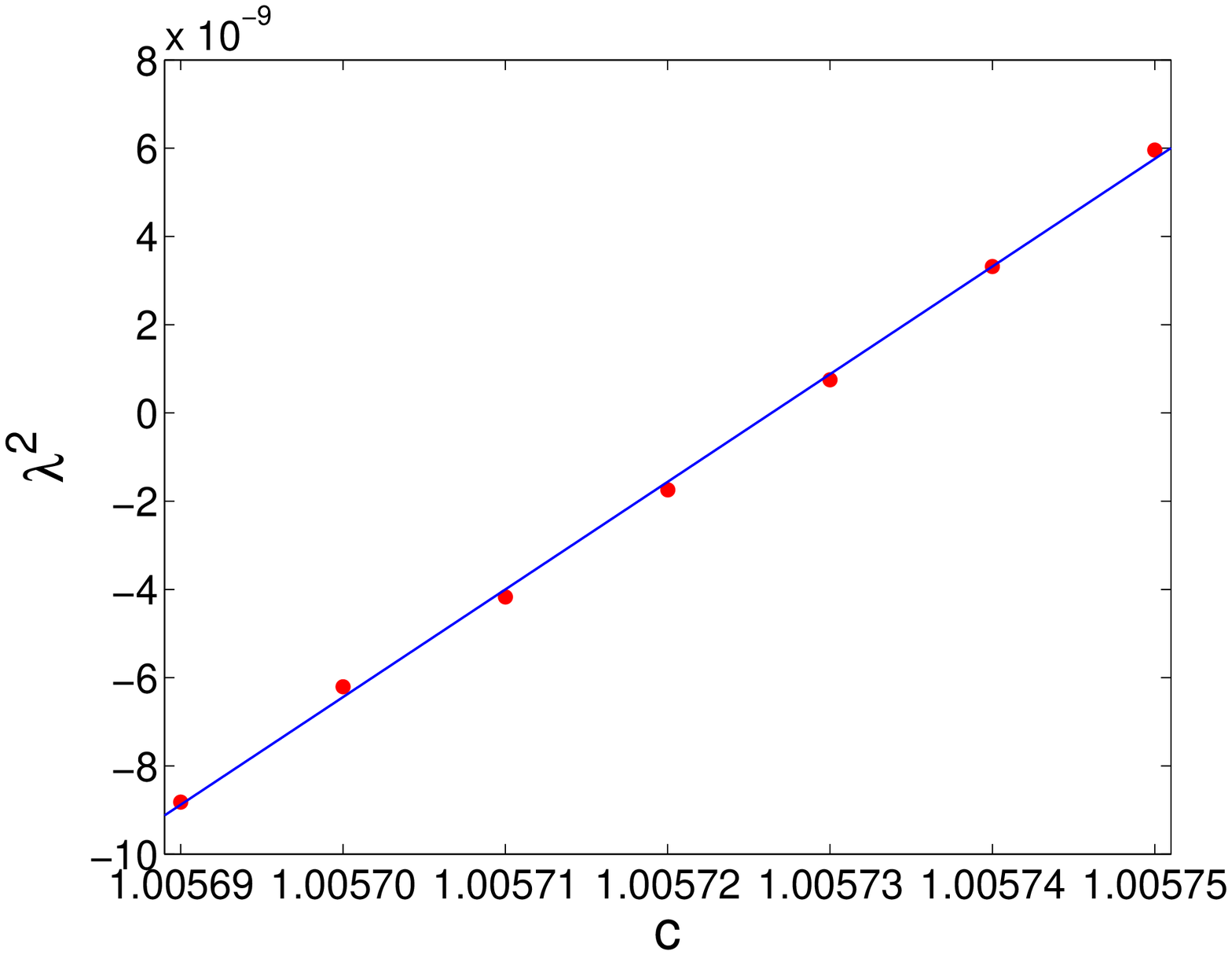} &
\includegraphics[width=6.0cm]{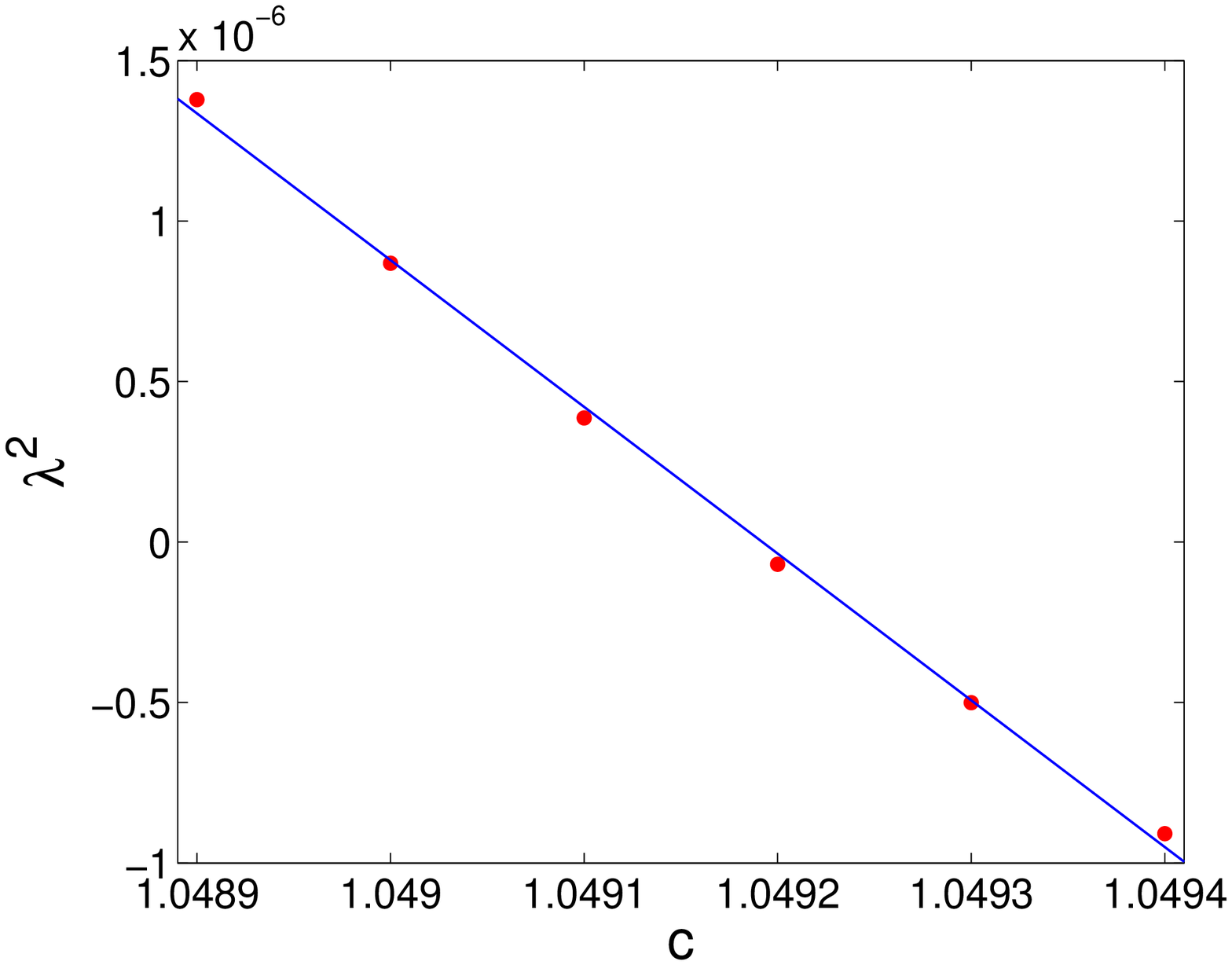}
\end{tabular}
\end{center}
\caption{Dependence of the squared linearization eigenvalue $\lambda^2$ on $c$ for solitary wave solutions for the regularized potential with $\varepsilon=0.25$, $\chi=4$, $r_c=1$ in the vicinity of bifurcations at $c_{0,max}$ (left panel) and $c_{0,min}$ (right panel). The straight line {in each panel} represents the best {linear} fit from which {the corresponding value of} $\beta$ is inferred numerically.}
\label{fig:regular4}
\end{figure}
\begin{table}[tb]
\caption{Comparison between the numerical slope $\beta$ and its theoretical prediction $\beta_t$ in \eqref{eq:beta} of the ${\lambda}^2$ vs $c$ dependence at the bifurcations $c_{0,max}$ and $c_{0,min}$ in the problem with the regularized potential ($\varepsilon=0.25$, $\chi=4$, $r_c=1$). Parameters involved in the theoretical calculations are also shown. Here the first column shows the values of $c_0$ at which $H'(c_0)=0$ ($c_{0,max}$ in the first row and $c_{0,min}$ in the second), and the second column shows the corresponding values of $c_0^l$ where an eigenvalue crosses zero (or a Floquet multiplier crosses $1$), together with $H''(c_0)$ and $\alpha_1$.}
\label{tab:regular}
\begin{center}
\begin{tabular}{cccccc}
\noalign{\smallskip}\hline\noalign{\smallskip}
$c_0$ & $c_0^l$ & $H''(c_0)$ & $\alpha_1$ & $\beta_t$ & $\beta$ \\
\noalign{\smallskip}\hline\noalign{\smallskip}
$1.00546$ & $1.00572$ & $-5.3912\times10^{4}$ & $-5.6315\times10^{7}$ & $9.5734\times10^{-4}$ & $2.4400\times10^{-4}$ \\
$1.0497$ & $1.0492$ & $3.2588\times10^{2}$ & $-7.0461\times10^{4}$ & $-4.6250\times10^{-3}$ &  $-4.5713\times10^{-3}$\\
\hline\noalign{\smallskip}
\end{tabular}
\end{center}
\end{table}

Despite the numerical issues discussed above, these results are generally in a very good agreement with the theoretical predictions. In addition to showing instability when $H'(c)<0$ and the linear dependence of the squared eigenvalue on $c-c_0$, in agreement with Theorem~\ref{lemma1}, they also suggest that $H'(c)>0$ corresponds to spectral stability. In particular, the solitary traveling waves are apparently stable at $1<c<c_{0,max}$, which is consistent with stability of near-sonic waves proved in \cite{pegof04b}. This example also provides a nice explanation of the instability of solitary waves near the lower sonic limit $c=1$ at $\varepsilon=0$, when $c_{0,max}=1$, and the problem no longer has a KdV limit (unless one allows $r_c$ and $\chi$ to depend on $c$ \cite{at}).

To complement these results by means of direct numerical simulations,
we now consider the dynamic evolution of unstable traveling waves by solving \eqref{eq:dyn} numerically (see Sec.~IV of the Supplementary Material for details) with initial conditions given by an unstable traveling wave perturbed along the direction of the Floquet mode causing the instability. The typical dynamical scenario is shown in Fig.~\ref{fig:dynamics1} for the unstable wave with $c=1.034$. The eigenvalue spectrum of this solution is shown in Fig.~\ref{fig:regular3}. One can see that linear waves are continuously being created, and the solitary wave degrades with time.

\begin{figure}
\begin{tabular}{cc}
\includegraphics[width=6.0cm]{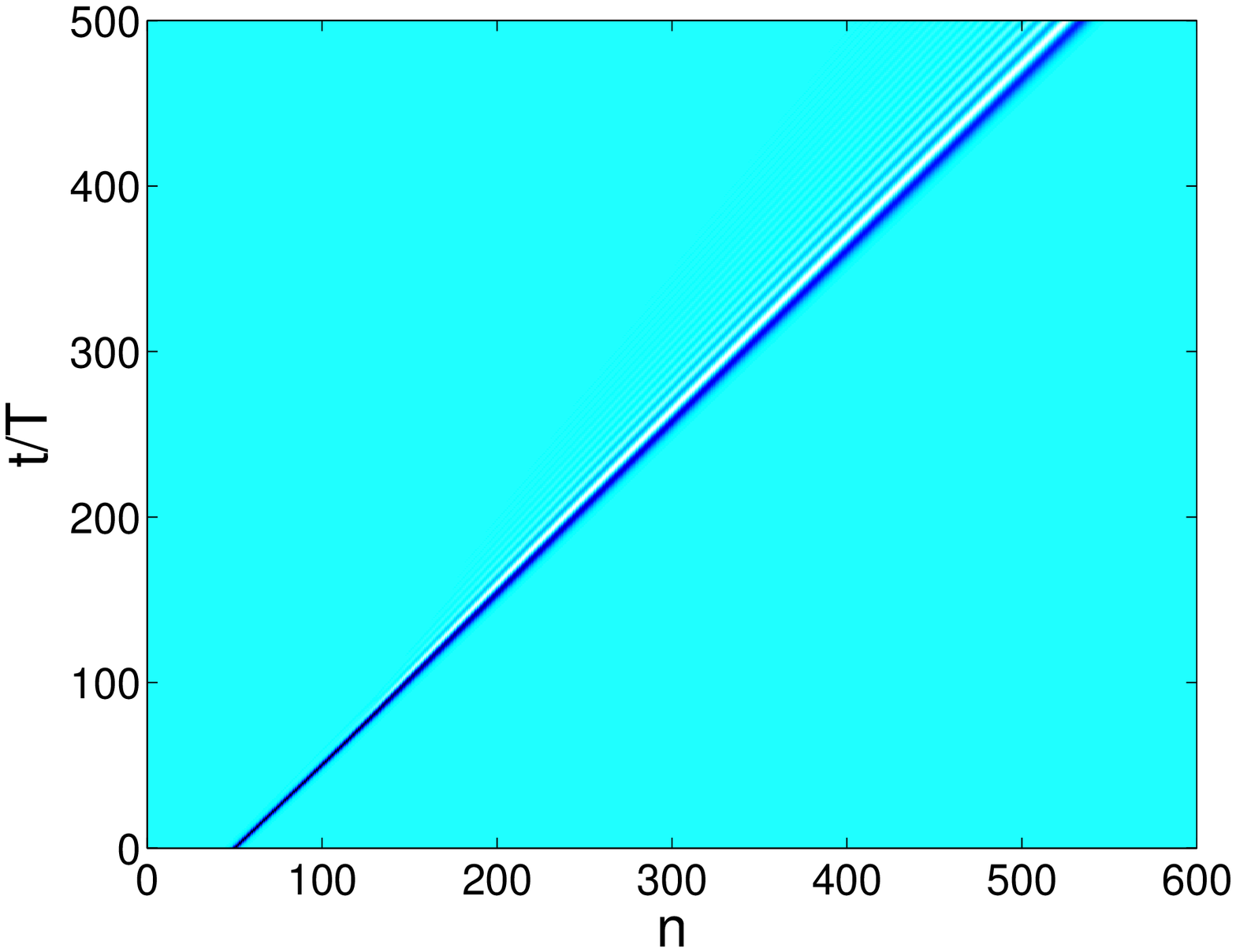} &
\includegraphics[width=6.0cm]{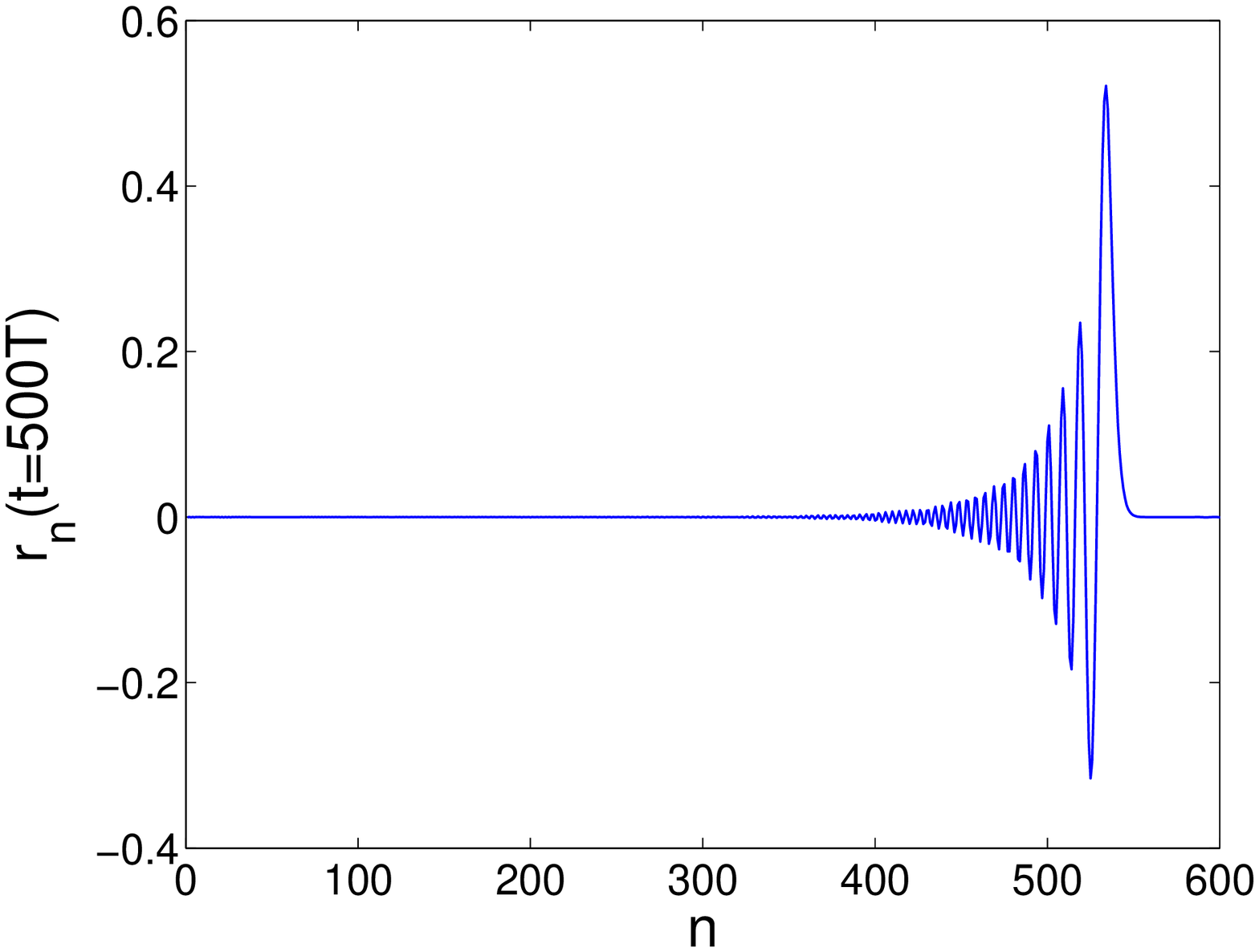}
\end{tabular}
\caption{Evolution of the unstable solitary traveling wave with $c=1.034$ for the regularized potential whose spectrum is represented in Fig.~\ref{fig:regular3}. Left panel shows the the space-time evolution dynamics of the strains, while the right panel shows the strain profile at $t=500T$, where $T=1/c$. Here $\varepsilon=0.25$, $\chi=4$ and $r_c=1$. The decay of the structure through
  the creation of a state reminiscent of a dispersive shock
  wave~\cite{EL201611} is
evident.}
\label{fig:dynamics1}
\end{figure}

Turning now to our second example, we consider the model studied in \cite{mertens2,Neuper94, Gaididei95,Gaididei97}, the $\alpha-$FPU lattice \cite{FPU55,Campbell05,Gallavotti07} with long-range Kac-Baker interactions that have exponentially decaying kernel. This corresponds to \eqref{eq:Ham} with
\beq
V(r)=\dfrac{r^2}{2}-\dfrac{r^3}{3}, \qquad \Lambda(m)=\rho(\e^\gamma-1)\e^{-\gamma|m|}(1-\delta_{m,0}).
\label{eq:example2}
\eeq
Long-range interactions of this type have been argued to be of relevance for modeling Coulomb interactions in DNA \cite{mertens2}.
As shown in \cite{mertens2}, when
\[
\rho_1<\rho<\rho_2,
\]
where
\begin{equation}
\label{eq:kacbakercond}
    \rho_1 \approx \begin{cases}0.23\frac{\gamma^4}{\gamma_1^2-\gamma^2}, & \gamma<\gamma_1\\
    \infty, & \gamma \geq \gamma_1,
    \end{cases}
    \qquad
    \rho_2 \approx \begin{cases}\frac{3}{8}\frac{\gamma^4}{\gamma_2^2-\gamma^2}, & \gamma<\gamma_2\\
    \infty, & \gamma \geq \gamma_2
    \end{cases}
\end{equation}
and $\gamma_1=0.25$ and $\gamma_2=0.16$, the energy $H(c)$ of the supersonic solitary traveling waves in this system is  non-monotone so that the curve presents a local maximum and a local minimum. Clearly, for this to hold it is necessary that $\gamma<\gamma_1$. The energy of the waves tends to zero when $c$ approaches the sound velocity
\[
c_s=\sqrt{1+\rho\frac{1+\exp(-\gamma)}{(1-\exp(-\gamma))^2}}.
\]

In what follows, we choose $\gamma=0.18$ and $\rho=0.02$, which places the system in the regime with non-monotone $H(c)$ and yields $c_s=1.5338$. The profiles of solitary traveling waves with different velocities $c$ are depicted in the right panel of Fig.~\ref{fig:profiles}. The resulting non-monotone $H(c)$ curve is shown in the left panel of Fig.~\ref{fig:lri1}. In particular, $H'(c)$ is negative for $c_{0,max} < c < c_{0,min}$, where $c_{0,max}=1.6958$ and $c_{0,min}=1.7185$ denote, as in the previous example, the velocity values at which $H(c)$ reaches a local maximum and a local minimum, respectively. Very close to this interval of velocities, an eigenvalue of the operator $\mathcal{L}$ crosses through ${\lambda}=0$ (dots in right panel of Fig.~\ref{fig:lri1}) and a Floquet mode pair crosses $(1,0)$ (solid line in right panel of Fig.~\ref{fig:lri1}). As discussed above, there are numerical effects caused by the finite length $l$ of the lattice; as a result, the interval of instabilities observed by means of the spectral stability analysis is $1.6971< c < 1.7137$ for $l=400$ and $1.6964 < c < 1.7161$ for $l=800$ (in both cases, we used $\Delta \xi=0.1$). Clearly, however, the instability interval approaches to the interval for which $H'(c)<0$.
\begin{figure}
\begin{tabular}{cc}
\includegraphics[width=6.5cm]{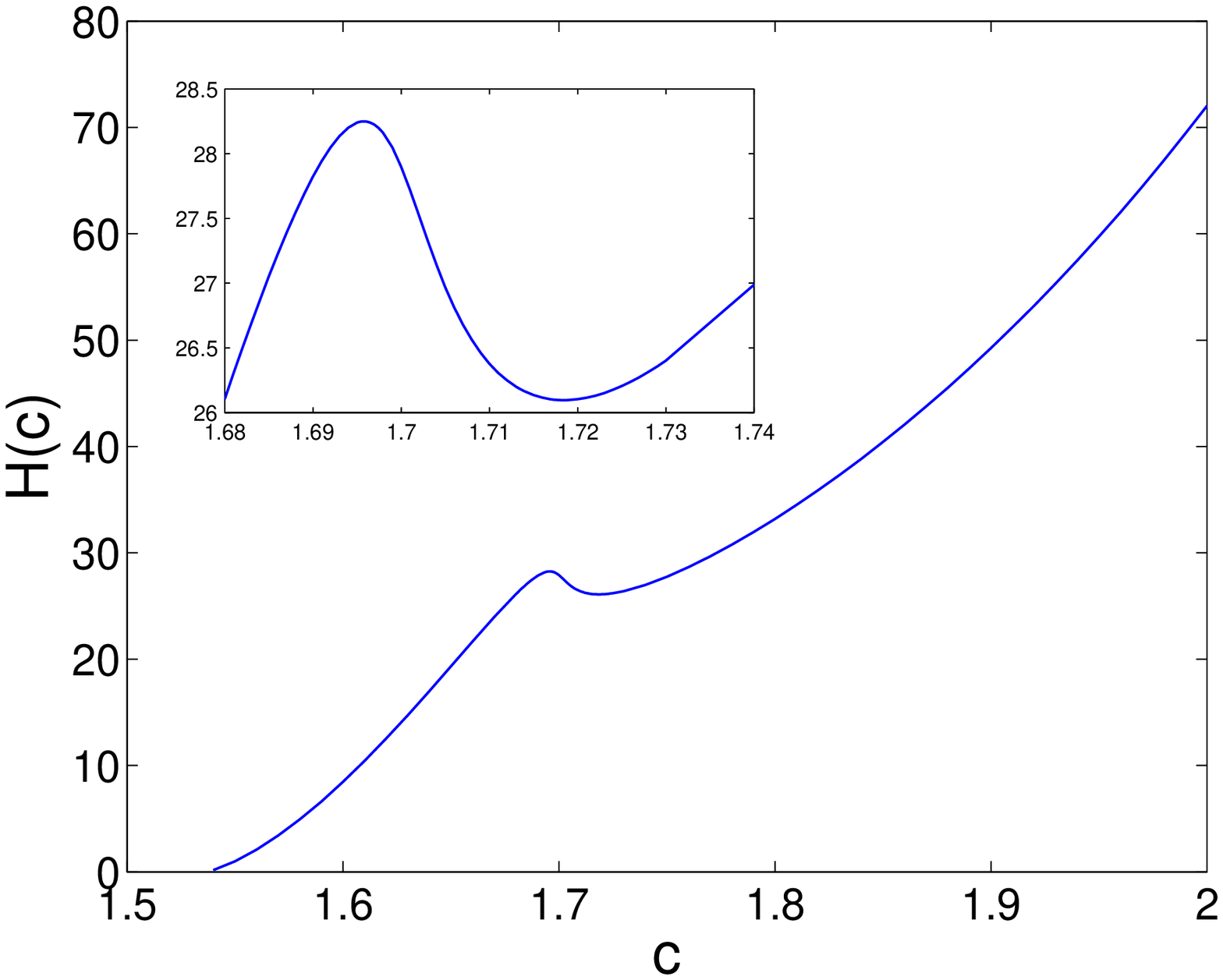} &
\includegraphics[width=6.5cm]{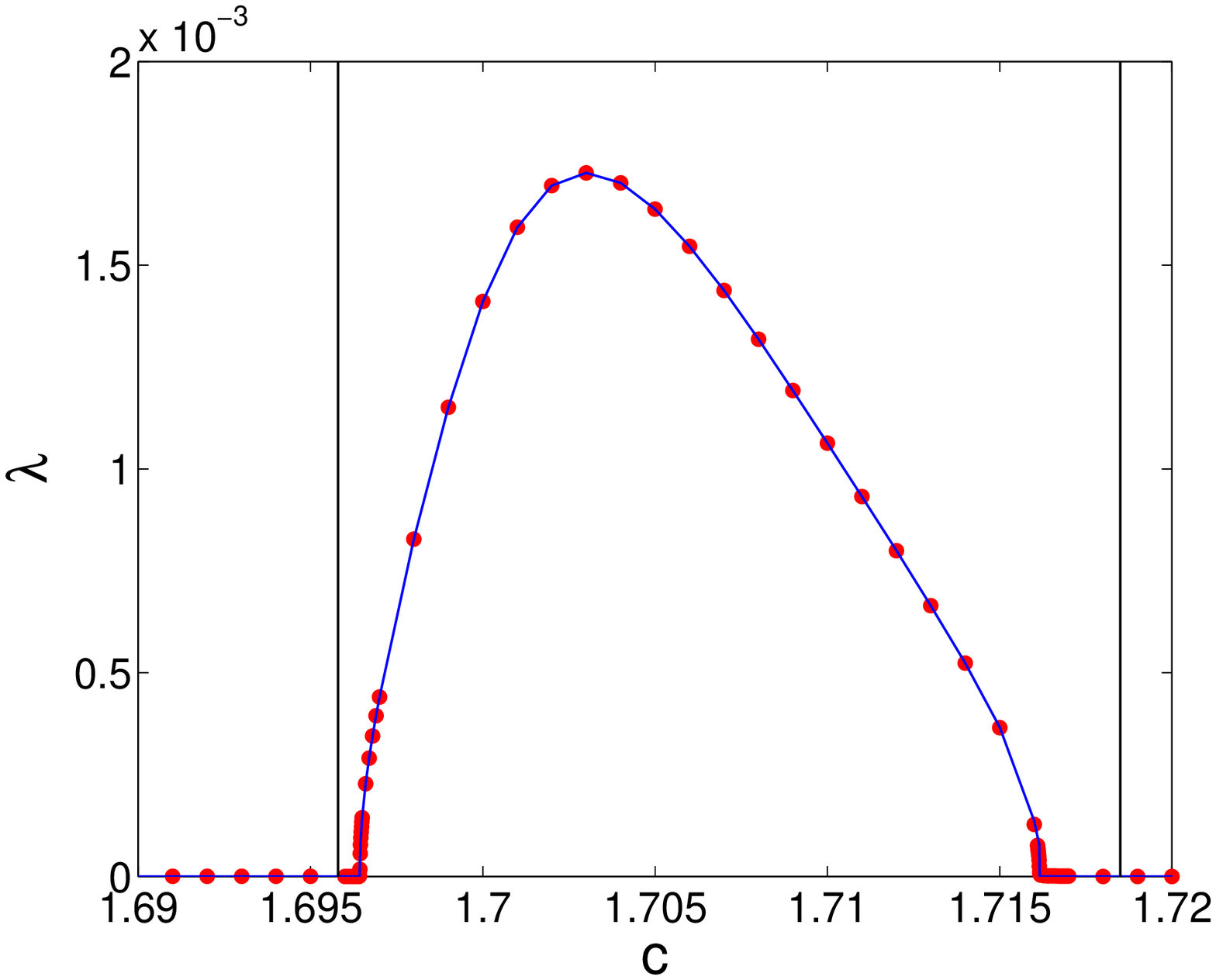}
\end{tabular}
\caption{Traveling solitary waves in the $\alpha$-FPU potential with Kac-Baker long-range interactions for $\gamma=0.18$ and $\rho=0.02$. Left panel shows the energy dependence on the velocity of the waves, with $H'(c)>0$ implying (spectral) stability, and $H'(c)<0$ implying instability. The inset zooms in on the portion of the curve where $H'(c)$ changes sign. The right panel confirms this by showing the {maximum real} eigenvalue obtained by diagonalizing the linearization operator $\mathcal{L}$ (dots) and transforming the relevant Floquet multiplier $\mu$ into a corresponding eigenvalue (for comparison) via the relation ${\lambda}=\log(\mu)$ (solid curve). {The vertical lines in the right panel indicate the {values of $c$ where} monotonicity of $H(c)$ {changes}.}}
\label{fig:lri1}
\end{figure}
Fig.~\ref{fig:lri2} shows the spectrum of $\mathcal{L}$ and the Floquet mode spectrum for an unstable traveling wave with velocity $c=1.7$ in this interval.
\begin{figure}
\begin{center}
\begin{tabular}{cc}
\includegraphics[width=6.5cm]{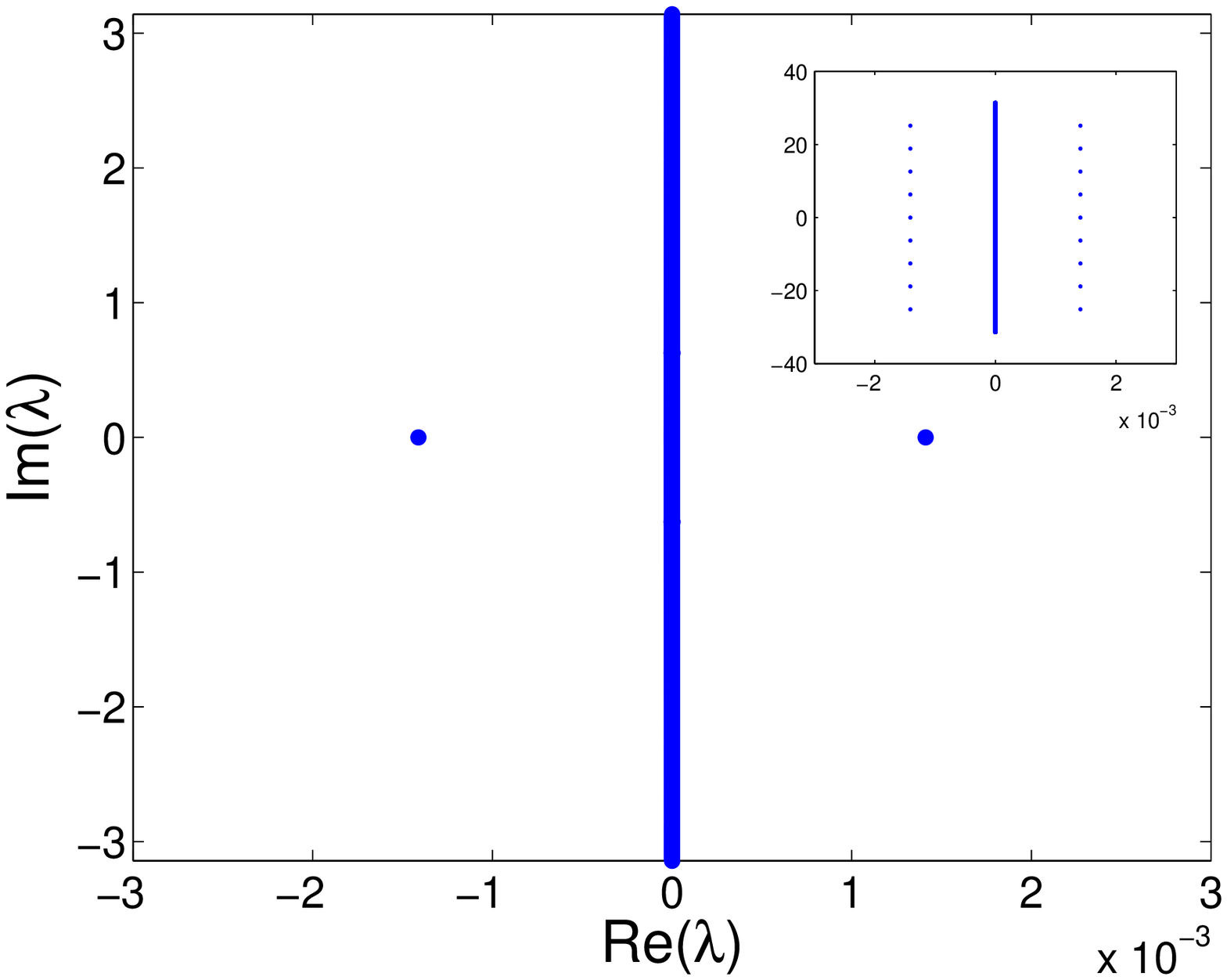} &
\includegraphics[width=6.5cm]{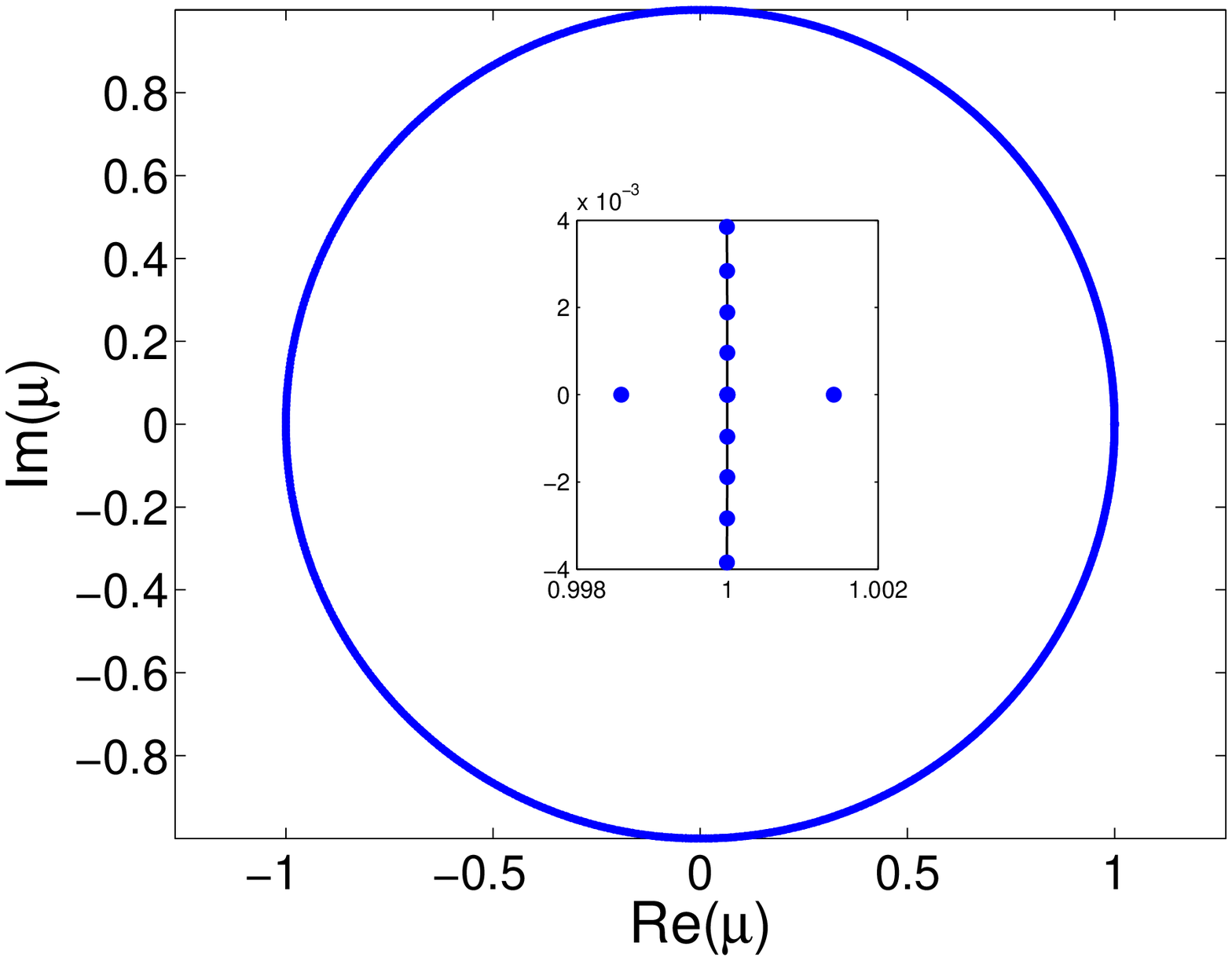}
\end{tabular}
\end{center}
\caption{Stability analysis of the unstable traveling wave with velocity $c=1.7$ for the $\alpha$-FPU potential with Kac-Baker long-range interactions for $\gamma=0.18$ and $\rho=0.02$. The left panel shows the spectrum of the linearization operator $\mathcal{L}$ for $\mathrm{Im}({\lambda})\in(-\pi ,\pi )$; the inset corresponds to the full spectrum. In the right panel, the Floquet multiplier spectrum is shown.}
\label{fig:lri2}
\end{figure}

The connection with the theoretical analysis of Theorem~\ref{lemma1} is presented in Fig.~\ref{fig:lri3}, where the dependence of ${\lambda}^2$ with respect to $c$ close to $c_0^l$ is shown in the vicinity of both bifurcations, while Table~\ref{tab:lri} compares the corresponding numerical values of the slope $\beta$ and its theoretical prediction \eqref{eq:beta}. For doing such comparisons, numerical computation has been performed using $L=800$ and $\Delta\xi=0.1$. There is a mismatch of $\sim9\%$ and $\sim5\%$ at the first and second bifurcation, respectively. The higher mismatch at the first bifurcation is a consequence of the slower decay of the solitary wave in that velocity interval. One should also keep in mind the potential contribution from the essential spectrum (which is expected to be of higher order) in the framework with unweighted spaces considered here. Generally,
however, we find the agreement between theoretical and numerical
results to be very good, which makes it reasonable to neglect the effects of the essential spectrum at the current order.

\begin{figure}
\begin{center}
\begin{tabular}{cc}
\includegraphics[width=6.5cm]{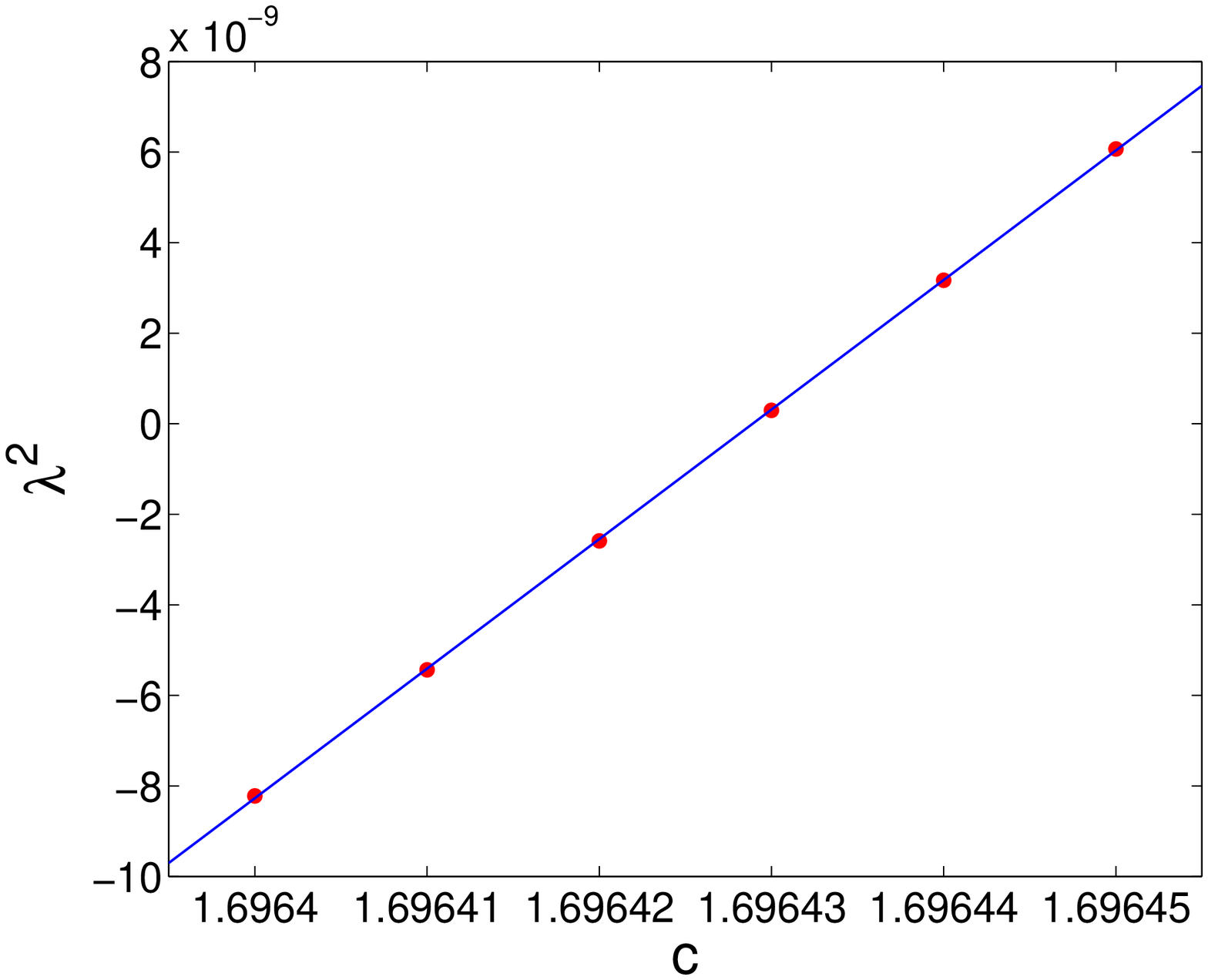} &
\includegraphics[width=6.5cm]{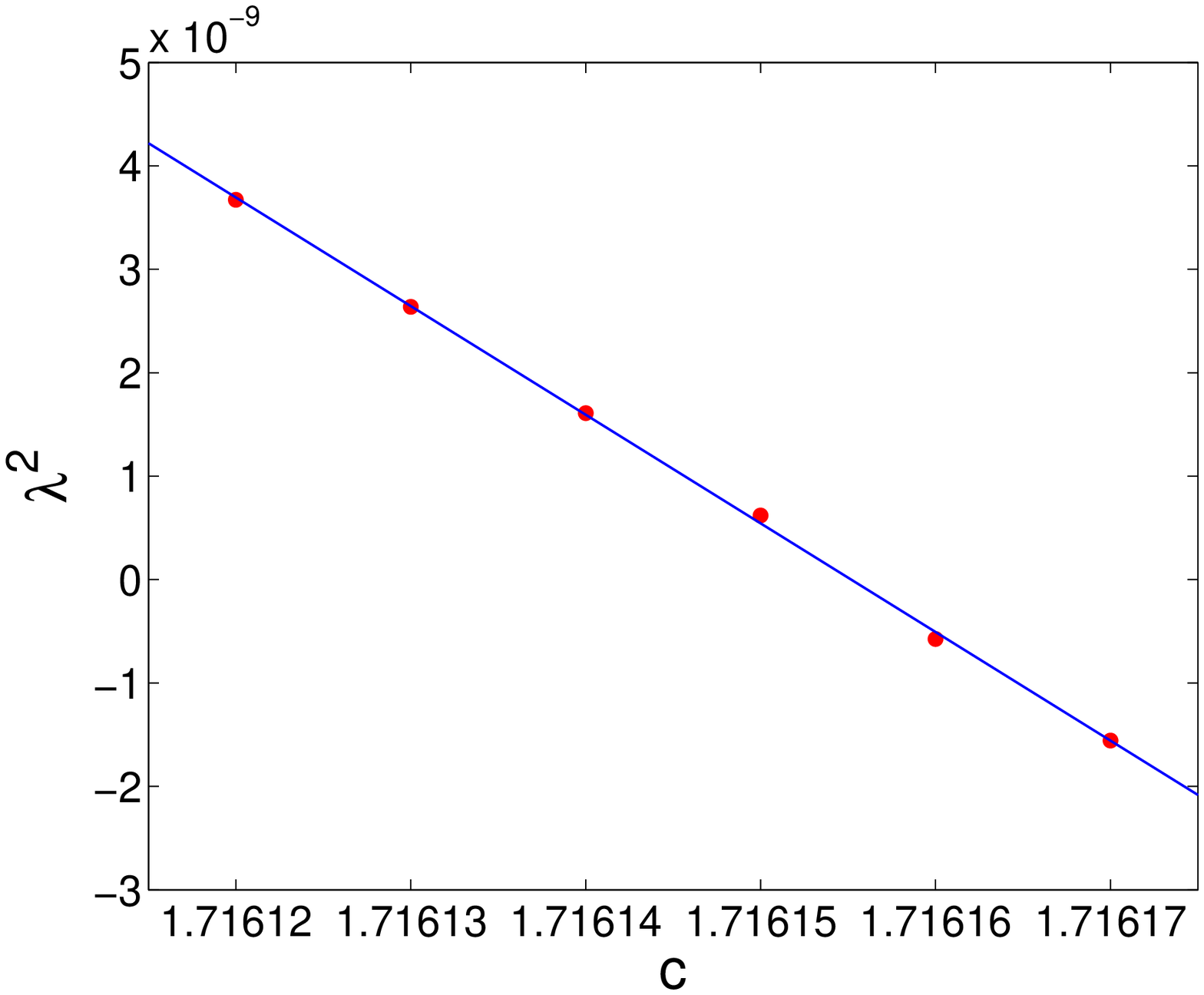}
\end{tabular}
\end{center}
\caption{Dependence of the squared linearization eigenvalue ${\lambda}^2$ on $c$ for solitary wave solutions for the $\alpha$-FPU potential with Kac-Baker long-range interactions for $\gamma=0.18$, $\rho=0.02$ in the vicinity of bifurcations at $c_{0,max}$ {(left panel)} and $c_{0,min}$ {(right panel)}. The solid line {in each panel} corresponds to the best linear fit from which the {corresponding} numerical value of $\beta$ is obtained.}
\label{fig:lri3}
\end{figure}

\begin{table}[tb]
\caption{Same as Table~\ref{tab:regular} but for the $\alpha$-FPU model with Kac-Baker long-range interactions for $\gamma=0.18$ and $\rho=0.02$.}
\label{tab:lri}
\begin{center}
\begin{tabular}{cccccc}
\noalign{\smallskip}\hline\noalign{\smallskip}
$c_0$ & $c_0^l$ & $H''(c_0)$ & $\alpha_1$ & $\beta_t$ & $\beta$ \\
\noalign{\smallskip}\hline\noalign{\smallskip}
$1.6958$ & $1.69642$ & $-3.4325\times10^{4}$ & $-1.0922\times10^{8}$ & $3.1428\times10^{-4}$ &  $2.8617\times10^{-4}$\\
$1.7185$ & $1.71615$ & $5.9857\times10^{3}$ & $-6.0313\times10^{7}$ & $-1.0505\times10^{-4}$ &  $-0.9924\times10^{-4}$\\
\hline\noalign{\smallskip}
\end{tabular}
\end{center}
\end{table}

Finally, we discuss the dynamics of unstable traveling waves perturbed along the direction of the Floquet mode causing the instability. As a typical example, we consider the unstable solution whose spectrum is shown in Fig.~\ref{fig:lri2}. One can see that a linear wave {moving at the speed of sound} is expelled from the {supersonic} solitary wave. {Subsequently, the solitary wave moves with a velocity $c=1.7478$, which corresponds to a stable wave} {(note that $H'(c)>0$ at this value, according to the inset in the left panel of Fig.~\ref{fig:lri1})}, as shown in Fig.~\ref{fig:dynamics2}.
 %slightly altering its shape (and without noticeable change in its velocity), as shown in Fig.~\ref{fig:dynamics2}. {\bf AV: this almost looks like the wave is stable. Is there a way to perhaps apply a larger kick in the initial conditions, to see if we can get the system to approach a stable wave with either larger or smaller velocity instead?}

%
\begin{figure}
\begin{tabular}{cc}
\includegraphics[width=6.5cm]{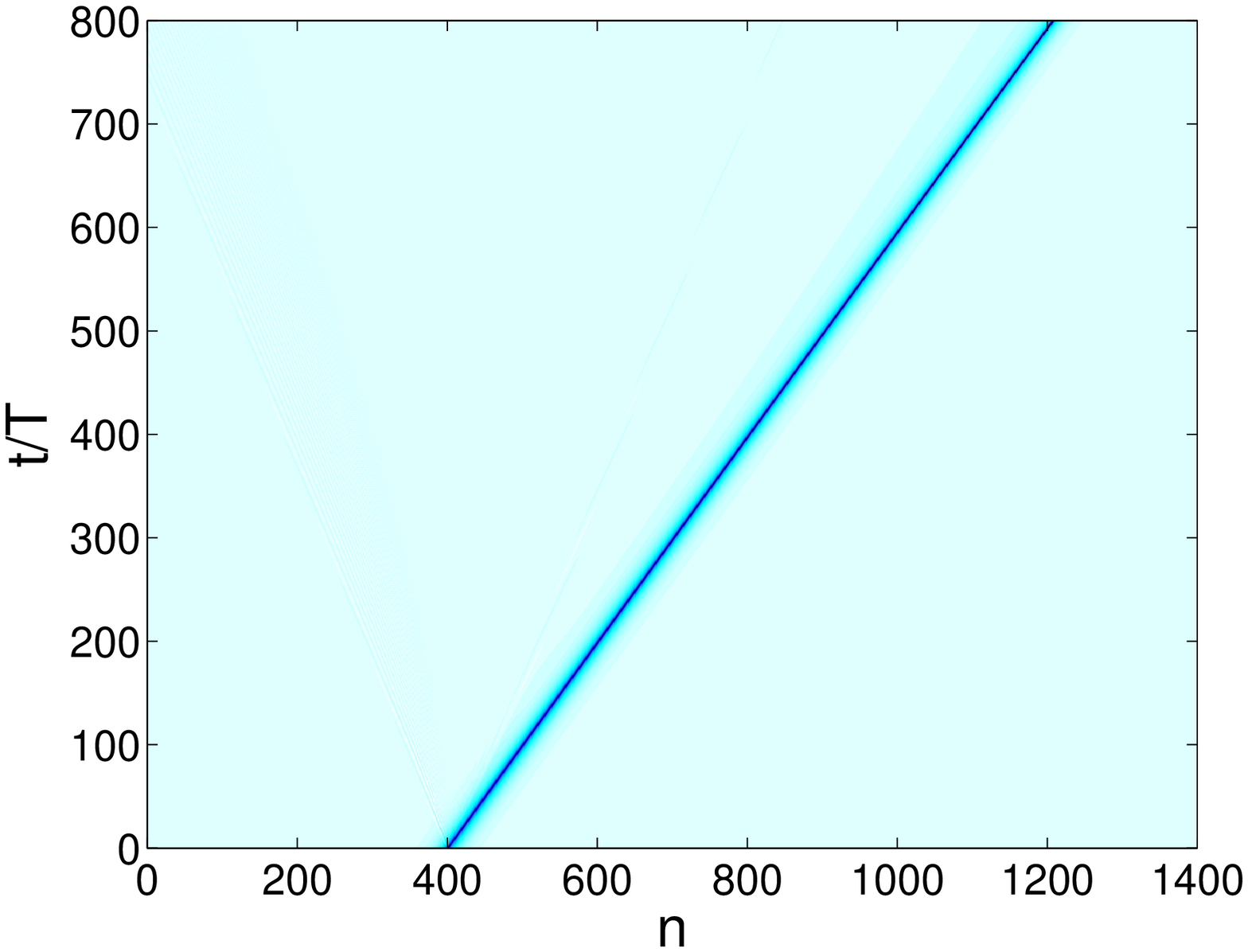} &
\includegraphics[width=6.5cm]{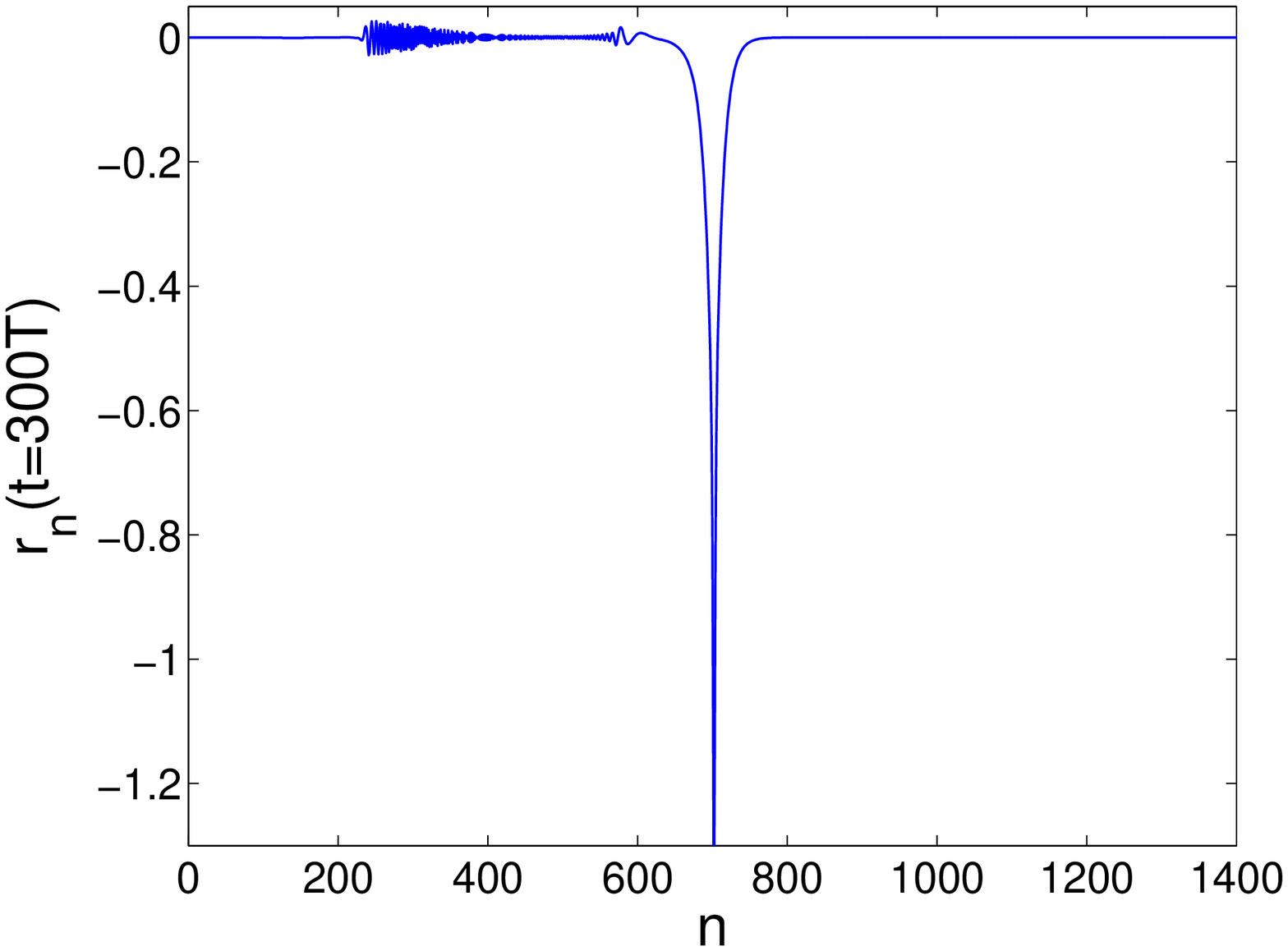}
\end{tabular}
\caption{Evolution of the unstable traveling wave with velocity $c=1.7$ for the $\alpha$-FPU model with Kac-Baker interaction for $\gamma=0.18$ and $\rho=0.02$. The spectrum of the corresponding linearization operator is shown in Fig.~\ref{fig:lri2}. The left panel shows the the space-time evolution dynamics of the strains, whereas the right panel shows the profile $t=300T$, where $T=1/c$.}
\label{fig:dynamics2}
\end{figure}

\section{Concluding remarks and future challenges}
\label{sec:conclude}

In {this} work, we have revisited the {stability} problem of {solitary} traveling
waves in {infinite-dimensional nonlinear Hamiltonian}
lattices. {We considered a class of problems where such waves exist for a range of velocities, with
our examples predominantly drawn from the pool of FPU-type models}. We have illustrated the useful {stability} criterion originally put forth
in the work of Friesecke and Pego~\cite{pegof04a} and have extended
it through a systematic analysis of the associated {linearized problem}.
In all {examples} that we have explored, a decreasing
$H(c)$ was associated with instability, and an increasing
one with stability. The formulation of the stability problem
was considered from the perspective of the co-traveling
frame steady state, whose linearization possesses {a pair of eigenvalues moving}
from imaginary to real {values} as the instability threshold
is traversed. It was also explored from the perspective
of a periodic orbit calculation with a {pair of Floquet multipliers}
(directly connected to the above {eigenvalues}) transitioning
from the unit circle to the real axis, as {the wave becomes unstable}. The use of time-translation invariance
and associated eigenvector and generalized eigenvectors
provided a framework for the analytical calculation of
the relevant {eigenvalues}, which is crucial towards identifying
the growth rate of the associated instability near the critical point {at which}
the traveling wave becomes unstable. Connections were also {established}
both with the stability theory of discrete breathers (where an intimately
connected criterion exists \cite{dmp}, provided that we {replace} the traveling
wave {velocity} with the breather frequency) and with the more general
theory of~\cite{GSS} (see Sec.~V of the Supplementary Material).

Nevertheless, {there are} numerous open {problems} in this direction {that} merit further investigation in the near future. {Below}
we mention a few of {representative questions}.
\begin{itemize}
\item We brought to bear perturbation techniques to study the motion of the zero eigenvalue in unweighted spaces and ignored the contribution from the essential spectrum, which based on the numerical results (at least
  in the finite domains considered) appears too small to affect our leading-order approximations. Nevertheless, it would certainly be useful to investigate the exact order of the contribution from the essential spectrum,
  as a step towards a full understanding of the perturbation mechanism.

\item In all {of} the examples {we} considered, the {parameter} $\alpha_1$ was found
  to be negative,
  {yielding transition from stability to instability when
    the sign of $H'(c)$ changes from positive to negative}. Are there
  examples of interest where $\alpha_1$ is
  positive and hence {these} stability conclusions can be reversed?
\item {We assumed the existence of a family of solitary traveling
  waves parameterized by velocity in some continuous range}. However,
  there are other examples, notably, e.g., of the Klein-Gordon variety,
  where traveling waves with vanishing tails are {\it isolated}
  {in terms of their velocities. In such cases}, traveling waves may exist
  as members of a family of solutions with
  nonvanishing tails~\cite{MELVIN2008551}. It is important to understand
  whether our criterion can still be applied in {these} cases and more
  broadly in the setting of traveling ``nanoptera'', where the
  non-vanishing nature of the tails would not allow the consideration
  of the spaces and assumptions made {in our analysis}.
\item Another question of interest concerns the case of
  {dissipative gradient} systems. It is worthwhile
  to examine whether in such settings there may exist a suitable
  modification or extension of the formulation presented {here that yields an}
  associated {stability criterion, perhaps based on the free energy}.
\end{itemize}
Some of these directions are presently under consideration and will
be reported in future publications.

%\dataccess{There were no experimental data collected for this publication. The algorithms and setups for numerical computing were explained accordingly in Sections~\ref{sec:examples} and~\ref{sec:numer}. The codes used to produce
%numerical results are available from the authors upon request. Please address relevant requests to:
%jcuevas@us.es.}

%\aucontribute{PGK and AV conceived the overall project and designed the research. HX was principally
%responsible for the mathematical analysis and JCM for the numerical computations. All
%authors discussed the results extensively, and all of them contributed to the
%writing of and approved the manuscript.}

%\competing{The authors declare that they have no competing interests.}

\acknowledgements {J.C.-M. is thankful for financial support from MAT2016-79866-R project (AEI/FEDER, UE). A.V. acknowledges support by the U.S. National Science Foundation through the grant DMS-1506904. P.G.K.~gratefully acknowledges support
from the Alexander von Humboldt Foundation, the Greek Diaspora Fellowship
Program,
the US-NSF under grant PHY-1602994.
as well as the ERC under FP7, Marie
Curie Actions, People, International Research Staff Exchange
Scheme (IRSES-605096).} {The authors thank Dmitry Pelinovsky for fruitful discussions.}

\appendix

\section{Proofs of some results in the main text}
In this section, we prove Propositions 4.1 and 4.2 and Lemma 5.1 in the main text.\\

\subsection{Proof of Proposition 4.1}

\begin{proof} Suppose $\tilde{S}_{+}\mathcal{E}_2(1,0)Y_0=\mu Y_0$. Letting $Y(\tau)=\mathcal{E}_2(\tau,0)Y_0$, we have
\[
\begin{split}
Y(\tau+1)&=\mathcal{E}_2(\tau+1,1)\mathcal{E}_2(1,0)Y_0=\mathcal{E}_2(\tau+1,1)\tilde{S}_{-}\mu Y_0
=\tilde{S}_{-}\mathcal{E}_2(\tau,0)\tilde{S}_+ \tilde{S}_- \mu Y_0\\
&=\tilde{S}_- \mu \mathcal{E}_2(\tau,0)Y_0
=\tilde{S}_{-} \mu Y(\tau),
\end{split}
\]
where we used the fact that $X_{tw}(\tau+1)=\tilde{S}_-X_{tw}(\tau)$ and that the operator $D\mathcal{A}$ in %\eqref{eq_floquet_2}
\beq
\label{eq_floquet_2}
\begin{split}
c\frac{d}{d\tau}Y(\tau)&=D\mathcal{A}(X_{tw}(\tau))Y(\tau), \quad \tau\in[\tau_0,\tau_1] \\
Y(\tau_0)&=Y_0.
\end{split}
\eeq
satisfies
%\beq
$
D\mathcal{A}(\tilde{S}_{\pm}x)=\tilde{S}_{\pm}D\mathcal{A}(x) \tilde{S}_{\mp}.
$
%\label{eq:DA_prop}
%\eeq
%\eqref{eq:DA_prop}.
Thus $Z_{tw}(\tau)=e^{-\log(\mu)\tau}Y(\tau)$ is a traveling wave solution of
\begin{equation}
\label{eq5}
\lambda Z_{tw}=\mathcal{L} Z_{tw}, \qquad \mathcal{L}=\frac{1}{c}D\mathcal{A}({X_{tw}})-\frac{d}{d\tau}= \frac{1}{c} J \frac{\partial^2\mathcal{H}}{\partial X^2}\bigg|_{X=X_{tw}}-\frac{d}{d\tau},
\end{equation}
with $\lambda=\log(\mu)$, satisfying $Z_{tw}(\tau+1)=\tilde{S}_-Z_{tw}(\tau)$.

Conversely, if $Y(\tau)=Z_{tw}(\tau)e^{\lambda\tau}$ solves
%\eqref{eq4}
\begin{equation}
\label{eq4}
\frac{d}{d\tau}{Y}(\tau)=\frac{1}{c}D\mathcal{A}({X_{tw}}(\tau))Y(\tau)=\frac{1}{c} J \frac{\partial^2\mathcal{H}}{\partial X^2}\bigg|_{X=X_{tw}}Y(\tau).
\end{equation}
and $Z_{tw}(\tau)$ solves \eqref{eq5}, we set $Y_0=Y(0)=Z_{tw}(0)$, so that
\[
\mathcal{E}_2(1,0)Y_0=Y(1)=e^{\lambda}Z_{tw}(1)
=e^{\lambda}\tilde{S}_- Z_{tw}(0)=\tilde{S}_- \mu Y_0,
\]
so $Y_0$ is an eigenfunction of $\tilde{S}_+ \mathcal{E}_2(\tau_1,\tau_0)$ associated with the eigenvalue $\mu=\e^{\lambda}$.
\end{proof}

%\begin{proposition}
%If $Z_0(\tau)$ is an eigenfunction of $\mathcal{L}$ associated with the eigenvalue $\lambda$ and $(\mathcal{L}-\lambda I)^k Z_{k}=(\mathcal{L}-\lambda I)^{k-1} Z_{k-1}=...=(\mathcal{L}-\lambda I) Z_1=Z_0$, then $Y_0=Z_0(0)$ is an eigenfunction of $\tilde{S}_+ \mathcal{E}_2(\tau_1,\tau_0)$ associated with the eigenvalue $\mu=e^{\lambda}$, and there exist $\{Y_j\}_{1\leq j\leq k}$ such that $(\tilde{S}_+ \mathcal{E}_2(\tau_1,\tau_0) -\mu I )^k Y_k=(\tilde{S}_+ \mathcal{E}_2(\tau_1,\tau_0) -\mu I )^{k-1} Y_{k-1}=...=(\tilde{S}_+ \mathcal{E}_2(\tau_1,\tau_0) -\mu I ) Y_1 =Y_0$. The converse is also true.
%\end{proposition}

\subsection{Proof of Proposition 4.2}

\begin{proof} Note that if $\lambda Z_0=\mathcal{L} Z_0$ and $(\mathcal{L}-\lambda I)Z_1=Z_0$, then $\tilde{Y}_1(\tau)=e^{\lambda\tau}(Z_1(\tau)+\tau Z_0(\tau))$ solves \eqref{eq4}. Similarly, if
\beq
(\mathcal{L}-\lambda I)^k Z_{k-1}=(\mathcal{L}-\lambda I)^{k-1} Z_{k-1}=...=(\mathcal{L}-\lambda I) Z_1=Z_0,
\label{eq:Z_gen}
\eeq
then $\tilde{Y}_k(\tau)=e^{\lambda\tau}(\sum_{j=0}^k \frac{1}{j!}\tau^j Z_{k-j}(\tau))$ solves \eqref{eq4}. Let $\tilde{Y}_1=\tilde{Y}_1(0)=Z_1(0)$, then
\[
\mathcal{E}_2(1,0)\tilde{Y}_1=\tilde{Y}_1(1)=e^{\lambda}(Z_0(1)+Z_1(1))=\mu\tilde{S}_- (Z_0(0)+Z_1(0)),
\]
where $\mu=e^\lambda$, and thus $(\tilde{S}_+\mathcal{E}_2(1,0)-\mu I)\tilde{Y}_1=\mu \tilde{Y}_0$.
Using mathematical induction, we can show that if $\tilde{Y}_j=\tilde{Y}_j(0)=Z_j(0)$ for $1\leq j\leq k$, then $(\tilde{S}_+\mathcal{E}_2(1,0)-\mu I)\tilde{Y}_j=\mu (\sum_{i=0}^{j-1} \frac{1}{(j-i)!}\tilde{Y}_i)$. We claim that there exists a (non-unique) lower triangular matrix $\Omega=(\Omega_{ij})_{0\leq i,j\leq k}$ such that for $\tilde{Y}_0=Z_0(0)$ and $Y_j=\sum_{i=0}^j \Omega_{ji}\tilde{Y_{i}}$ for $0\leq j\leq k$ we have
\beq
(\tilde{S}_+ \mathcal{E}_2(\tau_1,\tau_0) -\mu I )^k Y_k=(\tilde{S}_+ \mathcal{E}_2(\tau_1,\tau_0) -\mu I )^{k-1} Y_{k-1}=...=(\tilde{S}_+ \mathcal{E}_2(\tau_1,\tau_0) -\mu I ) Y_1 =Y_0.
\label{eq:Floquet_gen}
\eeq

To prove the existence of such $\Omega$, let $R_1=(R_{1,j,i})_{0\leq j,i\leq k}$ and $R_2=(R_{2,j,i})_{0\leq j,i\leq k}$ be such that
\[
R_{1,ji}=\begin{cases} 0, & i\geq j, \\
                       \frac{1}{(j-i)!}, & i < j
         \end{cases}, \qquad
R_{2,ji}=\begin{cases} 1, & i=j-1, \\
                       0, & i \neq j-1.
         \end{cases}
\]
Then we have
\[
(\tilde{S}_+\mathcal{E}_2(1,0)-\mu I) \tilde{Y} = \mu R_{1} \tilde{Y}, \quad (\tilde{S}_+\mathcal{E}_2(1,0)-\mu I) {Y} =  R_{2} {Y},
\]
where $Y=\Omega \tilde{Y}$, which implies
\[
\mu \Omega R_1 = R_2 \Omega.
\]
By comparing the terms on both sides, one can verify that given $\Omega_{00}$, one can compute the other diagonal terms $\Omega_{11}, \Omega_{22}, \dots, \Omega_{kk}$. Once all diagonal terms are known, given $\Omega_{10}$, one can solve for $\Omega_{21}, \Omega_{32}, ..., \Omega_{k(k-1)}$. Assuming $\Omega_{ji}$ are known for any $j-i<m$ and given $\Omega_{m0}$, one can then obtain $\Omega_{(m+1)1}, \Omega_{(m+2)2}, \dots, \Omega_{k(k-m)}$. Following this procedure, a matrix $\Omega$ can be found for given constants $\Omega_{j0}$, $0\leq j\leq k$. In particular, for $k=3$ one choice of $\Omega$ is
\[
\Omega=\left(\begin{array}{rrrr}
  \mu^3 & 0 & 0  & 0\\
  0 & \mu & 0 & 0\\
  0 & \frac{1}{2}\mu & \mu & 0 \\
  0 & \frac{1}{3} & -1 & 1
\end{array} \right).
\]

To prove the converse, we assume that \eqref{eq:Floquet_gen} holds and let $\tilde{Y}_j=\sum_{i=0}^k \Omega^{-1}_{j i}{Y_{i}}$ for $0\leq j\leq k$, with $\Omega$ defined above. Then for $1\leq j\leq k$, $(\tilde{S}_+\mathcal{E}_2(1,0)-\mu I)\tilde{Y}_j=\mu (\sum_{i=0}^{j-1} \frac{1}{(j-i)!}\tilde{Y}_i)$. Let $\tilde{Y}_0(\tau)=\mathcal{E}_2(\tau,0)\tilde{Y}_0$ and $\tilde{Y}_j(\tau)=\mathcal{E}_2(\tau,0)\tilde{Y}_j$ for $1\leq j\leq k$, then $\tilde{Y}_j(\tau+1)=\tilde{S}_- \mu (\sum_{i=0}^{j} \frac{1}{(j-i)!}\tilde{Y}_i(\tau))$. Using mathematical induction, one can show that $\tilde{Y}_j(\tau)$ has the form $\tilde{Y}_j(\tau)=e^{\lambda\tau}(\sum_{i=0}^j \frac{1}{i!}\tau^j Z_{j-i}(\tau))$, where $Z_j(\tau)$ are traveling waves, i.e., satisfy $Z_j(\tau+1)=\tilde{S}_- Z_j(\tau)$, for $0\leq j\leq k$. Moreover, since $\tilde{Y}_j(\tau)$ solves \eqref{eq4}, one can show $\lambda Z_0=\mathcal{L} Z_0$ and \eqref{eq:Z_gen} hold.
\end{proof}

%\begin{lemma}
%\label{lemma2}
%Suppose all the assumptions in Theorem~\ref{lemma1} hold except that ${\rm dim}({\rm gker}(\mathcal{L}_0))=n \geq 4$, where $n$ is even, and
%\[
%\mathcal{L}_0^n e_{n-1}=\mathcal{L}_0^{n-1} e_{n-2}=...=\mathcal{L}_0^2 e_1=\mathcal{L}_0 e_0=0.
%\]
%Then there exist $c_1<c_0$ and $c_2>c_0$ such that for $c\in(c_1,c_2)$, the leading-order terms of nonzero eigenvalues of $\mathcal{L}$ are
%\[
%\lambda=e^{j\frac{2\pi i}{(n-2)}} \biggl(\frac{H''(c_0)}{\alpha_1}(c-c_0)\biggr)^{1/(n-2)}+o(|c-c_0|^{1/(n-2)}), \quad j=0,1,\dots,n-3.
%\]
%where $\alpha_1=\langle J^{-1}e_{n-1}, e_0 \rangle$.
%\end{lemma}

\subsection{Proof of Lemma 5.1.}

\begin{proof}
{
Using the arguments similar to {the ones leading to Propositions}~5.1 and 5.2, one can show that $\lambda=\mathcal{O}(\epsilon^{1/(n-2)})$ in this case. To be more specific, using
\begin{equation}
\mathcal{L}(\partial_{\tau} X_{tw}(\tau;c))=0
\label{eq:constraint1}
\end{equation}
and
\begin{equation}
\mathcal{L}(c\partial_{c} X_{tw}(\tau;c))=\partial_{\tau} X_{tw}(\tau;c),
\label{eq:constraint2}
\end{equation}
we can similarly obtain the coefficients {$g_j$ and $f_j$ (as well as the analogs of $h_j$ defined in \eqref{eq:h})}. Moreover, from $0=Q_0 g_1+Q_1 g_0$ we can derive the {$n$-dimensional version of} $K_{00}=0$, which {eliminates} the possibility of $\lambda\sim \mathcal{O}(\epsilon^{1/n})$. Using symmetries or $K_{10}=K_{01}$ we rule out the case $\lambda\sim \mathcal{O}(\epsilon^{1/(n-1)})$. {Below} we show {that} $2K_{20}\neq K_{11}$ is satisfied, {so that} $\lambda=\mathcal{O}(\epsilon^{1/(n-2)})$ can be achieved.
}
Substituting
$\lambda=\epsilon^{1/(n-2)}\lambda_1+\epsilon^{2/(n-2)} \lambda_2+\epsilon^{3/(n-2)}\lambda_3+\dots$ and $Z=Z_0+\epsilon^{1/(n-2)} Z_1+\epsilon^{2/(n-2)} Z_2+\epsilon^{3/(n-2)} Z_3+\dots$ into $\lambda Z=\mathcal{L}Z$, we obtain
\begin{eqnarray*}
\label{eqn_lambda2_expand0}
\mathcal{L}_0 Z_j&=&\sum_{k=1}^{j} \lambda_k Z_{j-k}, \quad j=0,1,...,n-3 \\
\label{eqn_lambda2_expand1}
\mathcal{L}_0 Z_{n-2}+\mathcal{L}_1 Z_0&=&\sum_{k=1}^{n-2} \lambda_k Z_{n-2-k}\\
\label{eqn_lambda2_expand2}
\mathcal{L}_0 Z_{n-1}+\mathcal{L}_1 Z_1&=&\sum_{k=1}^{n-1} \lambda_k Z_{n-1-k}\\
\label{eqn_lambda2_expand3}
\mathcal{L}_0 Z_{n}+\mathcal{L}_1 Z_2&=&\sum_{k=1}^{n} \lambda_k Z_{n-k}.
\end{eqnarray*}
{Let} $Z_j=\sum_{k=0}^{n-1} (C_{jk}e_k)+(Z_j)^{\#}$, where $(Z_j)^{\#} \in G_n^{\perp}$. Then the above
equations are equivalent to
\begin{eqnarray}
\label{eqn_lambda2_expand0_2}
A_0 C_j&=&\sum_{k=1}^{j} \lambda_k  C_{j-k}, \quad j=0,1,...,n-3 \\
\label{eqn_lambda2_expand1_2}
A_0 C_{n-2}+M^{-1}Q_1 C_0&=&\sum_{k=1}^{n-2} \lambda_k  C_{n-2-k} \\
\label{eqn_lambda2_expand2_2}
A_0 C_{n-1}+M^{-1}Q_1 C_1&=&\sum_{k=1}^{n-1} \lambda_k  C_{n-1-k} \\
\label{eqn_lambda2_expand3_2}
A_0 C_{n}+M^{-1}Q_1 C_2&=&\sum_{k=1}^{n} \lambda_k  C_{n-k},
\end{eqnarray}
where $C_j$ and $n$-dimensional vectors with components $C_{jk}$ and $M=(\langle J^{-1}e_j, e_k\rangle)$, $Q_0=(\langle J^{-1}e_j, \mathcal{L}_0 e_k\rangle)$ and $Q_1=K=(\langle J^{-1}e_j, \mathcal{L}_1 e_k\rangle)$ are $n \times n$ matrices, with $j,k=0,\dots,n-1$. As before, we have $Q_0=M A_0$, where $A_0$ is the $n\times n$ matrix whose only nonzero entries are ones along the superdiagonal (generalization of the previous $4$-by-$4$ matrix $A_0$), $Q_0$ and $Q_1$ are symmetric, and $M$ is a skew-symmetric invertible matrix (recall that $n$ is even). Setting $g_0=(1,0,0,...,0)^T$, we obtain
\begin{eqnarray}
\label{eqn_lambda2_expand0_3}
C_j&=&\sum_{k=1}^{j} \lambda_k A_0^T C_{j-k}+C_{j,0}g_0, \quad j=0,1,...,n-3 \\
\label{eqn_lambda2_expand1_3}
C_{n-2}&=&\sum_{k=1}^{n-2} \lambda_k A_0^T C_{n-2-k}-A_0^T M^{-1}Q_1 C_0 + C_{n-2, 0}g_0, \\
\label{eqn_lambda2_expand2_3}
C_{n-1}&=&\sum_{k=1}^{n-1} \lambda_k A_0^T C_{n-1-k}-A_0^T M^{-1}Q_1 C_1 + C_{n-1, 0}g_0,\\
\label{eqn_lambda2_expand3_3}
C_{n}&=&\sum_{k=1}^{n} \lambda_k A_0^T C_{n-k}-A_0^T M^{-1}Q_1 C_2 + C_{n, 0}g_0.
\end{eqnarray}
With $C_{00}=1$, we get $C_{j,j}=\lambda_1^j$ and $C_{j,k}=0$ for $0\leq j\leq (n-3)$ and $j<k$.
{
The last row of (\ref{eqn_lambda2_expand1_3}) is $K_{00}=0$; the last row of (\ref{eqn_lambda2_expand2_3}) can be directly verified by using symmetries.
}

The last row of \eqref{eqn_lambda2_expand3_2} is the same as the last row of $M^{-1}Q_1 C_2=\lambda_1 C_{n-1}$, which yields
\beq
\alpha_1 \lambda_1^n + (2K_{20}-K_{11})\lambda_1^2=0.
\label{eq:lambda1}
\eeq
Similar to (5.32) in the main text, we can show that the coefficient in front of $\lambda_1^2$ equals $-H''(c_0)$. Indeed, using $\nabla\mathcal{H}$ and $\nabla^2\mathcal{H}$ to simplify notation and recalling the symmetry of $K$, we have
\beq
\label{eq_H_2nd_d_2}
\begin{split}
H''(c_0)&=2\langle\nabla\mathcal{H}, U_2 \rangle + \langle U_1, \nabla^2\mathcal{H} U_1 \rangle\\
&= 2 c_0 \langle J^{-1}e_0, U_2 \rangle + c_0\langle U_1, J^{-1}\mathcal{L}_0 U_1 \rangle + c_0\langle U_1, J^{-1} \partial_{\tau}U_1 \rangle \\
&= \langle e_1, J^{-1}(\partial_{\tau}U_1-c_0\mathcal{L}_1 U_1) \rangle+ \langle U_1, J^{-1}e_0 \rangle+c_0\langle (f_{10}e_0+\frac{1}{c_0}e_1), J^{-1}\partial_{\tau}U_1 \rangle \\
&= \langle e_2, J^{-1}\mathcal{L}_1 e_0 \rangle -\langle e_1, J^{-1}\mathcal{L}_1 e_1 \rangle  + 0 -c_0 f_{10}\langle e_1, J^{-1}\mathcal{L}_1 e_0 \rangle+\langle (c_0 f_{10}e_1+e_2), J^{-1}\mathcal{L}_1 e_0 \rangle \\
&= -K_{20}+K_{11}+f_{10}c_0 K_{10}-f_{10}c_0 K_{10}-K_{20}\\
&= K_{11}-2K_{20}.
\end{split}
\eeq
Thus, nonzero roots of \eqref{eq:lambda1} satisfy
\[
\lambda_1^{n-2}=\frac{K_{11}-2K_{20}}{\alpha_1}=\frac{H''(c_0)}{\alpha_1}.
\]
\end{proof}

\section{Projection coefficients}
\label{sec:proj}
In this section, we provide some additional expressions for the
projection coefficients arising in
\beq
\partial_{\tau} U_j=\sum_{k=0}^3 (g_{jk}e_k)+(\partial_{\tau} U_j)^{\#}, \quad (\partial_{\tau} U_j)^{\#} \in G_4^{\perp}
\label{eq:dU}
\eeq and
\beq
U_j=\sum_{k=0}^3 (f_{jk}e_k)+(U_j)^{\#}, \quad (U_j)^{\#} \in G_4^{\perp}.
\label{eq:U}
\eeq

We start with \eqref{eq:dU}. Recall that $g_{0j}=\delta_{1,0}$ and that $g_{1j}$ coefficients are given by (5.21). Using $0=Q_0 g_2 +Q_1 g_1+Q_2 g_0$, we obtain $\sum_{j=1}^3 g_{1j}K_{0j}+ L_{00}=0$, and, recalling $g_1=-A_0^T M^{-1} Q_1 g_0 + g_{10} g_0$, we have
\beq
\label{eq:g2}
\begin{split}
g_2&=-A_0^T M^{-1} Q_1 g_1 -A_0^T M^{-1} Q_2 g_0 +g_{20} g_0\\
&=[(-A_0^T M^{-1} Q_1)^2+(-A_0^T M^{-1} Q_2)] g_0 +(-A_0^T M^{-1} Q_1) g_{10} g_0+g_{20} g_0.
\end{split}
\eeq
This yields
\beq
\label{eq:g2comp}
\begin{split}
g_{21}&=\frac{\alpha_2^2 K_{10}K_{11}-\alpha_1\alpha_2(-K_{12}K_{20}+K_{11}K_{30}+2K_{10}K_{31})}{\alpha_1^4}\\
&+\frac{K_{30}K_{31}-K_{20}K_{32}+K_{10}K_{33}+\alpha_2( g_{10}K_{10}+L_{00})}{\alpha_1^2}-\frac{g_{10}K_{30}+L_{30}}{\alpha_1}\\
g_{22}&=\frac{\alpha_1\alpha_2 K_{10}K_{21}-\alpha_1^2(-K_{20}K_{22}+K_{10}K_{23}+K_{21}K_{30})}{\alpha_1^4}+\frac{g_{10}K_{20}+L_{20}}{\alpha_1} \\
g_{23}&=\frac{\alpha_2(-K_{10}K_{11})+\alpha_1(K_{10}K_{13}-K_{12}K_{20}+K_{11}K_{30})-\alpha_1^2(g_{10}K_{10}+L_{10})}{\alpha_1^3}.\\
\end{split}
\eeq
Condition $\sum_{j=1}^3 g_{2j}K_{0j}+ \sum_{j=0}^3 g_{1j}L_{0j}+B_{00}=0$ follows from $0=Q_0 g_3+Q_1 g_2+Q_2 g_1+Q_3 g_0$, which together with $g_1$ and $g_2$ also yields
\[
\begin{split}
g_3&=-A_0^T M^{-1} Q_1 g_2 -A_0^T M^{-1} Q_2 g_1-A_0^T M^{-1} Q_3 g_0 +g_{30} g_0 \\
&=[(-A_0^T M^{-1} Q_1)^3+(-A_0^T M^{-1} Q_1)(-A_0^T M^{-1} Q_2)\\
&+(-A_0^T M^{-1} Q_2)(-A_0^T M^{-1} Q_1)+(-A_0^T M^{-1} Q_3)] g_0 \\
&+ [(-A_0^T M^{-1} Q_1)^2+(-A_0^T M^{-1} Q_2)] g_{10} g_0+ (-A_0^T M^{-1} Q_1) g_{20} g_0 + g_{30} g_0,
\end{split}
\]
resulting in
\beq
\label{eq:g3comp}
\begin{split}
g_{31}&=-\frac{\sum_{j=0}^3 g_{2j}K_{3j}+\sum_{j=0}^3 g_{1j}L_{3j}+B_{30}}{\alpha_1} - \frac{\alpha_2}{\alpha_1}g_{33}\\
g_{32}&=\frac{\sum_{j=0}^3 g_{2j}K_{2j}+\sum_{j=0}^3 g_{1j}L_{2j}+B_{20}}{\alpha_1}\\
g_{33}&=-\frac{\sum_{j=0}^3 g_{2j}K_{1j}+\sum_{j=0}^3 g_{1j}L_{1j}+B_{10}}{\alpha_1} \\
\end{split}
\eeq
To summarize, (5.21), \eqref{eq:g2comp} and \eqref{eq:g3comp} determine the projection coefficients $g_{ij}$, $i,j=1,2,3$, in \eqref{eq:dU}.

Considering now the coefficients in \eqref{eq:U}, (5.26)-(5.28) yield
\begin{eqnarray*}
\label{eqn_U_expand2_1_3}
f_1&=&\frac{1}{c_0}A_0^T g_0 +f_{10}g_0, \\
\label{eqn_U_expand2_2_3}
f_2&=&\frac{1}{2c_0}[A_0^T g_1 -  A_0^T M^{-1} (Q_0 + c_0 Q_1) f_1 ]+f_{20}g_0 \\
&=&\frac{1}{2c_0}[A_0^T g_1 - A_0^T M^{-1} Q_1 A_0^T g_0 - c_0 A_0^T M^{-1} Q_1 f_{10} g_0- \frac{1}{c_0} A_0^T g_0 ]+f_{20}g_0, \\
\label{eqn_U_expand2_3_3}
f_3&=&\frac{1}{3c_0}[A_0^T g_2 - A_0^T M^{-1} (Q_1+c_0 Q_2) f_1- A_0^T M^{-1}(Q_0 + c_0 Q_1)2 f_2 ]+f_{30}g_0 \\
&=&\frac{1}{3 c_0}[A_0^T g_2-A_0^T M^{-1} (Q_1+c_0 Q_2)(\frac{1}{c_0}A_0^T g_0+f_{10} g_0)\\
& &-A_0^T M^{-1} (Q_0+c_0 Q_1)( \frac{1}{c_0}A_0^T g_1-\frac{1}{c_0}A_0^T M^{-1}Q_1 A_0^T g_0- A_0^T M^{-1} Q_1 f_{10} g_0\\
& &-\frac{1}{c_0^2}A_0^T g_0 +f_{20} g_0 )]+f_{30}g_0.
\end{eqnarray*}
Using (5.26), we find that $f_1=(f_{10}, \frac{1}{c_0}, 0, 0)^T$, which together with (5.27) yields
\[
\begin{split}
f_{21}&=\frac{\alpha_1^2(c_0 g_{10}-1)+\alpha_2 c_0(K_{11}+c_0 f_{10}K_{10})-\alpha_1 c_0(c_0 f_{10}K_{30}+K_{31})}{2\alpha_1^2 c_0^2} \\
f_{22}&=\frac{c_0 f_{10}K_{20}+K_{21}-K_{30}}{2\alpha_1 c_0}.
\end{split}
\]
and $f_{23}$ in (5.30).
Using (5.28) we obtain
\[
\begin{split}
f_{31}&=\frac{1}{3 c_0} [ g_{20} - \sum_{j=0}^3 \frac{1}{\alpha_1} (K_{3j}+c_0 L_{3j})  f_{1j} - \sum_{j=0}^3 c_0 \frac{1}{\alpha_1} K_{3j} 2f_{2j} -2 f_{21} -2\frac{\alpha_2}{\alpha_1}f_{23} ]-\frac{\alpha_2}{\alpha_1}f_{33}+\frac{g_{22}}{3 c_0}+f_{30}\\
f_{32}&=\frac{1}{3 c_0} [ g_{21} - \sum_{j=0}^3 (\frac{\alpha_2}{\alpha_1^2} (K_{0j}+c_0 L_{0j}) -\frac{1}{\alpha_1} (K_{2j}+c_0 L_{2j}) )  f_{1j} - \sum_{j=0}^3 c_0 (\frac{\alpha_2}{\alpha_1^2} K_{0j} -\frac{1}{\alpha_1} K_{2j}) 2f_{2j} -2 f_{22} ]  \\
f_{33}&=\frac{1}{3 c_0} [ g_{22} - \frac{1}{\alpha_1} \sum_{j=0}^3 (K_{1j}+c_0 L_{1j})f_{1j} - \frac{1}{\alpha_1} \sum_{j=0}^3 c_0 K_{1j}2f_{2j}-2 f_{23} ]
\end{split}
\]

To simplify some calculations, it is convenient to define $h_j=c_0 j f_j+(j-1)f_{j-1}$ for $j=1,2,3$, so that
\beq
\label{eq:h}
\begin{split}
h_1&=A_0^T g_0 +h_{10}g_0 \\
h_2&=A_0^T g_1 -  A_0^T M^{-1} Q_1 h_1 +h_{20}g_0 \\
h_3&=A_0^T g_2 - A_0^T M^{-1} Q_2 h_1 - A_0^T M^{-1} Q_1 h_2 + h_{30}g_0
\end{split}
\eeq

{
We note that the last rows of the equations (5.36)-(5.38) are closely connected to the constraints \eqref{eq:constraint1}--\eqref{eq:constraint2}, one of the reasons for which is as follows:
\begin{remark}
\label{rmk_contant_1}
If the constant term in $C_{2k}$ is given by $g_k$ with $C_{2k,0}=g_{k,0}$, then the constant term in $C_{2k+2}$ will be $g_{k+1}$ with $C_{2k+2,0}=g_{k+1,0}$.
\end{remark}
\begin{remark}
\label{rmk_coeff_1}
If the constant term in $C_{2k-2}$ is given by $g_{k-1}$ with $C_{2k-2,0}=g_{k-1,0}$ and the coefficient for $\lambda_1$ in $C_{2k-1}$ is given by $h_{k}$ with $h_{k,0}=0$, then the coefficient for $\lambda_1$ in $C_{2k+1}$ will be $h_{k+1}$ with $h_{k+1,0}=0$.
\end{remark}
By these two remarks, the constant terms in the last rows of  (5.36)-(5.38) {are} always zero. {A more involved calculation, omitted here, shows} that the coefficients for $\lambda_1$ in the last rows of these equations also vanish. {Similar} techniques can be used in other {cases}, including {the one considered in} Sec.~5, to show the constant terms and $\lambda_1$ coefficients vanish in the last rows of the equations.
}

\section{The degenerate case $H'(c_0)=H''(c_0)=0$}
\label{sec:degenerate}
In this section, we briefly discuss the degenerate case when $H'(c_0)=H''(c_0)=0$ but $H'''(c_0) \neq 0$. For simplicity, we assume
that ${\rm dim}({\rm gker}(\mathcal{L}_0))=4$. Following the arguments in Proposition~5.1 and Proposition~5.2, one can show that the near-zero eigenvalues are at most $O(\epsilon)$. Proceeding as before, one obtains the following equation for $\lambda_1$:
\[
\lambda_1^4-b \lambda_1^2=0,
\]
where
\begin{equation}
\label{eq:b}
\begin{split}
b&=-\frac{1}{\alpha_1}\biggl[\frac{ -2K_{01}K_{32}+2K_{02}K_{22}-K_{03}^2+K_{11}K_{31}-K_{12}^2 }{\alpha_1}\\
&+\frac{\alpha_2(2K_{21}K_{10}+K_{20}^2-K_{11}^2)}{\alpha_1^2}+2L_{02}-L_{11}\biggr].
\end{split}
\end{equation}
We omit the details. This yields
\begin{lemma}
\label{lemma3}
Suppose all the assumptions in Theorem~5.1 hold except that $\partial_{c}^3 (X_{tw,j}(\tau;c)) \in D^0([0,1])$ ($D^0_a([0,1])$ if weighted spaces are used), $H''(c_0)=0$ and $b$ defined in \eqref{eq:b} is nonzero. Then there exist $c_1<c_0$ and $c_2>c_0$ such that for $c\in(c_1,c_2)$, the leading-order terms of nonzero eigenvalues of $\mathcal{L}$ will be
\[
\lambda=\pm\sqrt{b}(c-c_0)+o(|c-c_0|).
\]
\end{lemma}
It is not clear whether the coefficient $b$ is related to $H'''(c_0)$, which is given by
\[
\begin{split}
H'''(c_0)&=2 \biggl[\frac{1}{\alpha_1} (-3K_{13} K_{20} + K_{12}^2 - 4 K_{20} K_{22} + 4 K_{10} K_{23}  +
    3 K_{21} K_{30} \\
    &+2c_0 f_{10} (K_{10}K_{13}-K_{12}K_{20}+K_{11}K_{30}) )  \\
 &+\frac{\alpha_2}{\alpha_1^2}(K_{11}^2 - 4 K_{10} K_{21}- 4 c_0 f_{10}K_{10}K_{11}) - 2 g_{10}K_{20} -\frac{K_{20}}{c_0}+L_{11}-4 L_{20}\\
 &+c_0(2f_{20}K_{10}-3f_{10}g_{10}K_{10}-2f_{10}L_{10})\biggr].
\end{split}
\]
We note that it is possible to make $f_{10}=f_{20}=0$ with a careful choice of $\{e_0, e_1, e_2, e_3\}$. {While this case has an} intrinsic theoretical value,
we have not {yet} identified {any Hamiltonian lattice models for which} this degeneracy
condition is satisfied.

\section{Numerical procedures}
\label{sec:numer}
In this section we describe the numerical procedures we used to obtain
the solitary traveling waves and analyze their stability.
More specifically, to identify the solitary wave structures, we use the procedure followed in \cite{Yasuda}. To this end, we seek solutions of (6.3) in the co-traveling frame corresponding to velocity $c$:
\begin{equation}
r_n(t)=\Phi(\xi,t), \quad \xi=n-ct,
\end{equation}
obtaining the advance-delay partial differential equation
\begin{equation}\label{eq:dynade}
\begin{split}
\Phi_{tt}&+c^2\Phi_{\xi\xi}-2c\Phi_{\xi t}+2V'(\Phi(\xi,t))-V'(\Phi(\xi+1,t))-V'(\Phi(\xi-1,t)) \\
&+\sum_{m=1}^\infty\Lambda(m)(2\Phi(\xi,t)-\Phi(\xi+m,t)-\Phi(\xi-m,t))=0.
\end{split}
\end{equation}
Solitary traveling waves $\phi(\xi)$ are stationary solutions of \eqref{eq:dynade}. They satisfy the advance-delay differential equation (6.4) and (6.5).

Following the approach in \cite{eilbeck}, we assume that $\phi(\xi)=o(1/\xi)$ and $\phi'(\xi)=o(1/\xi^2)$ as $|\xi| \rightarrow \infty$, multiply Eq.~(6.4) by $\xi^2$ and integrate by parts to derive the identity
\begin{equation}\label{eq:ade1}
    \left[c^2-\sum_{m=1}^{\infty}m^2\Lambda(m)\right]\int_{-\infty}^{\infty} \phi(\xi)\mathrm{d}\xi-\int_{-\infty}^{\infty} V'(\phi(\xi))\mathrm{d}\xi=0,
\end{equation}
which imposes the constraint (6.5) on the traveling wave solutions. Here we assume that $\Lambda(m)$ decays faster than $1/m^3$ at infinity, so that the series on the left hand side converges.

To solve Eq.~(6.4) numerically, we introduce a discrete mesh with step $\Delta\xi$, where $1/\Delta\xi$ is an integer, so that the advance and delay terms $\phi(\xi \pm m)$ are well defined on the mesh. We then use a Fourier spectral collocation method for the resulting system with periodic boundary conditions \cite{Trefethen} with large period $l$. Implementation of this method requires an even number $\mathcal{N}$ of collocation points $\xi_j\equiv j\Delta\xi$, with $j=-\mathcal{N}/2+1,\ldots,\mathcal{N}/2$, yielding a system for $\xi$ in the domain $(l/2,l/2]$, with $l=\mathcal{N}\Delta\xi$ being an even number, and the long-range interactions are appropriately truncated. To ensure that the solutions satisfy (6.5), we additionally impose
a trapezoidal approximation of \eqref{eq:ade1} on the truncated interval at the collocation points. This procedure is independent of the potential and the interaction range. However, the choices of $\Delta\xi$ and $l$ depend on the nature of the problem.

To investigate spectral stability of a traveling wave $\phi(\xi)$, we substitute
\begin{equation}
\Phi(\xi,t)=\phi(\xi)+\epsilon a(\xi)\exp({\lambda} c t),
\end{equation}
into \eqref{eq:dynade} and consider $O(\epsilon)$ terms resulting from this perturbation. This yields the following quadratic eigenvalue problem:
\begin{equation}\label{eq:perturbade}
\begin{split}
c^2{\lambda}^2a(\xi)&=-c^2a''(\xi)+2\lambda c^2 a'(\xi)
    -2V''(\phi(\xi))a(\xi)+V''(\phi(\xi+1))a(\xi+1)\\
    &+V''(\phi(\xi-1))a(\xi-1)-\sum_{m=1}^\infty\Lambda(m)(2a(\xi)-a(\xi+m)-a(\xi-m)).
\end{split}
\end{equation}

By defining $b(\xi)=c{\lambda} a(\xi)$, we transform this equation into the regular eigenvalue problem
\begin{equation}
c{\lambda}\left(
\begin{array}{c}
  a(\xi) \\
  b(\xi)
\end{array}
\right)
=
\mathcal{M}
\left(
\begin{array}{c}
  a(\xi) \\
  b(\xi)
\end{array}
\right)
\end{equation}
for the corresponding linear advance-delay differential operator $\mathcal{M}$. Note that this problem is equivalent to the eigenvalue problem (3.5) via the transformation $\xi=-\tau$ and $(a(\xi), b(\xi))=(\Gamma_a(-\tau), \Gamma_b(-\tau)-c \Gamma_a'(-\tau))$. Spectral stability can be determined by analyzing the spectrum of the operator $\mathcal{M}$ after discretizing the eigenvalue problem the same way as the nonlinear Eq.~(6.4) and again using periodic boundary conditions. A traveling wave solution is spectrally stable when the spectrum contains no real eigenvalues.

To analyze the dependence of the eigenvalues spectrum obtained using this procedure on the grid size $\Delta\xi$ and the size $l$ of the lattice, we consider as an example an $\alpha$-FPU lattice with $V(r)=r^2/2-r^3/3$ and without long-range interaction ($\Lambda(m)\equiv0$). In this case, $H'(c)>0$ for every $c$, and one expects all waves to be stable (see, e.g., \cite{ourTW}). However, the numerically obtained spectra show mild oscillatory instabilities, which are associated with complex-valued eigenvalues with nonzero imaginary parts and small but nonzero real parts. As one can see in the example shown in Fig.~\ref{fig:sri}, corresponding to $c=1.3$, the real part of the eigenvalues does not change when $\Delta \xi$ is varied (only their imaginary parts are affected, given that
smaller $\Delta \xi$ enables accessing higher wavenumbers); however, when the system's length $l$ is increased, the real parts of the eigenvalues decrease. This suggests that the oscillatory instabilities are a numerical artifact due to the finite length of the lattice that can, in principle, be expected to
disappear in the infinite lattice limit.
\begin{figure}
\begin{center}
\begin{tabular}{cc}
\includegraphics[width=6.5cm]{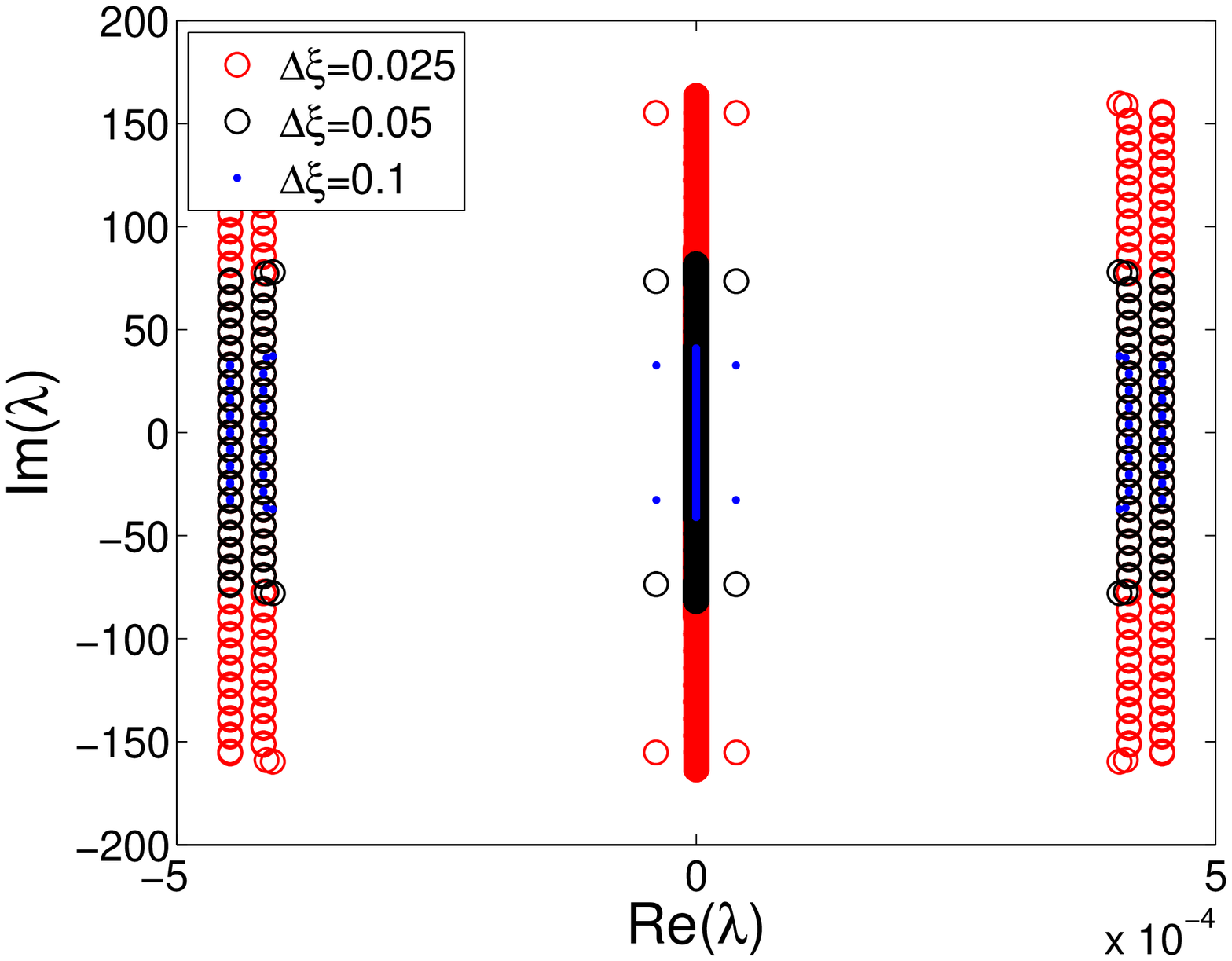} &
\includegraphics[width=6.5cm]{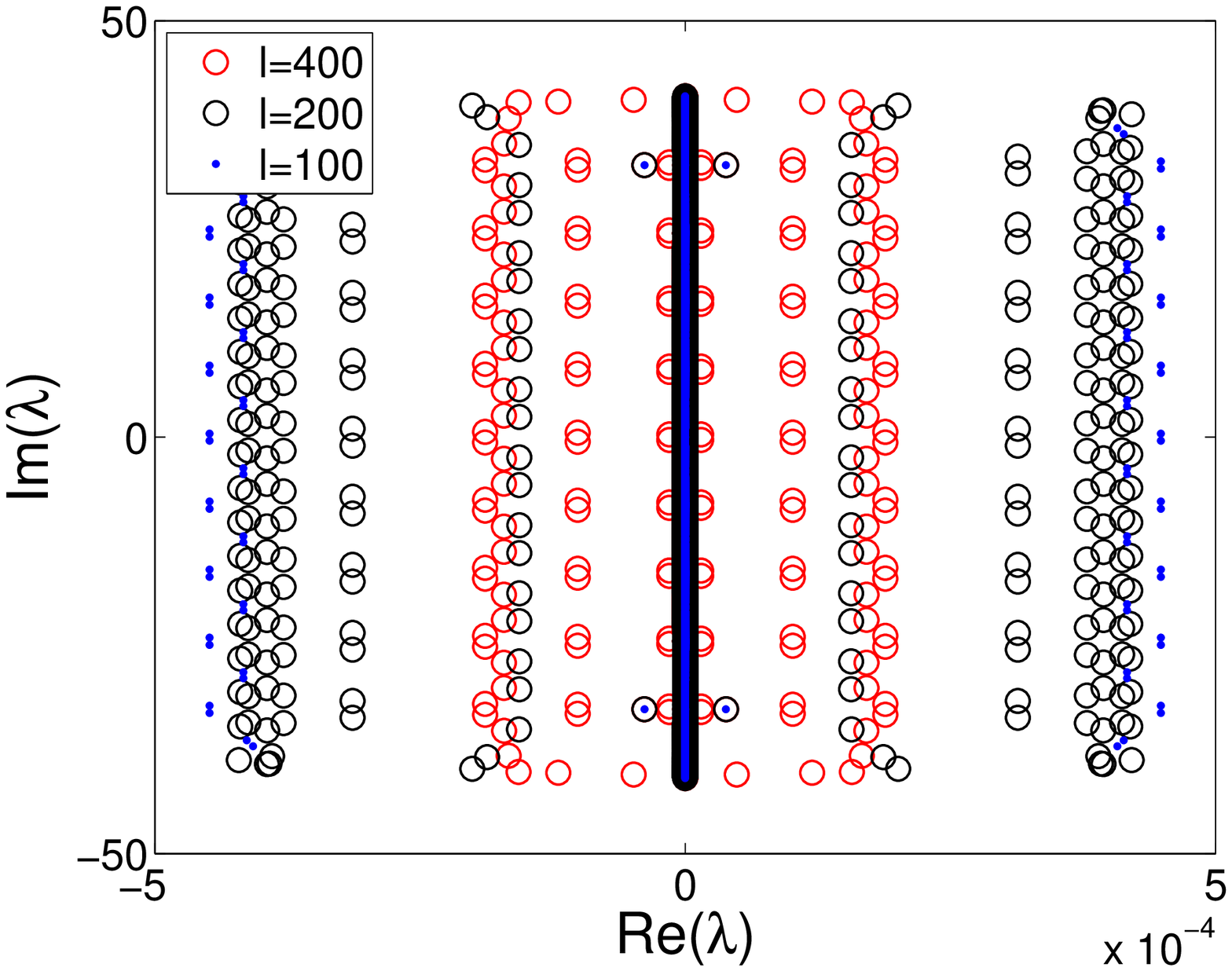}
\end{tabular}
\end{center}
\caption{Dependence of the linearization operator spectrum with the discretization parameter ({left panel, with fixed $l=100$}) and the system length ({right panel, with fixed $\Delta\xi=0.1$}) for a stable solitary traveling wave solution with $c=1.3$ of the problem with the $\alpha$-FPU potential governing nearest-neighbor interactions and no long-range interaction.}
\label{fig:sri}
\end{figure}
%\hx{Please reproduce Fig.11 using $\lambda$ instead of $\tilde{lambda}$}

An alternative method for determining the spectral stability of the traveling waves is to use Floquet analysis. To this end, we cast traveling waves $\phi(\xi)$ as fixed points of the map
\begin{equation}\label{eq:map1}
\left[\begin{array}{c}
  \{r_{n+1}(T)\} \\ \{\dot r_{n+1}(T)\} \\  \end{array}\right]
  \rightarrow
  \left[\begin{array}{c}
  \{r_{n}(0)\} \\ \{\dot r_{n}(0)\} \\  \end{array}\right],
\end{equation}
which is periodic modulo shift by one lattice point, with period $T=1/c$. Indeed, one easily checks that $\hat{r}_n(t)=\phi(n-ct)=\phi(n-t/T)$ satisfies $\hat{r}_{n+1}(T)=\hat{y}_n(0)=\phi(n)$ and $\dot{\hat r}_{n+1}(T)=\dot{\hat r}_{n}(0)=-c\phi'(n)$.  To apply the Floquet analysis, we trace time evolution of a small perturbation $\epsilon w_n(t)$ of the periodic-modulo-shift (traveling wave) solution $\hat{r}_n(t)$. This perturbation is introduced in (6.3) via $r_n(t)=\hat{r}_n(t)+\epsilon w_n(t)$. The resulting $O(\epsilon)$ equation reads
\begin{equation}\label{eq:perturb}
\begin{split}
    &\ddot w_n+2V''(\hat{r}_n)w_n-V''(\hat{r}_{n+1})w_{n+1}-V''(\hat{r}_{n-1})w_{n-1}\\
    &+\sum_{m=1}^\infty\Lambda(m)(2w_n-w_{n+m}-w_{n-m})=0.
\end{split}
\end{equation}
Then, in the framework of Floquet analysis, the stability properties of periodic orbits are resolved by diagonalizing the monodromy matrix $\mathcal{F}$ (representation of the Floquet operator at finite systems), which is defined as:
\begin{equation}\label{eq:Floquet}
\left[\begin{array}{c}
  \{w_{n+1}(T)\} \\ \{\dot w_{n+1}(T)\} \\  \end{array}\right]
  =\mathcal{F}
  \left[\begin{array}{c}
  \{w_{n}(0)\} \\ \{\dot w_{n}(0)\} \\  \end{array}\right] .
\end{equation}
For the symplectic Hamiltonian systems we consider in this work, the linear stability of the solutions requires that the monodromy eigenvalues $\mu$ (also called Floquet multipliers) lie on the unit circle. The Floquet multipliers can thus written as $\mu=\exp(i\theta)$, with Floquet exponent $\theta$ ({note that it coincides with $-i\lambda$}).

The results were complemented by numerical simulations of the ODE system (6.3), performed using the fourth-order explicit and symplectic Runge-Kutta-Nystr\"om method developed in \cite{Calvo}, with time step equal to $10^{-3}$. In particular, to explore the evolution of unstable waves, we used initial conditions with such waves perturbed along the direction of the Floquet mode causing the instability.

\section{Discussion: connection to existing energy criteria}
\label{sec:discuss}
In this section, we discuss connections between the energy-based stability criterion we derived and analyzed for solitary traveling waves in Hamiltonian lattices and related criteria for such waves in continuum systems and for discrete breathers.
\subsection{Energy criteria for discrete breathers in Hamiltonian lattices}
As we discuseed in Section~4, solitary traveling wave solutions in lattices can be viewed as time-periodic solutions (or discrete breathers,
if they are exponentially localized) modulo an integer shift. In particular, suppose (2.1) has a family of time-periodic solutions $x_{per}(t;\omega)$  parametrized by the frequency $\omega$ and satisfying $x_{per}(t+\frac{1}{\omega};\omega)=x_{per}(t;\omega)$. Setting $\tau=\omega t$ and $X_{per}(\tau; \omega)=x_{per}(t; \omega)$, one can similarly define the linear operator
\[
\mathcal{L}=\frac{1}{c}D\mathcal{A}({X_{per}})-\frac{d}{d\tau}= \frac{1}{c}J \frac{\partial^2\mathcal{H}}{\partial X^2}|_{X=X_{per}}-\frac{d}{d\tau} .
\]
 With $\omega$ playing the role of $c$ for traveling waves, almost all of the results in this work, including Theorem~5.1, Lemma~5.1 and Lemma~\ref{lemma3}, can also be applied to discrete breathers; see~\cite{dmp} for details and a number of examples.

\subsection{Energy criteria for solitary waves in continuum Hamiltonian systems}
Stability of solitary waves in continuum Hamiltonian systems with symmetry (e.g $U(1)$-invariance) was analyzed in \cite{GSS,VK}. Solitary traveling waves in such systems possess translational invariance in space, which breaks down in the case of lattice dynamical systems. However, in what follows, we rewrite the energy-based stability criteria obtained in~\cite{GSS,VK} to identify the similarities with our criterion for solitary traveling waves in lattices.

Following the formulation in~\cite{GSS}, we consider a continuum Hamiltonian system
\begin{equation}
\label{eq0_cont}
\dfrac{\partial u}{\partial t}=J E'(u), \quad u(x,t)=\left(\begin{array}{c}
  {q(x,t)} \\
  {p(x,t)}
\end{array} \right), \quad J=\left(\begin{array}{cc}
0 & I \\
-I & 0
\end{array} \right).
\end{equation}
where $E$ is the energy functional, while $q(x,t)$ and $p(x,t)$ are the displacement and momentum fields, respectively. Let $\mathcal{T}(s)$ be a one-parameter unitary operator group with the infinitesimal generator $\mathcal{T}'(0)$, i.e. $\mathcal{T}(s)={\rm exp}(s\mathcal{T}'(0))$. We assume that $E$ is invariant under $\mathcal{T}(s)$ for any $s$, i.e., $E(u)=E(T(s)u)$. This invariance corresponds to the symmetry of the system. In particular, $\mathcal{T}(s)u(x)=e^{s\partial_x}u(x)=u(x+s)$ corresponds to the translational symmetry.

Assuming that $\mathcal{T}(s)J \mathcal{T}^*(s)=J$ and differentiating it with respect to $s$ at $s=0$ gives $\mathcal{T}'(0)J=-J(\mathcal{T}'(0))^*$, so that $J^{-1}\mathcal{T}'(0)$ is self-adjoint. As a result, there exists a self-adjoint bounded linear operator $\mathcal{B}$ such that $J\mathcal{B}$ extends $\mathcal{T}'(0)$. This implies that $Q(u)=\frac{1}{2}\langle u, \mathcal{B} u \rangle$ is also invariant under $\mathcal{T}(s)$, i.e., $Q(u)=Q(\mathcal{T}(s)u)$.

 Observe that if $\phi_{\omega}(x)$ satisfies $E'(\phi_{\omega})=\omega Q'(\phi_{\omega})$, then $u(x,t)=\mathcal{T}(\omega t)\phi_{\omega}(x)$ is a solitary solution of \eqref{eq0_cont}. In fact, the converse also holds. Indeed, if $u(x,t)=\mathcal{T}(\omega t)\phi_{\omega}(x)$ solves \eqref{eq0_cont}, then
\beq
\begin{split}
J E'(\phi_{\omega})&= J \mathcal{T}(\omega t)^* E'(\mathcal{T}(\omega t)\phi_{\omega})= J \mathcal{T}(\omega t)^* J^{-1}\frac{\partial}{\partial t}(\mathcal{T}(\omega t)\phi_{\omega})\\
&= J \mathcal{T}(\omega t)^* J^{-1}\frac{\partial}{\partial t}(\mathcal{T}(\omega t)\phi_{\omega})=\mathcal{T}(-\omega t) \omega \mathcal{T}'(0) \mathcal{T}(\omega t)\phi_{\omega}\\
&= \omega \mathcal{T}'(0) \phi_{\omega}
= J\omega Q'(\phi_{\omega}).
\end{split}
\eeq
Here we used the fact that $\mathcal{T}'(0)\mathcal{T}(\omega t)=\mathcal{T}(\omega t)\mathcal{T}'(0)$.

We now define the ``free energy'' $d(\omega)=E(\phi_{\omega})-\omega Q(\phi_{\omega})$ and the associated Hessian
$H_{\omega}=E''(\phi_{\omega})-\omega Q''(\phi_{\omega})$.
 Observe that they satisfy
\begin{eqnarray}
\label{eq1_cont}
\begin{split}
& H_{\omega} (\mathcal{T}'(0)\phi_{\omega} )=0, \quad H_{\omega} (\partial_{\omega} \phi_{\omega})=Q'(\phi_{\omega})=J^{-1}\mathcal{T}'(0)\phi_{\omega},\quad
 d'(\omega)=-Q(\phi_{\omega}), \\
& d''(\omega)=-\langle \partial_{\omega}\phi_{\omega}, Q'(\phi_{\omega})\rangle=-\langle \partial_{\omega}\phi_{\omega},  H_{\omega}(\partial_{\omega}\phi_{\omega}) \rangle=-\frac{d}{d\omega}Q(\phi_{\omega})=-\frac{1}{\omega}\frac{d}{d\omega}E(\phi_{\omega}).
\end{split}
\end{eqnarray}
As stated in~\cite{GSS}, if $H_{\omega}$ has at most one negative eigenvalue, $\ker(H_{\omega})$ is spanned by $\mathcal{T}'(0)\phi_{\omega}$ and the rest of its spectrum is bounded below from zero, then $d''(\omega)>0$ implies that $\phi_{\omega}$-orbit $\{\mathcal{T}(\omega t)\phi_{\omega},\, t\in\mathbb{R}\}$ is stable. On the other hand, $d''(\omega)<0$ implies instability. In other words,
\[
d''(\omega)=0=-\frac{1}{\omega}\frac{d}{d\omega}E(\phi_{\omega})
\]
implies the change of stability, which is similar to our energy-based criterion $H'(c_0)=0$ (and to the corresponding one for discrete breathers).

On the one hand, the solutions in our lattice problems are discrete in space and the translation operator $\mathcal{T}(s)=e^{s \partial_x}$
  cannot be directly applied -- except in the quasi-continuum, advance-delay
  variant of the problem. However,  the nature of the traveling waves
  implies that the time dynamics on all sites is connected to a continuous profile function and one can easily {replace} the translation in space {by} translation in time by going to the co-moving frame. In fact, {replacing} $\mathcal{T}(s)=e^{s\partial_x}$ {by} $\mathcal{T}(s)=e^{\frac{1}{\omega} s\partial_t}$, we can similarly define
\[
\begin{split}
Q(u)=\frac{1}{2}\langle u, \mathcal{B}u  \rangle = \frac{1}{2} \langle u,\frac{1}{\omega} J^{-1}\partial_t u  \rangle, \quad
H_{\omega}=E''(\phi_{\omega})-\omega Q''(\phi_{\omega})=E''(\phi_{\omega})-J^{-1}\partial_t
\end{split}
\]
and derive \eqref{eq1_cont} where $u(t)=\mathcal{T}(\omega t)\phi_{\omega}$ is a solitary traveling wave solution of the lattice system with  $E'(\phi_{\omega})=\omega Q'(\phi_{\omega})$.  Note also the relation $\mathcal{L}=\omega J H_{\omega}$, with $\omega=c$ and $x_{tw}(c)=\phi_{\omega}$.

          A more complete
          understanding of the relationship between the energy-based criteria in discrete and continuum systems remains a challenging problem to be considered in the future work.

%\bibliography{refs}

\end{document}